\documentclass[fleqn,useAMS]{Papers}
\usepackage{newtxtext,newtxmath}
\usepackage[T1]{fontenc}
\usepackage{subfloat}
\usepackage{makecell}
\usepackage[square,sort,numbers]{natbib}
\usepackage{graphicx}	
\usepackage{amsmath}	
\usepackage{comment}
\usepackage{xcolor}  
\newcommand{\rev}[1]{{\color{black} #1}} 

\title[Velocity and density distributions of dark matter]{On the statistical theory of self-gravitating collisionless dark matter flow: Scale and redshift variation of velocity and density distributions}

\author[Z. Xu]{Zhijie (Jay) Xu,$^{1}$\thanks{E-mail: \href{mailto:zhijie.xu@pnnl.gov}{zhijie.xu@pnnl.gov}; \href{mailto:zhijiexu@hotmail.com}{zhijiexu@hotmail.com}}
\\
$^{1}$Physical and Computational Sciences Directorate, Pacific Northwest National Laboratory; Richland, WA 99352, USA\\
}

\date{Accepted XXX. Received YYY; in original form ZZZ}
\pubyear{2024}

\begin{document}
\label{firstpage}
\pagerange{\pageref{firstpage}--\pageref{lastpage}}
\maketitle

\begin{abstract}
The statistics of velocity and density fields are crucial for cosmic structure formation and evolution. This paper extends our previous work on the two-point second-order statistics for the velocity field [Phys. Fluids 35, 077105 (2023)] to one-point probability distributions for both density and velocity fields. The scale and redshift variation of density and velocity distributions are studied by a halo-based non-projection approach. First, all particles are divided into halo and out-of-halo particles so that the redshift variation can be studied via generalized kurtosis of distributions for halo and out-of-halo particles, respectively. Second, without projecting particle fields onto a structured grid, the scale variation is analyzed by identifying all particle pairs on different scales $r$. We demonstrate that: i) Delaunay tessellation can be used to reconstruct the density field. The density correlation, spectrum, and dispersion functions were obtained, modeled, and compared with the N-body simulation; ii) the velocity distributions are symmetric on both small and large scales and are non-symmetric with a negative skewness on intermediate scales due to the inverse energy cascade on small scales with a constant rate $\varepsilon_u$; iii) On small scales, the even order moments of pairwise velocity $\Delta u_L$ follow a two-thirds law $\propto{(-\varepsilon_ur)}^{2/3}$, while the odd order moments follow a linear scaling $\langle(\Delta u_L)^{2n+1}\rangle=(2n+1)\langle(\Delta u_L)^{2n}\rangle\langle\Delta u_L\rangle\propto{r}$; iv) The scale variation of the velocity distributions was studied for longitudinal velocities $u_L$ or $u_L^{'}$, pairwise velocity (velocity difference) $\Delta u_L$=$u_L^{'}$-$u_L$ and velocity sum $\Sigma u_L$=$u^{'}_L$+$u_L$. Fully developed velocity fields are never Gaussian on any scale, despite that they can initially be Gaussian; v) On small scales, $u_L$ and $\Sigma u_L$ can be modeled by a $X$ distribution to maximize the entropy of the system. The distribution of $\Delta u_L$ can be different; vi) On large scales, $\Delta u_L$ and $\Sigma u_L$ can be modeled by a logistic or a $X$ distribution, while $u_L$ has a different distribution; vii) the redshift variation of the velocity distributions follows the evolution of the $X$ distribution involving a shape parameter $\alpha(z)$ decreasing with time.
\end{abstract}


\begingroup
\let\clearpage\relax
\tableofcontents
\endgroup

\section{Introduction}
\label{sec:1}
\begin{figure}
\includegraphics*[width=\columnwidth]{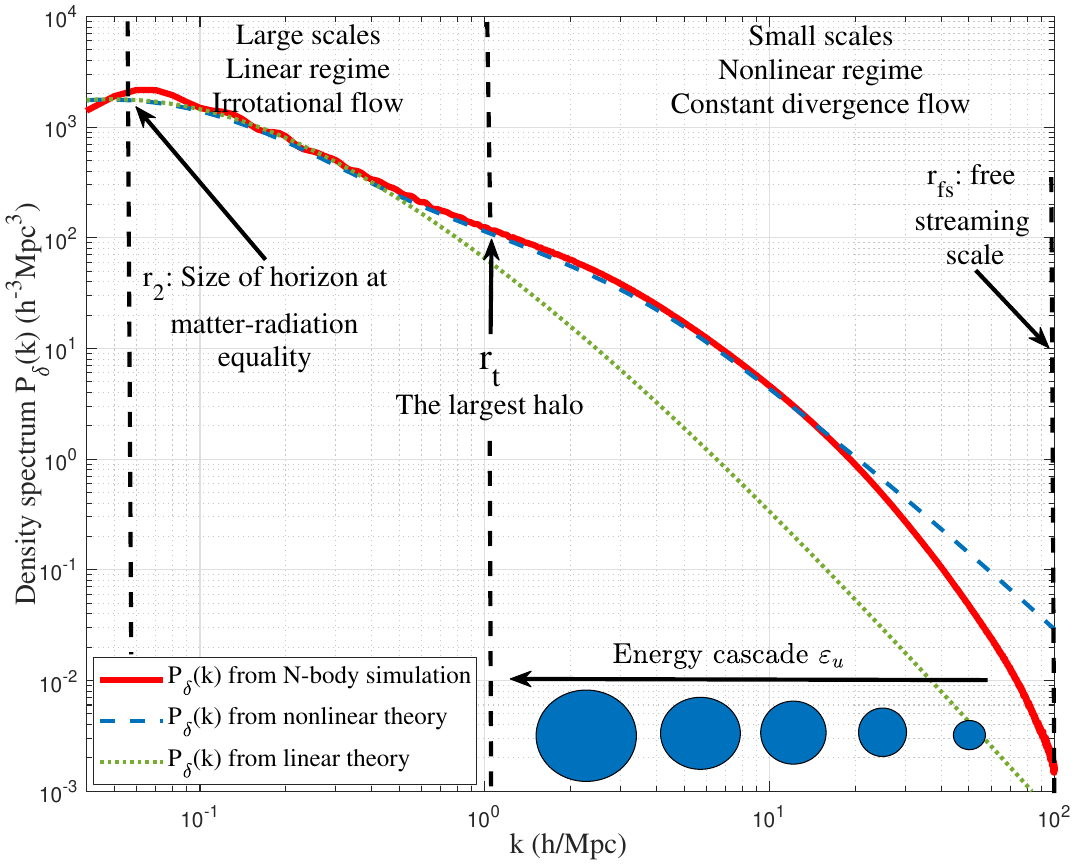}
\caption{The variation of density power spectrum $P_{\delta}(k)$ with comoving wavenumber $k$ at \textit{z}=0 from a N-body simulation. Linear and non-linear theory predictions are also presented for comparison. The pivot wavenumber $k_{\max}$ (or pivot scale $r_2$=21.3 Mpc/h \citep{Xu:2023-On-the-statistical-theory-of-self-gravitating}) denotes the size of the horizon at the matter-radiation equality. The scale $r_t\approx$1Mpc/h is roughly the size of the largest halo. The scale $r_{fs}$ represents the free streaming scales for the smallest haloes that depend on dark matter particle properties. Dark matter flow is irrotational on large scales $r>r_t$ (linear regime) and constant divergence on small scales $r_{fs}<r<r_t$ (nonlinear regime), along with a smooth transition around scale $r_t$ \citep{Xu:2023-On-the-statistical-theory-of-self-gravitating}. On small scales $r<r_t$ (nonlinear regime), there exists a halo-mediated energy cascade with a constant rate of $\varepsilon_u$, which can be used to derive halo mass functions, density profiles \citep{Xu:2023-Dark-matter-halo-mass-functions-and}, and dark matter particle properties \citep{Xu:2022-Postulating-dark-matter-partic}. This paper focuses on the one-point probability distributions of dark matter velocity and density fields and their variation with redshifts and scales (e.g., distributions on small, intermediate, and large scales).}
\label{fig:119}
\end{figure}

Many astronomical observations support the existence of dark matter. In standard $\Lambda$CDM cosmology, the amount of dark matter is about five times that of baryonic matter \citep{Spergel:2003-First-Year-Wilkinson-Microwave-Anisotropy,Komatsu:Seven-year-Wilkinson-Microwave-Anisotropy-Probe, Aghanim:2021-Planck-2018-results--VI--Cosmo}. Therefore, the flow of dark matter has a wide presence in our universe. In contrast to the conventional collisional hydrodynamics, self-gravitating collisionless fluid dynamics (SG-CFD) concerns the dynamics and statistics of the velocity and density fields of collisionless dark matter, which provides valuable information for the large-scale structure formation and constraining cosmological parameters \citep{Ma:2015-Constraining-cosmology-with-pa}. Our previous work mainly focuses on the two-point second-order statistical correlations and kinematic relations for the velocity field of dark matter flow \citep{Xu:2023-On-the-statistical-theory-of-self-gravitating}. Later, high-order two-point statistics and relevant kinematic and dynamic relations were developed in \citep{Xu:2024-On-the-statistical-theory-of-self-gravitating-high-order}. We demonstrate a scale-dependent nature of dark matter flow, e.g., a constant divergence flow on small scales and an irrotational flow on large scales (Fig. \ref{fig:119}). In this paper, the statistical analysis is extended to the one-point probability distributions of velocity and density fields and their scale and redshift variation, e.g., distributions at different scales and redshifts. 

The cosmic velocity and density fields contain rich information for structure formation and evolution. Velocities of galaxies are a robust probe for the search of dark matter on large scales \citep{Courtois:2023-Gravity-in-the-local-Universe}, where statistical analysis can be performed based on galaxy number weighted statistics \citep{Juszkiewicz:1998-Skewed-Exponential-Pairwise-Velocities,Seto:1999-On-the-Statistical-Analyses}. Statistical analysis of velocity fields was also applied to describe the evolution of a system of self-gravitating collisionless particles using BBGKY equations \citep{Davis:1977-Integration-of-Bbgky-Equations}. Pairwise velocity has been introduced to probe the cosmological density parameter \citep{Ferreira:1999-Streaming-velocities-as-a-dyna, Juszkiewicz:2000-Evidence-for-a-low-density-uni}, and the two-point correlation function has been introduced to quantify the cosmic velocity field from the real dataset \citep{Gorski:1988-On-the-Pattern-of-Perturbation, Gorski:1989-Cosmological-Velocity-Correlat}. The density statistics, including the matter density distributions, are important for gravitational lensing and nonlinear clustering. The study of matter density has a long history dating back to the 1930s when Hubble found that the matter distribution is non-Gaussian and can be approximated by a log-normal distribution \citep{Hubble:1934-The-distribution-of-extra-gala}. Interests and efforts are still ongoing both theoretically and numerically \citep{Bernardeau:1995-Properties-of-the-Cosmological, Klypin:2018-Density-distribution-of-the-co}.

In addition, the velocity statistics have profound implications for direct/indirect detection. For predicted DM-nucleon scattering in direct detection \citep{Kuhlen:2010-Dark-matter-direct-detection-w, Ullio:2001-Velocity-distributions-and-ann}, the detection rate of the scattering is proportional to the inverse moment of distribution (or order of -1). That rate is very sensitive to the high-velocity tail of the velocity distribution. For indirect search \citep{Zhao:2018-Constraint-on-the-velocity-dep, Petac:2018-On-velocity-dependent-dark-mat}, the annihilation cross section depends directly on the distribution of relative velocity. The velocity distribution of dark matter particles is expected to be different from Maxwell-Boltzmann. This can be confirmed by N-body simulations \citep{Kazantzidis:2004-Generating-equilibrium-dark-ma, Wojtak:2008-The-distribution-function-of-d} and by our previous work on the maximum entropy distributions in dark matter flow \citep{Xu:2023-Maximum-entropy-distributions-of-dark-matter}. 

Cosmic velocity and density fields exhibit a different nature on different scales. To illustrate this, Figure \ref{fig:119} presents the density power spectrum $P_{\delta}(k)$ as a function of the comoving wavenumber $k$ from a N-body simulation (solid red line) \citep{Frenk:2000-Public-Release-of-N-body-simul}. The corresponding predictions from linear theory (dotted green line) and nonlinear theory (dashed blue line) are also presented \citep{Jenkins:1998-Evolution-of-structure-in-cold}. There are three distinct scales in this figure. The comoving length scale $r_2$ or the pivot wavenumber $k_{max}$ ($k_{max}r_2=\sqrt{2}$ \citep{Xu:2023-On-the-statistical-theory-of-self-gravitating}) is related to the horizon size at the matter-radiation equality, where $k_{max}\propto \Omega_m h^2$ is proportional to the content of matter $\Omega_m$. The length scale $r_t\approx 1Mpc/h$ is roughly the size of the largest halo at $z=0$ \citep{Xu:2023-On-the-statistical-theory-of-self-gravitating}. The linear theory is valid only on large scales $r>r_t$ and underestimates the power spectrum on small scales. The free streaming scale $r_{fs}$ represents the scale of the smallest structure due to the thermal velocity of dark matter particles. This scale depends heavily on the properties and nature of dark matter. A detailed analysis of the free streaming scale is presented in a separate work \citep{Xu:2022-Postulating-dark-matter-partic}. On small scales ($r_{fs}<r<r_t$) in the highly nonlinear regime, haloes have a vanishing (proper) radial flow. The peculiar velocity $\textbf{u}$ satisfies a (spatially) constant divergence or $\nabla \cdot \textbf{u}=-3Ha$, where $H$ is the Hubble parameter and $a$ is the scale factor \citep{Xu:2022-The-mean-flow--velocity-disper}. On large scales ($r>r_t$) in the linear regime, the flow becomes irrotational or $\nabla \times \textbf{u}=0$. The different nature of flow on various scales is a unique feature of the self-gravitating collisionless dark matter flow (SG-CFD) that is different from conventional hydrodynamic turbulence \citep{Xu:2023-On-the-statistical-theory-of-self-gravitating}. In this paper, we focus on the velocity and density distributions on small ($r<r_t$), intermediate ($r\approx r_t$), and large scales ($r>r_t$), respectively. 

Directly measuring velocity and density fields from real samples is still very challenging. However, tremendous information can be obtained from N-body simulations, an invaluable tool for studying the dynamics of collisionless dark matter flow in both linear and non-linear regimes \citep{Angulo:2012-Scaling-relations-for-galaxy-c, Springel:2005-The-cosmological-simulation-co, Peebles:1989-A-Model-for-the-Formation-of-t, Efstathiou:1985-Numerical-Techniques-for-Large}. However, it is not trivial to extract and characterize the statistics of velocity and density fields from N-body simulations. There is a fundamental problem as velocity and density are only sampled at discrete locations of the particle position in N-body simulations. That sampling has poor quality at locations with low particle density \citep{Jennings:2011-Modelling-redshift-space-disto}. The standard approach computes the power spectrum of the velocity and density fields (and gradients) in Fourier space \citep{Hahn:2015-The-properties-of-cosmic-veloc, Pueblas:2009-Generation-of-vorticity-and-ve, Jelic_Cizmek:2018-The-generation-of-vorticity-in}, where cloud-in-cell (CIC) \citep{Hockney:1988-Computer-Simulation-Using-Part} or triangular-shaped-cloud (TSC) schemes are used to project the density and velocity fields onto regular structured grids. This will unavoidably introduce sampling errors in the density and velocity fields \citep{Baugh:1994-A-Comparison-of-the-Evolution-, Baugh:1995-A-Comparison-of-the-Evolution-}. 

Since real-space and Fourier-space data contain the same information, directly working with real-space data can avoid information loss due to field projection and the conversion between Fourier-space and real-space. In this paper, a halo-based non-projection approach is applied for statistical analysis of density and velocity fields: 
\begin{enumerate}
\item \noindent Instead of projecting particle fields onto the structured grid, analysis is performed in real space by computing the statistics (second or high orders) for all particle pairs with a given separation $r$ (shown in Fig. \ref{fig:13}). The scale-dependent nature of these statistics can be studied efficiently as a function of scale $r$. This will maximally preserve and utilize the information contained in N-body simulations; 
\item \noindent Based on the halo description of the N-body system, dark matter particles continuously form halo structures. In this paper, all haloes in the N-body system are identified, and all particles are divided into halo and out-of-halo particles. We focus on relevant statistics for all halo particles and out-of-halo particles, respectively. Since the critical scale $r_t$ stands for the size of the largest haloes, the statistics for all halo particles represent the average statistics on small scales $r<r_t$. While the statistics for out-of-halo particles represent the average statistics on large scales $r>r_t$.
\end{enumerate}

From this practice, a huge amount of knowledge can be learned for self-gravitating collisionless fluid dynamics (SG-CFD) when compared with the isotropic, homogeneous, and incompressible turbulence \citep{Taylor:1935-Statistical-theory-of-turbulan, Taylor:1938-Production-and-dissipation-of-,de_Karman:1938-On-the-statistical-theory-of-i, Batchelor:1953-The-Theory-of-Homogeneous-Turb}. An example is the pairwise velocity ($\Delta u_L=u_L^{'}-u_L$, see Fig. \ref{fig:13}). For incompressible isotropic turbulence, there exists an inertial range of scales with a constant energy flux $\varepsilon _{u}$, followed by a dissipation range dominated by the viscous force due to the viscosity of the liquid. A simple but universal form has been established for the \textit{m}th order longitudinal velocity structure function (or the \textit{m}th moment of pairwise velocity in cosmology terms) in the inertial range \citep{Kolmogorov:1962-A-Refinement-of-Previous-Hypot}. For incompressible isotropic turbulence, the \textit{m}th moment of pairwise velocity reads
\begin{equation}
\label{eq:1}
\begin{split} 
&S_{m}^{lp} \left(r\right)=\left\langle \Delta u_L^{m} \right\rangle = \left\langle \left(u_{L}^{'} -u_{L} \right)^{m} \right\rangle =\beta _{m} \left(\varepsilon _{u} \right)^{{m/3} } r^{{m/3} } ,       \\
&S_{2}^{lp} \left(r\right)=\beta _{2} \varepsilon _{u} r^{{2/3} } \quad \textrm{and} \quad S_{3}^{lp} \left(r\right)=-{4/5} \varepsilon _{u} r,
\end{split}
\end{equation} 
where $u_{L}^{'} $ and $u_{L} $ are longitudinal velocities in Fig. \ref{fig:13}, $\beta _{m} $ is a universal constant, and $\varepsilon _{u} $ is the rate of energy cascade (energy flux across different scales). 
Specifically, for $m=2$, constant $\beta _{2} \approx 2$. This is known as the two-thirds law, where the second-order structure function (or pairwise velocity dispersion in cosmology) satisfies $S_{2}^{lp} \left(r\right)\propto r^{2/3} $. A different scaling $S_{2}^{lp} \left(r\right)\propto r^{2} $ works in the dissipation range where the effects of viscosity are dominant. For $m=3$, $\beta _{3} =-{4/5}$. The third-order structure function satisfies $S_{3}^{lp} \left(r\right)=-{4/5} \varepsilon _{u} r$. This is known as the four-fifths law that can be precisely derived from the Navier-Stokes equation \citep{de_Karman:1938-On-the-statistical-theory-of-i}. Similarly, a different scaling $S_{3}^{lp} \left(r\right)\propto r^{3} $ works in the dissipation range. 

However, the dark matter flow exhibits different behaviors due to its collisionless nature and long-range gravitational interactions, which leads to different scaling laws and behaviors on different scales. Using the halo-based non-projection approach, we can 
\begin{enumerate}
\item \noindent study the scale and redshift variation of velocity and density distributions via the variation of generalized kurtosis (Eq. \eqref{eq:37}); 
\item \noindent demonstrate that velocity fields are non-Gaussian on all scales despite that they can be initially Gaussian (Fig. \ref{fig:15}); 
\item \noindent analytically model velocity distributions on small and large scales, respectively; 
\item \noindent identify a universal two-thirds law for even order structure functions in Fig. \ref{fig:22} and a liner scaling law for odd order structure functions in Fig. \ref{fig:23} (generalized stable clustering hypothesis GSCH); 
\item \noindent obtain the redshift evolution of one-point density and log-density distributions for halo and out-of-halo particles and analytical models for two-point density correlations; 
\end{enumerate}

The paper is organized as follows: Section \ref{sec:2} introduces the N-body simulation, followed by the statistical measures of the velocity field in Section \ref{sec:4}. The redshift and scale dependence of the velocity distributions are presented and modeled in Section \ref{sec:5}, as well as a comparison with N-body simulations. The statistical measures and distributions for the density field are presented and modeled in Section \ref{sec:3}.

\section{N-body simulations and numerical data}
\label{sec:2}
The numerical data was publicly available and generated from the N-body simulations carried out by the Virgo consortium. A comprehensive description of the data can be found in \citep{Frenk:2000-Public-Release-of-N-body-simul, Jenkins:1998-Evolution-of-structure-in-cold}. As a first step, the current study was carried out using simulation runs with $\Omega$=1 and the standard CDM power spectrum (SCDM) that started from \textit{z} = 50 to focus on the matter-dominant self-gravitating flow of collisionless dark matter. Similar analysis can be extended to other simulations with different cosmological assumptions and parameters in the future. The same set of data has been widely used in studies from clustering statistics \citep{Jenkins:1998-Evolution-of-structure-in-cold} to formation of cluster haloes in a large-scale environment \citep{Colberg:1999-Linking-cluster-formation-to-l}, and in the testing of models for halo abundances and mass functions \citep{Sheth:2001-Ellipsoidal-collapse-and-an-im}. Some key numerical parameters of the N-body simulation are listed in Table \ref{tab:1}. 

In this paper, a new approach is applied to directly compute the real-space statistics by pairwise averaging over all particle pairs on a given scale $r$. This approach can maximally employ the information contained in N-body simulation and generate complete statistics on all scales without involving any projection kernels (e.g., CIC). However, it is also computationally intensive to identify all pairs of particles on all scales. The selected N-body simulation has a relatively low resolution compared with other recent cosmological simulations. This enables a computationally affordable direct extraction of real-space two-point statistics. With increasing computing power, the same approach can be similarly extended to other simulations with higher resolution and higher-order statistics.

\begin{table}
\caption{Numerical parameters of SCDM N-body simulation}
\begin{tabular}{p{0.25in}p{0.05in}p{0.05in}p{0.05in}p{0.05in}p{0.05in}p{0.4in}p{0.1in}p{0.35in}p{0.35in}} 
\hline 
Run & $\Omega_{0}$ & $\Lambda$ & $h$ & $\Gamma$ & $\sigma _{8}$ & \makecell{L\\(Mpc/h)} & $N$ & \makecell{$m_{p}$\\$M_{\odot}/h$} & \makecell{$l_{soft}$\\(Kpc/h)} \\ 
\hline 
SCDM1 & 1.0 & 0.0 & 0.5 & 0.5 & 0.51 & \centering 239.5 & $256^{3}$ & 2.27$\times 10^{11}$ & \makecell{\centering 36} \\ 
\hline 
\end{tabular}
\label{tab:1}
\end{table}

\section{Statistical measures of velocity field}
\label{sec:4}

\begin{figure}
\includegraphics*[width=\columnwidth]{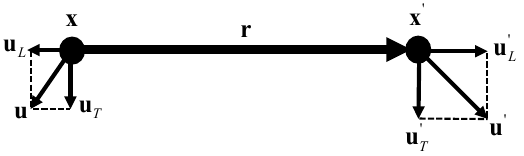}
\caption{Two particles form a pair with a separation of scale $r$. The longitudinal ($u_L$) and transverse ($u_T$) velocities on scale \textit{r} can be calculated from the projection of particle velocity ($\boldsymbol{u}$) on to vector of separation $\boldsymbol{r}$ (Eq. \eqref{ZEqnNum658459}). In this paper, we focus on the scale dependence (r-dependence) of various statistical measures and distributions of the velocity field. }
\label{fig:13}
\end{figure}

To understand how velocity distributions vary with scale \textit{r} and redshift \textit{z}, we are interested in three types of velocities on different scales, i.e. the longitudinal velocity $u_{L} $ or $u_{L}^{'} $, the velocity difference (or pairwise velocity) $\Delta u_{L} =u_{L}^{'} -u_{L} $, and the velocity sum $\Sigma u_{L} =u_{L} +u_{L}^{'}$. In Fig. \ref{fig:13}, for a pair of particles with velocities $\boldsymbol{\mathrm{u}}$ and $\boldsymbol{\mathrm{u}}^{'}$, and a vector of separation $\boldsymbol{\mathrm{r}}=\boldsymbol{\mathrm{x}}^{'} -\boldsymbol{\mathrm{x}}$, the longitudinal and transverse velocities are 

\begin{equation}
\begin{split}
&u_{L} =\boldsymbol{\mathrm{u}}\cdot \hat{\boldsymbol{\mathrm{r}}} \quad \textrm{and} \quad \boldsymbol{\mathrm{u}}_{T} =-\left(\boldsymbol{\mathrm{u}}\times \hat{\boldsymbol{\mathrm{r}}}\times \hat{\boldsymbol{\mathrm{r}}}\right)=\boldsymbol{\mathrm{u}}-\left(\boldsymbol{\mathrm{u}}\cdot \hat{\boldsymbol{\mathrm{r}}}\right)\hat{\boldsymbol{\mathrm{r}}},\\  
&u_{L}^{'} =\boldsymbol{\mathrm{u}}^{'} \cdot \hat{\boldsymbol{\mathrm{r}}} \quad \textrm{and} \quad \boldsymbol{\mathrm{u}}_{T}^{'} =-\left(\boldsymbol{\mathrm{u}}^{'} \times \hat{\boldsymbol{\mathrm{r}}}\times \hat{\boldsymbol{\mathrm{r}}}\right)=\boldsymbol{\mathrm{u}}^{'} -\left(\boldsymbol{\mathrm{u}}^{'} \cdot \hat{\boldsymbol{\mathrm{r}}}\right)\hat{\boldsymbol{\mathrm{r}}}, 
\end{split}
\label{ZEqnNum658459}
\end{equation}
where $\hat{\boldsymbol{\mathrm{r}}}={\boldsymbol{\mathrm{r}}/r}$ is the normalized unit vector.  

For any given scale \textit{r}, all particle pairs with a separation between \textit{r} and \textit{r+dr} ($dr=0.001{Mpc/h} $ in this work) are identified, and the positions and velocities of the particles are recorded. Statistical quantities can be calculated by averaging that quantity on all pairs of particles on the same scale \textit{r}. By this approach, the information contained in N-body simulations is maximally preserved without projecting particle velocity onto a structured grid.

In addition to three types of longitudinal velocities that are scale-dependent ($u_{L}$ or $u_{L}^{'}$, $\Delta u_{L}$, and $\Sigma u_{L}$), we are also interested in the distributions of four velocities based on the halo description. These velocities are the velocity of all particles in the entire system ($\boldsymbol{\mathrm{u}}_{p} $), the velocity of all halo particles ($\boldsymbol{\mathrm{u}}_{hp} $), the velocity of all out-of-halo particles ($\boldsymbol{\mathrm{u}}_{op} $), and the velocity of all haloes identified in the system ($\boldsymbol{\mathrm{u}}_{h}$, the mean velocity of all particles in the same halo). The distributions of these velocities are dependent on the redshift $z$. In addition, the velocity of halo particles represents the velocity on small scales $r<r_t$, while the velocity of out-of-halo particles and haloes represents the velocity on large scales $r>r_t$ (Fig. \ref{fig:119}). The redshift and scale variation of these distributions will significantly improve our understanding of dark matter flow.  

\subsection{Pair conservation equation and algorithm validation}
\label{sec:4.3-2}
To validate the algorithm that identifies all particle pairs with a given separation $r$, we compare the mean pairwise velocity $\left\langle\Delta u_{L}\right\rangle$ to the pair conservation equation that relates the mean pairwise velocity with the density correlation \citep{Peebles:1980-The-Large-Scale-Structure-of-t}, 
\begin{equation} 
\label{ZEqnNum143209} 
\frac{\left\langle \Delta u_{L} \right\rangle }{Har} =-\frac{\left(1+\bar{\xi }\left(r,a\right)\right)}{3\left(1+\xi \left(r,a\right)\right)} \frac{\partial \ln \left(1+\bar{\xi }\left(r,a\right)\right)}{\partial \ln a},       
\end{equation} 
where $\bar{\xi }\left(r,a\right)=3r^{-3} \int _{0}^{r}\xi \left(y,a\right) y^{2} dy$ is the volume averaged density correlation (Eq. \eqref{ZEqnNum197404}). On large scales in the linear regime, $\bar{\xi }\ll 1$ and ${\partial \ln \bar{\xi }/\partial \ln a} =2$, Eq. \eqref{ZEqnNum143209} reduces to
\begin{equation} 
\label{ZEqnNum538076} 
\frac{\left\langle \Delta u_{L} \right\rangle }{Har} =-\frac{2\bar{\xi }\left(r,a\right)\left(1+\bar{\xi }\left(r,a\right)\right)}{3\left(1+\xi \left(r,a\right)\right)} \approx -\frac{2}{3} \bar{\xi }\left(r,a\right).      
\end{equation} 
On small scales in the nonlinear regime with $\bar{\xi }\gg 1$ and assuming the scaling with scale factor as $\xi \left(r,a\right)\propto a^{\alpha } $ (Fig. \ref{fig:9}) and ${\partial \ln \bar{\xi }/\partial \ln a} =\alpha$, the pair conservation Eq. \eqref{ZEqnNum143209} reduces to
\begin{equation} 
\label{ZEqnNum464521} 
\frac{\left\langle \Delta u_{L} \right\rangle }{Har} =-\frac{\alpha \left(1+\bar{\xi }\left(r,a\right)\right)}{3\left(1+\xi \left(r,a\right)\right)} .          
\end{equation} 
On small scales, if stable clustering hypothesis ($\left\langle \Delta u_{L} \right\rangle =-Har$) (demonstrated in \citep{Xu:2021-A-non-radial-two-body-collapse}) is assumed and considering a self-similar gravitational clustering $\xi \left(r,a\right)\propto a^{\alpha } r^{\gamma } $, we have
\begin{equation} 
\label{eq:44} 
\frac{\left\langle \Delta u_{L} \right\rangle }{Har} =-1 \quad \textrm{and} \quad \alpha =\gamma +3,    
\end{equation}
where the exponents $\alpha$ and $\gamma$ are related to each other.

\begin{figure}
\includegraphics*[width=\columnwidth]{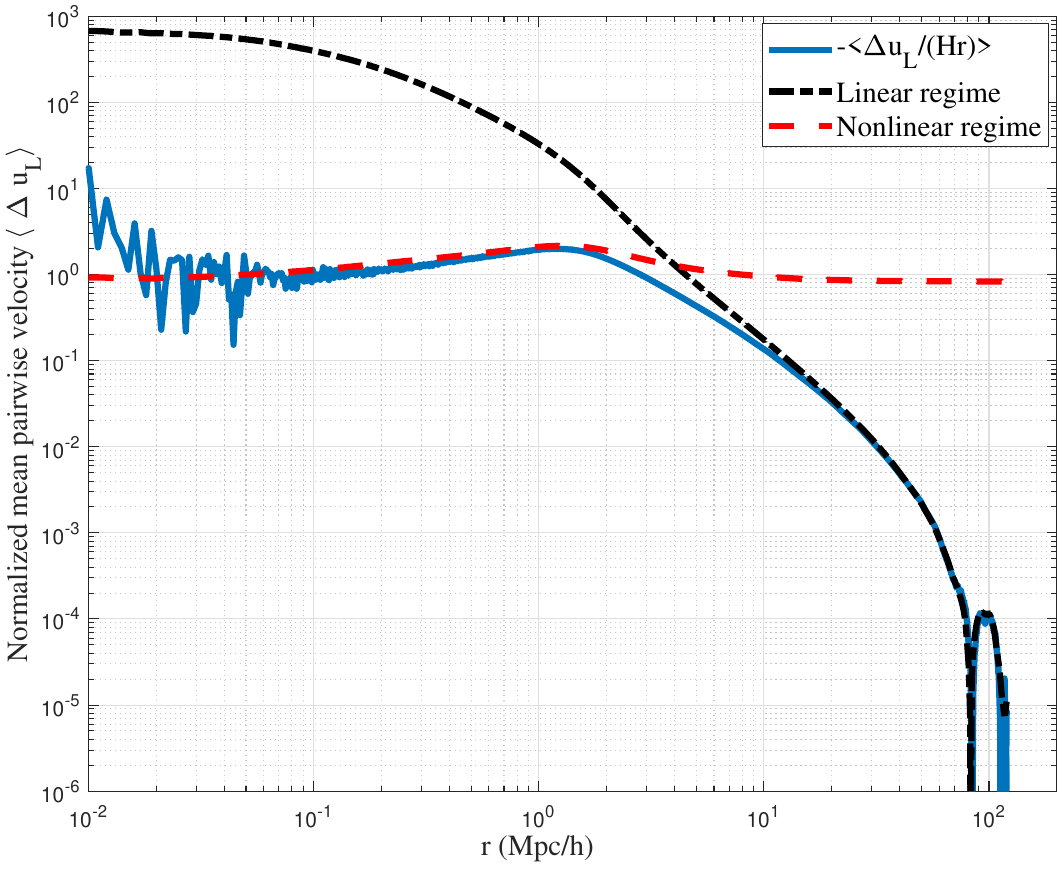}
\caption{The variation of mean pairwise velocity $\langle \Delta u_{L}\rangle$ with scale \textit{r} at \textit{z}=0 (normalized by Hubble constant \textit{H}) from N-body simulation (blue solid). Results are compared with predictions of pair conservation equation for both linear (black dashed dashed from Eq. \eqref{ZEqnNum538076}) and nonlinear regime (red dash from Eq. \eqref{ZEqnNum464521}). Predictions are made with the density correlation $\xi \left(r\right)$ obtained from the same N-body simulation.}
\label{fig:17}
\end{figure}

Figure \ref{fig:17} plots the variation of the mean pairwise velocity $\langle \Delta u_{L}\rangle$ with scale \textit{r} (normalized by the Hubble constant) at \textit{z}=0. The results are compared with the prediction of the pair conservation equation for both linear (black dashed line from Eq. \eqref{ZEqnNum538076}) and nonlinear regime (red dashed line from Eq. \eqref{ZEqnNum464521} with $\alpha ={5/2} $ from Fig. \ref{fig:9}) using the density correlation $\xi \left(r,a\right)$ obtained from the N-body simulation (Fig. \ref{fig:8}). The blue line is the normalized pairwise velocity computed directly from the N-body simulation by identifying all particle pairs and associated velocities. A good match with the pair conservation equation validates our numerical implementation to identify all pairs of particles on any fixed scale $r$. 

\subsection{Generalized kurtosis, moments, and structure functions}
\label{sec:4.1}
The velocity distributions can be best characterized by dimensionless generalized kurtosis. An example is the generalized kurtosis for the distribution of velocity difference $\Delta u_{L} $ (or pairwise velocity). On any scale of $r$, the generalized kurtosis reads
\begin{equation}
\label{eq:37} 
K_{n} \left(\Delta u_{L} ,r\right)=\frac{\left\langle \left(\Delta u_{L} -\left\langle \Delta u_{L} \right\rangle \right)^{n} \right\rangle }{\left\langle \left(\Delta u_{L} -\left\langle \Delta u_{L} \right\rangle \right)^{2} \right\rangle ^{{n/2} } } =\frac{S_{n}^{cp} \left(\Delta u_{L} ,r\right)}{S_{2}^{cp} \left(\Delta u_{L} ,r\right)^{{n/2} } } ,      
\end{equation} 
where the central moment of order \textit{n} for $\Delta u_{L} $ reads
\begin{equation} 
\label{eq:38} 
S_{n}^{cp} \left(\Delta u_{L} ,r\right)=\left\langle \left(\Delta u_{L} -\left\langle \Delta u_{L} \right\rangle \right)^{n} \right\rangle.         
\end{equation} 
On the same scale $r$, the \textit{n}th order longitudinal structure function of $\Delta u_{L} $ is defined as the $n$th order moment 
\begin{equation} 
\label{ZEqnNum250774} 
S_{n}^{lp} \left(r\right)=\left\langle \left(\Delta u_{L} \right)^{n} \right\rangle =\left\langle \left(u_{L}^{'} -u_{L} \right)^{n} \right\rangle.        
\end{equation} 
Specifically, the first order structure function is just the mean pairwise velocity $\langle \Delta u_{L}\rangle$.

\textbf{Remarks:} For incompressible hydrodynamics, the mean velocities $\langle u_{L}\rangle =\langle \Delta u_{L}\rangle =\langle \Sigma u_{L}\rangle =0$ on all scales of \textit{r} such that $S_{n}^{cp} (\Delta u_{L} ,r)=S_{n}^{lp} (r)$. The central moment of $\Delta u_{L} $ equals the structure function defined in Eq. \eqref{ZEqnNum250774}. However, for self-gravitating collisionless dark matter flow (SG-CFD), two particles tend to approach each other under gravity that leads to a non-zero mean longitudinal velocity $\langle u_{L}\rangle ={-\langle \Delta u_{L}\rangle /2} >0$. Therefore, the central moment $S_{n}^{cp} (\Delta u_{L} ,r)\ne S_{n}^{lp} (r)$ for SG-CFD. The distributions of longitudinal velocity $u_{L}$ and pairwise velocity $\Delta u_{L}$ are symmetric on small and large scales ($\langle\Delta u_{L}\rangle=0$ on scales $r\ll r_t$ and $r\gg r_t$). Distributions of $\Delta u_{L}$ and $u_L$ can be asymmetric with non-vanishing odd-order moments on intermediate scales $r\approx r_t$ (Fig. \ref{fig:26}). Due to symmetry, the mean $\langle \Sigma u_{L}\rangle =0$. The distribution of $\Sigma u_{L}$ is always symmetric on any scale \textit{r} with vanishing odd-order moments. 

\subsection{Generalized kurtosis from N-body simulation}
\label{sec:4.2}
\begin{figure}
\includegraphics*[width=\columnwidth]{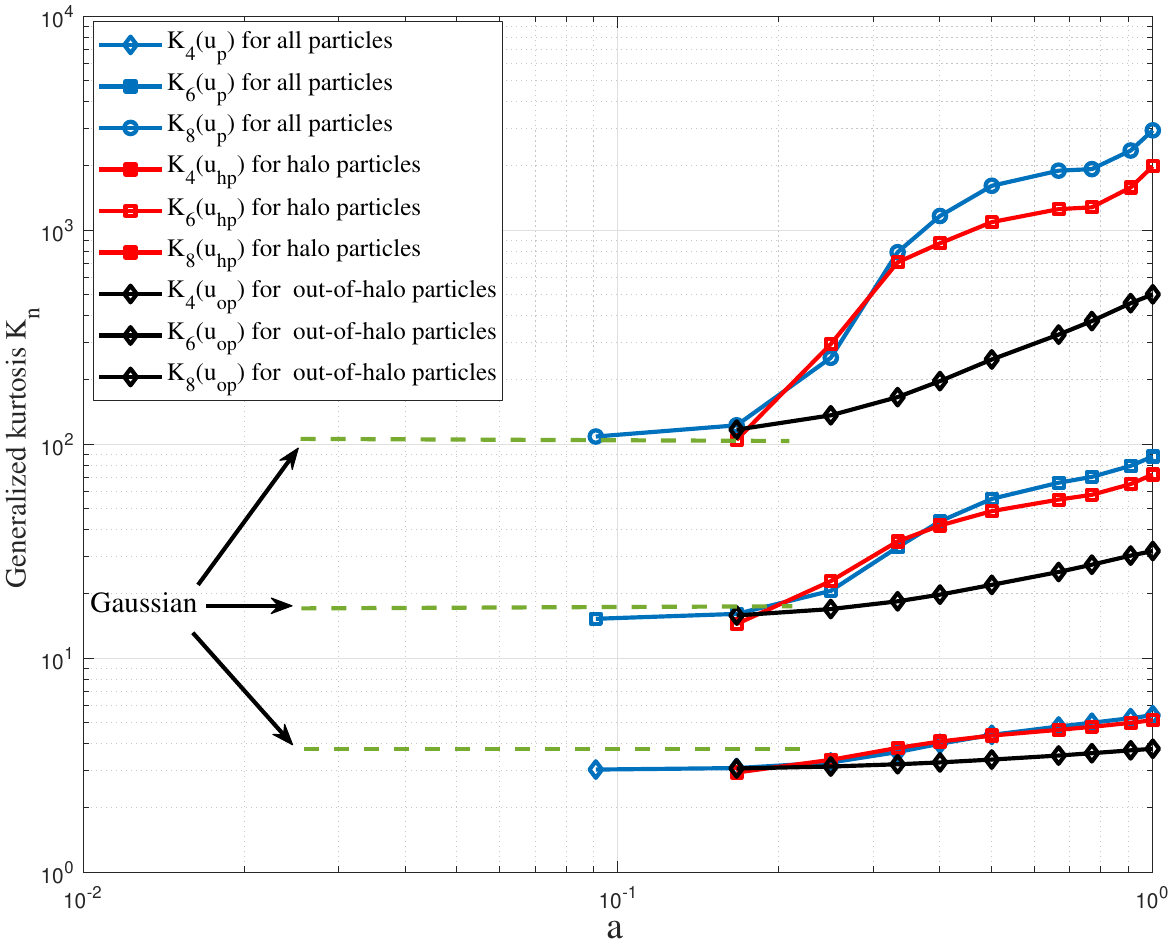}
\caption{The variation of generalized kurtosis (even order) with scale factor $a$ for the velocity of all particles ($u_p$: blue), all halo particles ($u_{hp}$: red), and all out-of-halo particles ($u_{op}$: black). Gaussian distribution is presented as green dashed lines. Since the Gaussian distribution is not the maximum entropy distribution of dark matter flow \citep{Xu:2023-Maximum-entropy-distributions-of-dark-matter}, all velocities are initially Gaussian and quickly become non-Gaussian with increasing kurtosis to maximize entropy. The evolution of the distribution of the out-of-halo particle velocity is much slower than that of the halo particles due to weak gravity on large scales.}
\label{fig:14}
\end{figure}

Figure \ref{fig:14} presents the time variation of generalized kurtosis for velocity $\boldsymbol{\mathrm{u}}_{p}$ (for all particles), $\boldsymbol{\mathrm{u}}_{hp}$ (for halo particles) and $\boldsymbol{\mathrm{u}}_{op} $ (for out-of-halo particles). Different orders of kurtosis for the Gaussian distribution are plotted as dashed green lines for comparison. All velocities are initially Gaussian. The distribution of the halo particle velocity $\boldsymbol{\mathrm{u}}_{hp} $ deviates from the Gaussian much faster than the distribution of the out-of-halo particle velocity $\boldsymbol{\mathrm{u}}_{op} $ due to much stronger gravitational interactions in the haloes than the gravity between the haloes. All velocities become non-Gaussian with time to maximize the entropy of the system \citep{Xu:2023-Maximum-entropy-distributions-of-dark-matter}. 
 
\begin{figure}
\includegraphics*[width=\columnwidth]{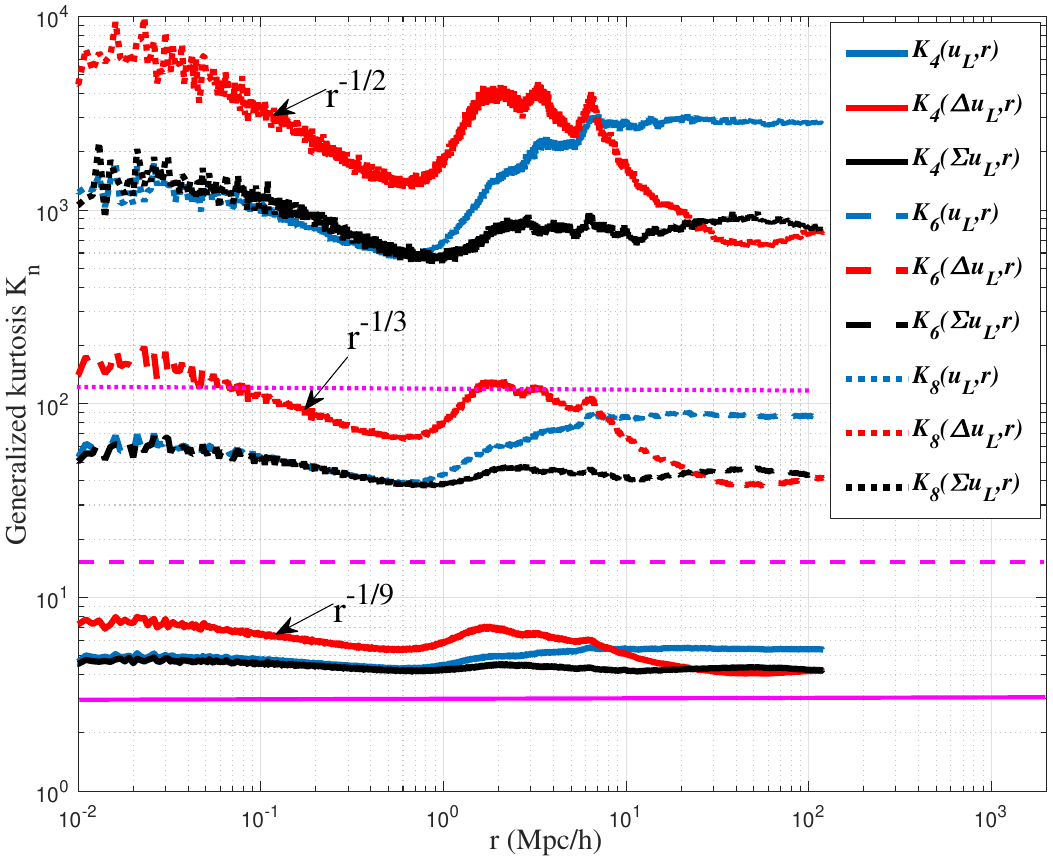}
\caption{The even order generalized kurtosis (4${}^{th}$, 6${}^{th}$, and 8${}^{th}$ order) of three velocities varying with scale \textit{r} at \textit{z}=0. The generalized kurtosis of Gaussian distribution is plotted in magenta for comparison. All velocity distributions are non-Gaussian on all scales due to the long-range gravitational interaction, despite the fact that they can be initially Gaussian. The distribution of $\Sigma u_{L}$ approaches that of $u_{L}$ on small scales, while the distribution of $\Sigma u_{L}$ approaches that of $\Delta u_{L}$ on large scales. There exist limiting probability distributions for all velocities on both small and large scales.} 
\label{fig:15}
\end{figure}

Figure \ref{fig:15} plots the even-order generalized kurtosis (4${}^{th}$ order -- bottom, 6${}^{th}$ order -- middle, and 8${}^{th}$ order -- top) of three velocities ($u_{L} $, $\Delta u_{L}$ and $\Sigma u_{L}$) at \textit{z}=0. The 4${}^{th}$, 6${}^{th}$, and 8${}^{th}$ order kurtosis of the Gaussian distribution (magenta) is also plotted in the same figure with $K_{4} =3$, $K_{6} =15$, and $K_{8} =105$. Clearly, distributions of three velocities are non-Gaussian on all scales due to the long-range nature of gravity. This is important as it poses serious challenges to any theory that assumes the Gaussianity of velocity fields. The velocity field of fully developed self-gravitating collisionless dark matter flow (SG-CFD) is non-Gaussian on any scale, despite the fact that they can initially be Gaussian. In contrast, for incompressible hydrodynamics with short-range interactions, the distribution of velocity $u_{L}$ is nearly Gaussian on large scales. Distributions of $\Delta u_{L}$ are also Gaussian on large scales and only become non-Gaussian in the dissipation range because of the viscous force (Table \ref{tab:3}). 

In Fig. \ref{fig:15}, the distribution of $\Sigma u_{L}$ approaches the distribution of $u_{L}$ on small scales with a limiting correlation (between $u_{L}$ and $u_{L}^{'}$), where $\rho _{L} = \langle u_{L}^{'}u_{L}\rangle /\langle u_{L}^{2}\rangle =0.5$ between two velocities \citep[see ref.][Fig. 17]{Xu:2023-On-the-statistical-theory-of-self-gravitating}. As $r\to 0$, and the sum velocity ${\mathop{\lim }\limits_{r\to 0}} \Sigma u_{L} ={\mathop{\lim }\limits_{r\to 0}} (u_{L}^{'}+u_{L})$ will become the total velocity $\boldsymbol{\mathrm{u}}$ at location $x$. Longitudinal velocities $u_{L}$ and $u_{L}^{'} $ along many different directions will simply collapse into velocity $\boldsymbol{\mathrm{u}}$, and this also requires $\rho_{L}=0.5$, i.e.
\begin{equation} 
\label{ZEqnNum571209} 
\begin{split}
{\mathop{\lim }\limits_{r\to 0}} \left\langle \left(u_{L}^{'} +u_{L} \right)^{2} \right\rangle &={\mathop{\lim }\limits_{r\to 0}} \left(\langle u_{L}^{'2} \rangle+ \langle u_{L}^{2} \rangle+2 \langle u_{L}^{'}u_{L}\rangle\right)\\
&={\mathop{\lim }\limits_{r\to 0}} \left|\boldsymbol{\mathrm{u}}\left(\boldsymbol{\mathrm{x}}\right)\right|^{2} =3{\mathop{\lim }\limits_{r\to 0}} \left\langle u_{L}^{2} \right\rangle.
\end{split}
\end{equation} 

On large scales, the distribution of $\Sigma u_{L}$ approaches the distribution of $\Delta u_{L}$ with correlation $\rho _{L}=0$ between $u_{L}$ and $u_{L}^{'}$. This is also expected, as the sum and difference of two independent random variables with symmetric distributions should follow the same distribution. Finally, on both small and large scales, generalized kurtosis approaches a constant so that there exist unique (limiting) probability distributions that are independent of scale \textit{r} when $r\to 0$ or $r\to \infty $. On an intermediate scale of around $r_t$=1Mpc/h, all three velocity distributions exhibit the greatest value of generalized kurtosis. The velocity distributions on small, intermediate, and large scales from simulation and theory are presented in Section \ref{sec:5}. 

\begin{figure}
\includegraphics*[width=\columnwidth]{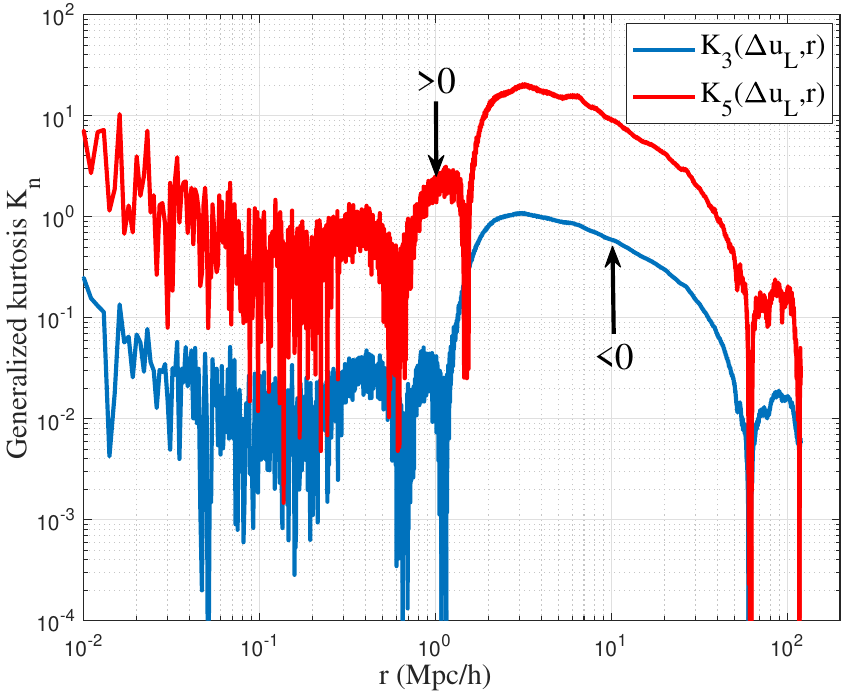}
\caption{The odd-order generalized kurtosis of pairwise velocity $\Delta u_{L}$varying with scale \textit{r} at \textit{z}=0. The third order kurtosis $K_{3}(\Delta u_{L},r)$ (skewness) vanishes on both small and large scales, where the distribution of $\Delta u_{L}$is symmetric. The skewness $K_{3}(\Delta u_{L},r)<0$ on the intermediate scale $r\approx r_t$ (distribution skews toward positive side). This negative skewness on the intermediate scale should be a result of an inverse cascade of kinetic energy from small to large scales up to the scale $r_t$ of the largest haloes. The energy cascade on small scales $r<r_t$ \citep{Xu:2021-Inverse-mass-cascade-mass-function} might also lead to the negative "effective" viscosity on large scales $r>r_t$ \citep{Xu:2024-On-the-statistical-theory-of-self-gravitating-high-order}.} 
\label{fig:16}
\end{figure}

Figure \ref{fig:16} plots the scale variation of odd-order generalized kurtosis ($K_{3}$ and $K_{5}$) with scale \textit{r} at \textit{z}=0 for pairwise velocity $\Delta u_{L}$. Odd-order kurtosis vanishes on both small and large scales, where the distribution of $\Delta u_{L}$ is symmetric. The skewness $K_{3}(\Delta u_{L},r)<0$ on the intermediate scale $r=r_t$ (the distribution of $\Delta u_{L}$ skews toward the positive side; see Fig. \ref{fig:26}). Negative skewness is an important signature of the inverse cascade of kinetic energy on small scales $r<r_t$ (Fig. \ref{fig:119}) that might lead to the negative "effective" viscosity on large scales $r>r_t$ \citep{Xu:2024-On-the-statistical-theory-of-self-gravitating-high-order}. For hydrodynamic turbulence, the negative skewness takes place in the dissipation range, where energy is cascaded from large to small scales and destroyed by viscosity.

\subsection{First order moment of velocity}
\label{sec:4.3}
While generalized kurtosis can be used to characterize the distributions of different velocities, the moments of velocity distributions can be studied in detail to provide more insight. Due to symmetry, the first-order moment of the velocity sum $\langle \Sigma u_{L}\rangle =0$ vanishes on all scales. This section focuses on the first-order moment of pairwise velocity, e.g., the first-order structure function $S_1^{lp}(r)=\langle \Delta u_{L}\rangle$ in Eq. \eqref{ZEqnNum250774}. On small scales, an exact expression can be identified from the stable clustering hypothesis, 
\begin{equation}
\left\langle \Delta u_{L} \right\rangle =-Har  \quad \textrm{and} \quad  \left\langle u_{L} \right\rangle ={Har/2}.  
\label{eq:45}
\end{equation}
For nonlinear regime below the critical scale $r_{t}=1.3a^{{1/2}}${Mpc/h} where the longitudinal velocity correlation equals the transverse velocity correlation \citep[see ref.][Figs. 3, 4, and 5]{Xu:2023-On-the-statistical-theory-of-self-gravitating}, a better relation to fit the simulation data reads
\begin{equation} 
\label{ZEqnNum115620} 
\left\langle \Delta u_{L} \right\rangle =-Har-ua^{-{5/3} } \left(\frac{r}{r_{t} } \right)^{{5/2} },         
\end{equation} 
where $u(a)$ is one-dimension velocity dispersion in Table \ref{tab:2}. 

On large scales, from the pair conservation Eq. \eqref{ZEqnNum538076}, the mean pairwise velocity can be written as
\begin{equation} 
\label{eq:47)} 
\left\langle \Delta u_{L} \right\rangle \approx -\frac{2}{3} Har\bar{\xi }\left(r,a\right)=-\frac{2Ha}{r^{2} } \int _{0}^{r}\xi \left(y\right)y^{2}  dy.      
\end{equation} 
With Eq. \eqref{ZEqnNum197404} for mean correlation $\bar{\xi }\left(r,a\right)$, the mean pairwise velocity is simply the derivative of the velocity correlation $R_{2} =\langle \boldsymbol{\mathrm{u}}\cdot \boldsymbol{\mathrm{u}}_{}^{'} \rangle $ \citep[see ref.] [Eq. (124)]{Xu:2023-On-the-statistical-theory-of-self-gravitating}. Therefore, we should have
\begin{equation} 
\label{ZEqnNum164953} 
\left\langle \Delta u_{L} \right\rangle =\frac{2}{aHf\left(\Omega _{m} \right)} \frac{\partial R_{2} }{\partial r} =\frac{2a_{0} u^{2} }{aHr_{2} } \exp \left(-\frac{r}{r_{2} } \right)\left(\frac{r}{r_{2} } -4\right).      
\end{equation} 

\begin{figure}
\includegraphics*[width=\columnwidth]{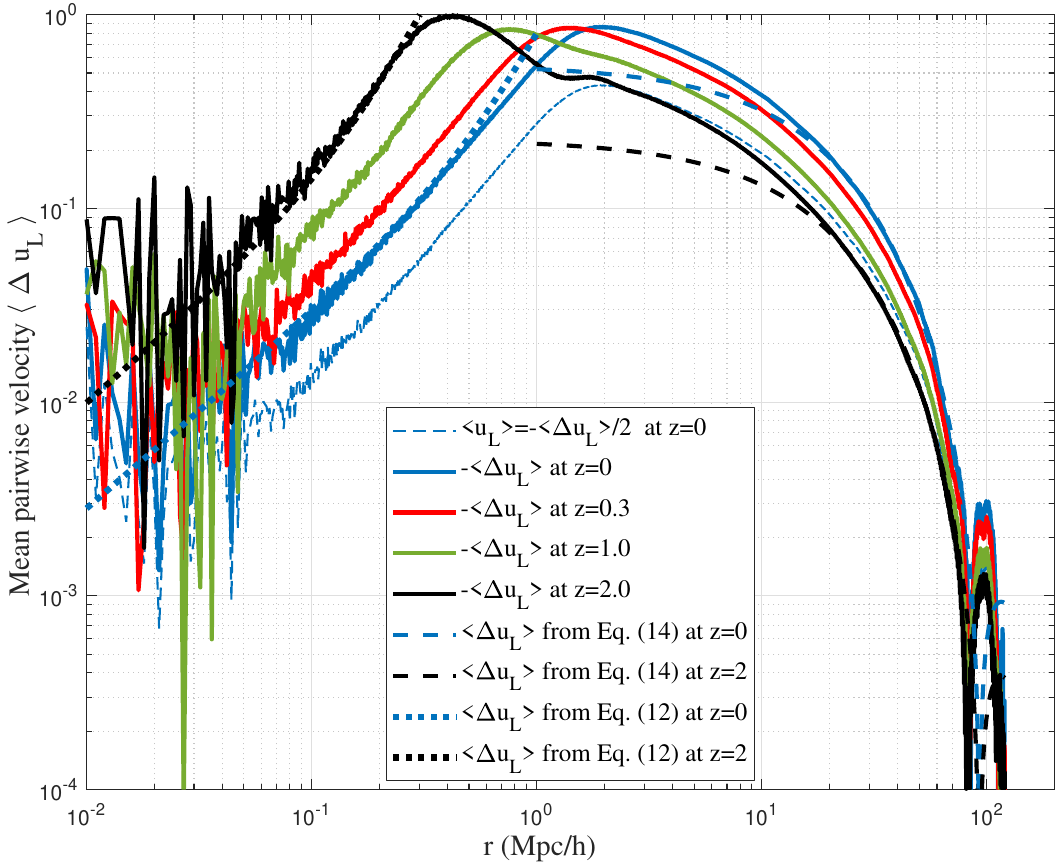}
\caption{The variation of the mean (or first-order moment) longitudinal velocity $\langle u_{L} \rangle $ and mean pairwise velocity $|\langle \Delta u_{L}\rangle|$ with scale \textit{r} at different redshifts \textit{z}, normalized by velocity dispersion $u(a)$ in Table \ref{tab:2}. Note that $\langle \Delta u_{L}\rangle =-2\langle u_{L}\rangle $ does not vanish on intermediate scale $r_t$ and approaches zero on both small and large scales. The longitudinal velocity $\langle u_{L}\rangle =-\langle u_{L}^{'}\rangle$ and the velocity sum $\langle \Sigma u_{L}\rangle =0$ on all scales. For SG-CFD, the velocity field \textbf{u} and vector \textbf{r} between two particles are correlated leading to a nonzero longitudinal velocity $\langle u_{L}\rangle =\langle \boldsymbol{\mathrm{u}}\cdot \boldsymbol{\mathrm{r}}\rangle >0$ on all scales.} 
 \label{fig:18}
\end{figure}

Figure \ref{fig:18} plots the mean longitudinal velocity $\langle u_{L}\rangle $ and pairwise velocity $-\langle \Delta u_{L}\rangle $ at different redshift \textit{z}. Note that $\langle \Delta u_{L}\rangle =-2\langle u_{L}\rangle $ vanishes on both small and large scales. Since $\langle u_{L}\rangle =-\langle u_{L}^{'} \rangle $, mean velocity sum $\langle \Sigma u_{L}\rangle =0$ on all scales. Here, $\langle u_{L}\rangle >0$ and $\langle \Delta u_{L}\rangle <0$ reflect that two particles are moving toward each other due to gravity. In contrast, $\langle u_{L}\rangle =\langle \Delta u_{L}\rangle =\langle \Sigma u_{L}\rangle =0$ on all scales for incompressible collisional hydrodynamics, where \textbf{u} and \textbf{r} are independent of each other. Models for pairwise velocity on small and large scales (Eqs. \eqref{ZEqnNum115620} and \eqref{ZEqnNum164953}) are also presented for comparison.

\subsection{Second order moment of velocity}
\label{sec:4.4}

\begin{figure}
\includegraphics*[width=\columnwidth]{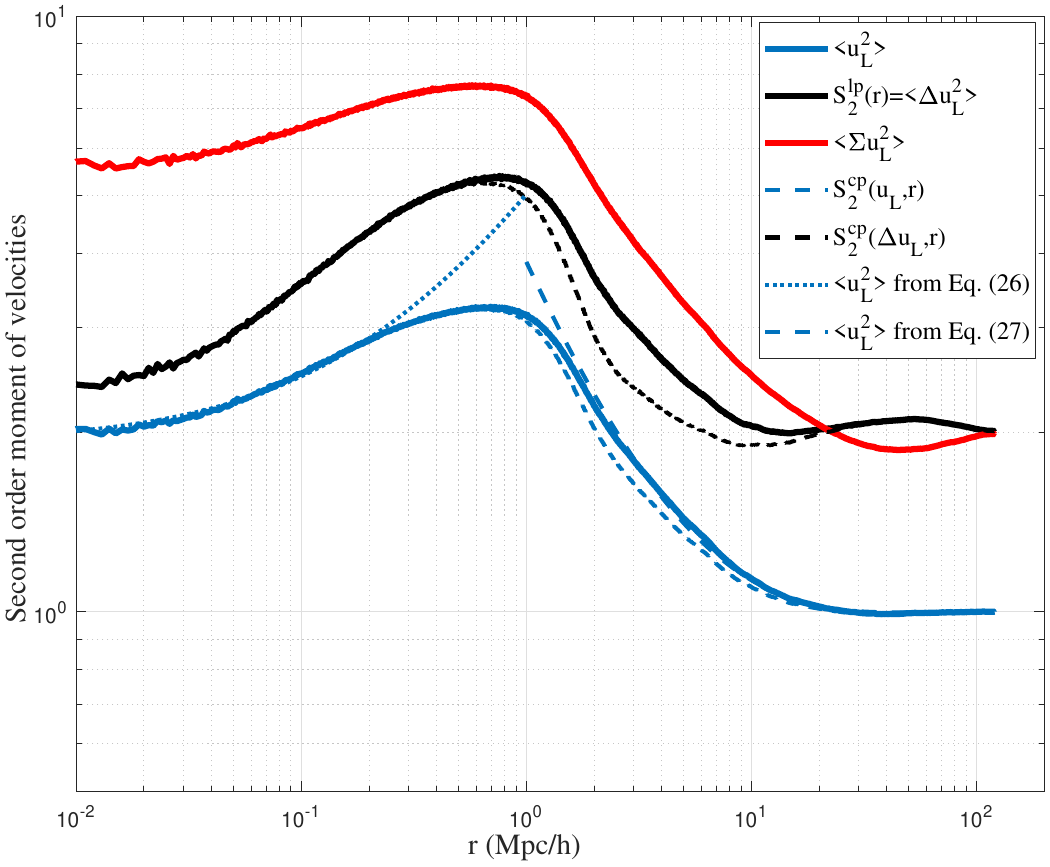}
\caption{The variation of second-order moment $\langle u_{L}^{2}\rangle $, $\langle \Delta u_{L}^{2}\rangle $ and $\langle \Sigma u_{L}^{2}\rangle $ with scale $r$ at \textit{z}=0, normalized by velocity dispersion $u_{0}^{2} $ of entire system in Table \ref{tab:2}. On small scales, $\langle \Delta u_{L}^{2}\rangle =\langle u_{L}^{2}\rangle ={\langle \Sigma u_{L}^{2}\rangle /3} =2u^{2} $, while on large scales $\langle \Delta u_{L}^{2} \rangle =\langle \Sigma u_{L}^{2} \rangle =2\langle u_{L}^{2} \rangle =2u^{2} $. The difference between the second-order longitudinal structure function $S_{2}^{lp} (r)=\langle \Delta u_{L}^{2}\rangle $ and the central moment $S_{2}^{cp}(\Delta u_{L},r)$ is due to the nonzero $\langle \Delta u_{L}\rangle $ on intermediate scales (Eq. \eqref{eq:38}). On the contrary, $S_{2}^{lp} =S_{2}^{cp} $ and $\langle u_{L}^{2}\rangle =u^{2}$ on all scales for incompressible hydrodynamics. Models for longitudinal velocity dispersion $\langle u_{L}^{2}\rangle$ in Eq. \eqref{ZEqnNum864517} (small scales) and Eq. \eqref{ZEqnNum864804} (large scales) are also plotted.} 
\label{fig:19}
\end{figure}

Figure \ref{fig:19} plots the second-order moments and the central moments (normalized by $u_0^{2}$) of velocities $u_{L} $, $\Delta u_{L}$, and $\sum u_{L} $ on all scales at \textit{z}=0. Longitudinal velocities ($u_{L}$ and $u_{L}^{'} $) must be strongly correlated on small scales due to gravitational interaction and uncorrelated on large scales. The correlation between $u_{L}$ and $u_{L}^{'}$ leads to
\begin{equation}
\left\langle \Delta u_{L}^{2} \right\rangle =2\left\langle u_{L}^{2} \right\rangle \left(1-\rho _{L} \right), \quad \left\langle \Sigma u_{L}^{2} \right\rangle =2\left\langle u_{L}^{2} \right\rangle \left(1+\rho _{L} \right),
\label{49}
\end{equation}
\noindent where $\rho _{L} $ is the correlation coefficient. For small $r$ and on small scales, $\rho _{L} ={1/2} $ for $r\to 0$ \citep[see ref.][Fig. 17]{Xu:2023-On-the-statistical-theory-of-self-gravitating}. Longitudinal velocities of particle pairs in small haloes are fully correlated with $\rho _{L}\rightarrow 1$, while the longitudinal velocities of particle pairs in large haloes are uncorrelated with $\rho _{L}\rightarrow 0$. Therefore, the average correlation is roughly 1/2 \citep[see ref.][Fig. 17]{Xu:2023-On-the-statistical-theory-of-self-gravitating}. From Eq. \eqref{49}, on small scales with $\rho _{L} = 1/2$, we should have
\begin{equation} 
\label{50} 
\left\langle \Delta u_{L}^{2} \right\rangle =\left\langle u_{L}^{2} \right\rangle ={\left\langle \Sigma u_{L}^{2} \right\rangle /3} =2u^{2} .         
\end{equation} 
On large scales with $\rho _{L} =0$ when $r\to \infty $, 
\begin{equation} 
\label{ZEqnNum388521} 
\left\langle \Delta u_{L}^{2} \right\rangle =\left\langle \Sigma u_{L}^{2} \right\rangle =2\left\langle u_{L}^{2} \right\rangle =2u^{2} .        
\end{equation} 
By contrast, for incompressible hydrodynamics, we have $\langle \Delta u_{L}^{2} \rangle =0$ and $\langle \Sigma u_{L}^{2} \rangle =4u^2$ on small scale with $\rho _{L} =1$ when $r\to 0$, and $\langle u_{L}^{2} \rangle =u^{2} $ on all scales (See comparison in Table \ref{tab:3}). 

The difference between the second-order moments and the central moments of $u_{L} $ and $\Delta u_{L} $ on intermediate scales is due to the nonzero first moments $\langle u_{L} \rangle $ and $\langle \Delta u_{L} \rangle $, as shown in Fig. \ref{fig:19} and Eq. \eqref{eq:38}. All second-order moments increase with \textit{r} initially and decrease when $r>r_{t} $. The models for the second-order moment $\langle u_{L}^{2} \rangle $ on small and large scales are presented in Eqs. \eqref{ZEqnNum864517} and \eqref{ZEqnNum864804}.  

Identifying all pairs of particles with different separation \textit{r}, we can compute the velocity variance on different scales \textit{r}, namely the total variance $\langle u^{2} \rangle =\langle \boldsymbol{\mathrm{u}}\cdot \boldsymbol{\mathrm{u}}\rangle $, the longitudinal variance $\langle u_{L}^{2} \rangle $ and the transverse variance $\langle u_{T}^{2} \rangle =\langle \boldsymbol{\mathrm{u}}_{T} \cdot \boldsymbol{\mathrm{u}}_{T} \rangle $, where  
\begin{equation} 
\label{eq:52} 
\left\langle u^{2} \right\rangle =\left\langle \boldsymbol{\mathrm{u}}\cdot \boldsymbol{\mathrm{u}}\right\rangle =\left\langle u_{L}^{2} \right\rangle +\left\langle \boldsymbol{\mathrm{u}}_{T} \cdot \boldsymbol{\mathrm{u}}_{T} \right\rangle.        
\end{equation} 

\begin{figure}
\includegraphics*[width=\columnwidth]{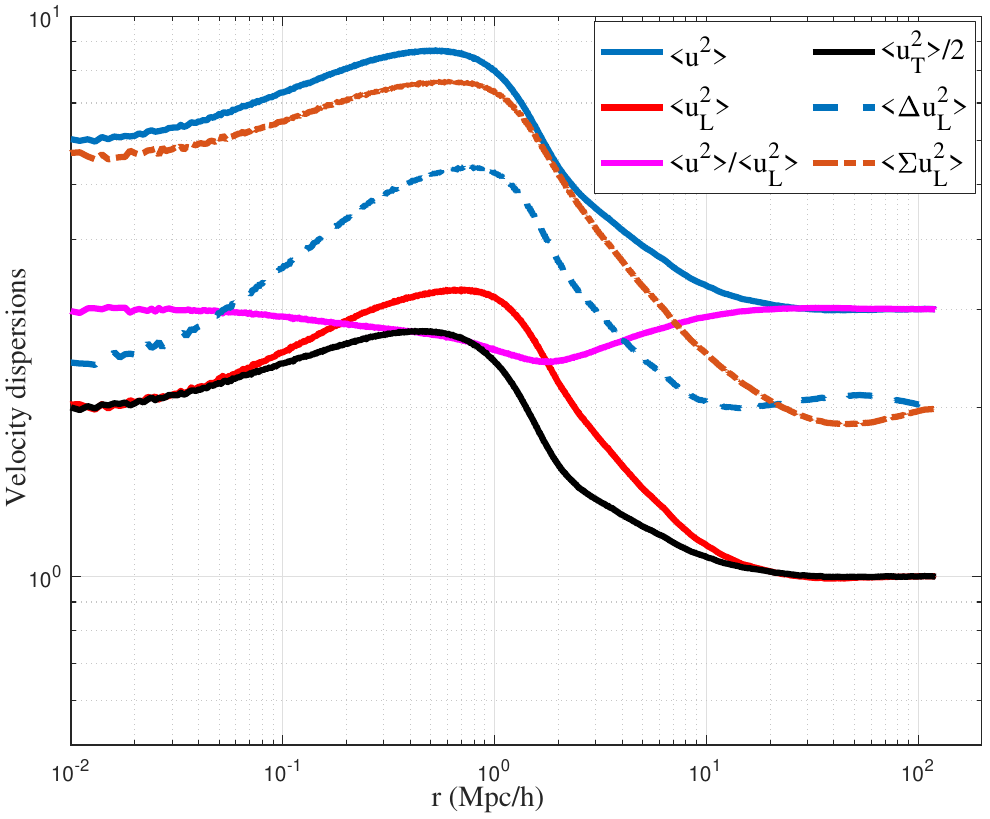}
\caption{The variation of velocity dispersions $\langle u^{2} \rangle =\langle \boldsymbol{\mathrm{u}}\cdot \boldsymbol{\mathrm{u}}\rangle $, $\langle u_{L}^{2} \rangle $, and $\langle u_{T}^{2} \rangle =\langle \boldsymbol{\mathrm{u}}_{T} \cdot \boldsymbol{\mathrm{u}}_{T} \rangle $ with scale \textit{r} at \textit{z}=0 (normalized by $u_{0}^{2} $). The initial increase of all dispersions with \textit{r} for $r<r_{t} $ is mainly due to the increasing velocity dispersion with the size of the halo on small scales. With more pairs of particles from different haloes on larger scales $r>r_{t} $, the dispersion starts to decrease with \textit{r}. With all pairs of particles from different haloes, the velocity dispersion reaches a plateau with $\langle u^{2} \rangle =3\langle u_{L}^{2} \rangle =3u^{2}$. The variation of $\langle u^{2} \rangle$ can be related to the density correlation $\xi(r)$ through dynamic relations on large scales \citep[see ref.][Eq. (120)]{Xu:2024-On-the-statistical-theory-of-self-gravitating-high-order}.}
\label{fig:20}
\end{figure}

Figure \ref{fig:20} plots three velocity dispersions $\langle u^{2} \rangle $, $\langle u_{L}^{2} \rangle$, and $\langle u_{T}^{2} \rangle $ on different scale \textit{r} at \textit{z}=0. The initial increase of the three dispersions with \textit{r} for $r<r_{t} $ (the pair of particles are more likely to be from the same haloes) is mostly due to the increase in the velocity dispersion with the size of the halo. With two particles forming a pair in Fig. \ref{fig:13} that are from different haloes on scale $r>r_{t}$, the velocity dispersions decrease sharply with \textit{r}. At some large scale \textit{r}, almost all pairs of particles are from different haloes, where velocity dispersions reach a plateau with $\langle u^{2} \rangle =3\langle u_{L}^{2} \rangle =3u^{2}$, where dispersion $u^2$ for all particles is listed in Table \ref{tab:2}. The variation of $\langle u^{2} \rangle$ can be related to the density correlation $\xi(r)$ through dynamic relations on large scales \citep[see ref.][Eq. (120)]{Xu:2024-On-the-statistical-theory-of-self-gravitating-high-order}

For particle pairs separated by scale \textit{r}, the second-order moments of longitudinal and transverse velocities are comparable on both small and large scales. However, $\langle u_{L}^{2} \rangle >{\langle u_{T}^{2} \rangle /2} $ on intermediate scales with $\langle u_{L}^{2} \rangle >{\langle u^{2} \rangle /3} $, that is, energy is not equipartition on intermediate scales. More kinetic energy is associated with the longitudinal velocity than with the transverse velocity. The velocity dispersion on small scales $\left. \langle u_{L}^{2} \rangle \right|_{r=0} \approx 2\left. \langle u_{L}^{2} \rangle \right|_{r=\infty }$, i.e. the kinetic energy on small scales is twice the kinetic energy on large scales due to the finite velocity correlation $\rho_L=1/2$ (Eqs. \eqref{50} and \eqref{ZEqnNum388521}). 

The variation of pairwise velocity dispersion (or the second-order longitudinal structure function) 
\begin{equation}
\label{eq:53}
\begin{split}
&S_{2}^{lp} =\left\langle \left(\Delta u_{L} \right)^{2} \right\rangle =\left\langle \left(u_{L}^{'} -u_{L} \right)^{2} \right\rangle  \\& \textrm{and the velocity sum}\\ &\left\langle \left(\sum u_{L} \right)^{2} \right\rangle =\left\langle \left(u_{L}^{'} +u_{L} \right)^{2} \right\rangle
\end{split}
\end{equation}
are also plotted in the same figure for comparison.

\subsection{Even order moments and two-thirds law}
\label{sec:4.5}
Now we focus on the second order structure function $S_{2}^{lp} $ (pairwise velocity dispersion in Eqs. \eqref{ZEqnNum250774} and \eqref{eq:53}) which is defined as
\begin{equation} 
\label{ZEqnNum530715} 
S_{2}^{lp} \left(r\right)=\left\langle \left(\Delta u_{L} \right)^{2} \right\rangle =2\left(\left\langle u_{L}^{2} \right\rangle -L_{2} \left(r\right)\right),        
\end{equation} 
and a modified version of longitudinal structure function $S_{2}^{l} \left(r\right)$
\begin{equation} 
\label{ZEqnNum955110} 
S_{2}^{l} \left(r\right)=2\left(u^{2} -L_{2} \left(r\right)\right).          
\end{equation} 
With model for $\langle u_{L}^{2} \rangle$ (Eq. \eqref{ZEqnNum864804}), model for $\langle u^{2} \rangle$ \citep[see ref.][Eq. (120)]{Xu:2024-On-the-statistical-theory-of-self-gravitating-high-order}, and longitudinal correlation function $L_{2} =\langle u_{L} u_{L}^{'} \rangle $ \citep[see ref.][Eq. (111)]{Xu:2023-On-the-statistical-theory-of-self-gravitating}, structure functions $S_{2}^{lp}(r)$ and $S_{2}^{l}(r)$ on large sales can be completely modelled. On small scales, $S_{2}^{l}(r)$ was also determined to follow a one-fourth law $\propto r^{{1/4} } $ \citep[see ref.][Eq. (137)]{Xu:2023-On-the-statistical-theory-of-self-gravitating}. However, the model for structure functions $S_{2}^{lp} (r)$ on small scales is still lacking. 

\begin{figure}
\includegraphics*[width=\columnwidth]{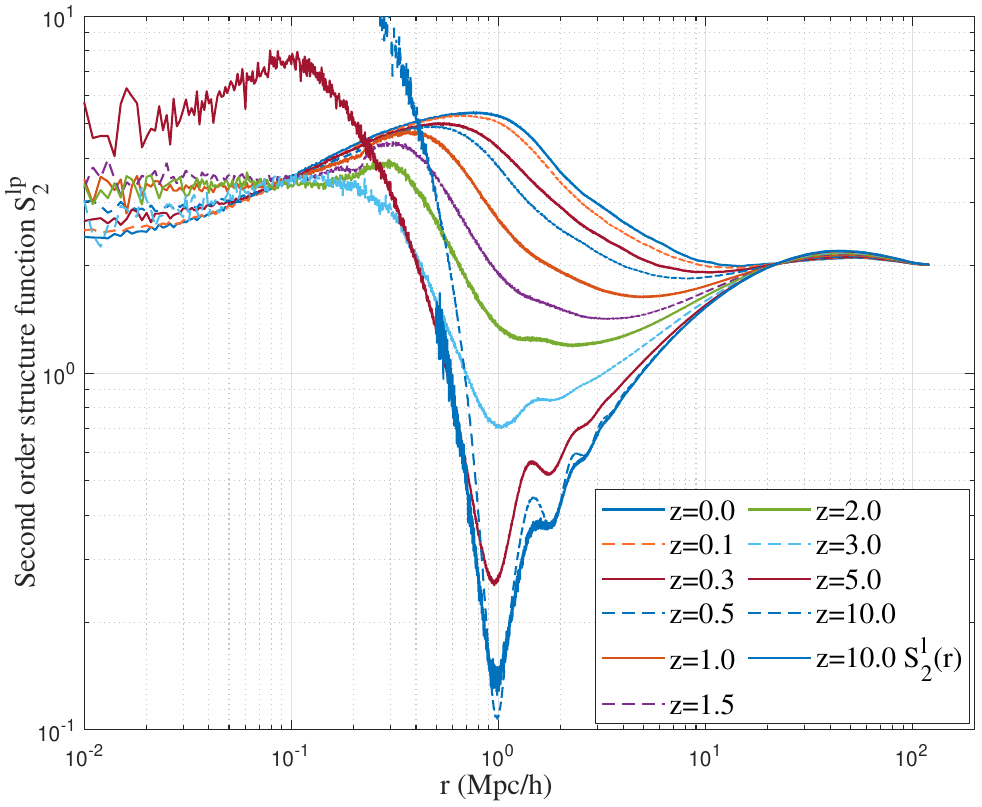}
\caption{The variation of second order longitudinal structure function (or pairwise velocity dispersion) ${S_{2}^{lp} (r)}$ with scale \textit{r} and redshifts \textit{z} (normalized by velocity dispersion $u^{2}$ in Table \ref{tab:2}). The two limits $S_{2}^{lp} (r\to 0)=S_{2}^{lp} (r\to \infty )=2u^{2} $ due to correlation coefficient (between longitudinal velocities $u_{L} $ and $u_{L}^{'}$) $\rho _{L}={1/2} $ and $\rho _{L}=0$ on small and large scales, respectively. Two second-order structure functions $S_{2}^{lp} (r)\approx S_{2}^{l} (r)$ at high redshift \textit{z} ($z=10$ in the figure) when velocity is still Gaussian and small scale structures are not developed (see Eqs. \eqref{ZEqnNum530715} and \eqref{ZEqnNum955110}).} 
\label{fig:21}
\end{figure}

Figure \ref{fig:21} presents the variation of the pairwise velocity dispersion $S_{2}^{lp}(r)$ with scale \textit{r} and redshift \textit{z} with limits $S_{2}^{lp}(r\to 0)=S_{2}^{lp} (r\to \infty )=2u^{2}$ due to correlation coefficient $\rho _{L} ={1/2}$ and $\rho _{L}=0$ on small and large scales. Also, $S_{2}^{lp} (r)\approx S_{2}^{l} (r)$ for high redshift \textit{z}, when the velocity distribution is nearly Gaussian, halo structures are not formed and $\langle u_{L}^{2} \rangle \approx u^{2}$ on all scales (Eqs. \eqref{ZEqnNum530715} and \eqref{ZEqnNum955110}). 

The two-thirds and four-fifths laws for second- and third-order structure functions in incompressible hydrodynamics in Eq. \eqref{eq:1} are no longer valid for SG-CFD due to the collisionless nature of the flow. Since the peculiar velocity is of constant divergence on small scales \citep{Xu:2023-On-the-statistical-theory-of-self-gravitating}, the second-order structure and correlation functions for the peculiar velocity should satisfy the same kinematic relations as if the peculiar velocity field is incompressible \citep{Xu:2023-On-the-statistical-theory-of-self-gravitating}. Furthermore, similarly to the direct energy cascade in 3D turbulence and the inverse energy cascade in 2D turbulence, there exists also a constant energy flux $\varepsilon_{u}<0$ in SG-CFD for the inverse kinetic energy cascade from small to large mass scales \citep{Xu:2023-Universal-scaling-laws-and-density-slope, Xu:2023-Dark-matter-halo-mass-functions-and}. Therefore, we expect that the second order structure function $S_{2}^{lp} (r)$ on small scales in SG-CFD should also be related to the constant energy flux $\varepsilon _{u}$ in some way that is different from Eq. \eqref{eq:1} for incompressible turbulence.

Since the viscous force is not present in SG-CFD, a reduced structure function $S_{2r}^{lp} =S_{2}^{lp} -2u^{2} $ can be introduced with a vanishing limit ${\mathop{\lim }\limits_{r\to 0}} S_{2r}^{lp} =0$. The limiting pairwise velocity dispersion is inherent to all pairs of particles with $r\to 0$ and equals the kinetic energy on small scales, that is, ${\mathop{\lim }\limits_{r\to 0}} S_{2}^{lp} ={\mathop{\lim }\limits_{r\to 0}} \langle u_{L}^{2} \rangle =2u^{2}$. This part of kinetic energy reflects the collective motion of particles (the mean velocity of two particles) in the same halo that is not related to the energy cascade. Only the reduced structure function $S_{2r}^{lp}$ reflects the excess pairwise velocity dispersion. This portion of kinetic energy is completely due to the random motion of particles of two particles that is relevant to the energy cascade \citep{Xu:2022-Postulating-dark-matter-partic}. This portion of the kinetic energy should be determined only by the constant energy flux $\varepsilon _{u}$ (${m^{2}/s^{3}} $) and the scale \textit{r}. By a simple dimensional analysis, $S_{2r}^{lp} $ must follow a two-thirds law, i.e., $S_{2r}^{lp} \propto (-\varepsilon _{u})^{{2/3} } r^{{2/3}}$, which can also be derived from the scale independence of $\varepsilon_u$ \citep{Xu:2021-Inverse-mass-cascade-mass-function}. 

\begin{figure}
\includegraphics*[width=\columnwidth]{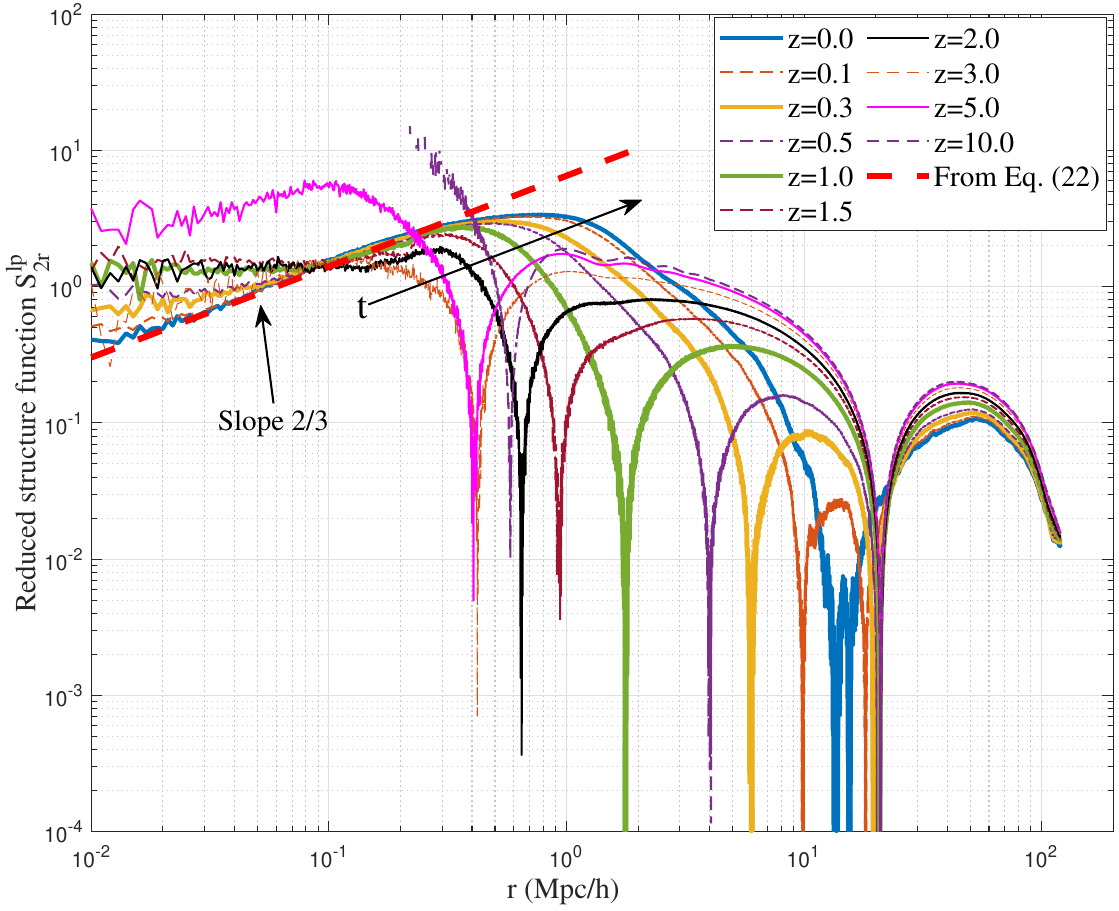}
\caption{The variation of reduced longitudinal structure functions $S_{2r}^{lp} =(S_{2}^{lp} -2u^{2})$ with scale \textit{r} at different redshifts \textit{z}, normalized by velocity dispersion $u^{2}(a)$ in Table \ref{tab:2}. A scaling of $S_{2r}^{lp} \propto (-\varepsilon _{u} )^{{2/3} } r^{{2/3} } $ (two-thirds law) can be clearly identified in a range that is gradually expanding with time, where $\varepsilon _{u} <0$ is the constant rate of energy cascade. The model of Eq. \eqref{ZEqnNum836912} is also presented for comparison. When Combined with the structure formation and evolution in radiation and matter eras, this relation might be useful for postulating dark matter particle mass and properties on small scales \citep{Xu:2022-Postulating-dark-matter-partic}.}
\label{fig:22}
\end{figure}

Here, to test this idea, Figure \ref{fig:22} plots the variation of reduced second-order structure function $S_{2r}^{lp} $ with scale \textit{r} at different redshifts \textit{z}. A range with scaling $S_{2r}^{lp} \propto r^{{2/3}}$ can be clearly identified due to the formation of halo structures on small scales. This range gradually extends to larger scales due to the increasing critical scale $r_t$. The interesting finding is that the constant energy flux $\varepsilon _{u}$ determines a new two-thirds law for a reduced second-order structure function $S_{2r}^{lp} $ in SG-CFD. As expected, the reduced structure function quickly converges to $S_{2r}^{lp} \propto (-\varepsilon _{u})^{{2/3} } r^{{2/3} } $ with the development of halo structures. The length scale at which $S_{2}^{lp} $ is at its maximum is approximately $r_{d} \approx 0.7a{Mpc/h} $, the same as the length scale for $\langle u_{L}^{2}\rangle $ \citep[see ref.][Fig. 24]{Xu:2023-On-the-statistical-theory-of-self-gravitating}. 

Therefore, the second-order longitudinal structure function on small scales can be finally modeled as
\begin{equation} 
\label{ZEqnNum836912} 
\begin{split}
&S_{2r}^{lp} \left(r\right) = a^{{3/2} } \beta _{2}^{*} \left(-\varepsilon _{u} \right)^{{2/3} } r^{{2/3} },\\
&S_{2}^{lp} \left(r\right)=u^{2} \left[2+\beta _{2}^{*} \left(\frac{r}{r_{s} } \right)^{{2/3} } \right]=2u^{2} +a^{{3/2} } \beta _{2}^{*} \left(-\varepsilon _{u} \right)^{{2/3} } r^{{2/3} },  
\end{split}
\end{equation} 
where the length scale $r_{s}$ is purely determined by $u_{0}$ and $\varepsilon _{u}$ with
\begin{equation} 
\label{eq:56} 
r_{s} =-\frac{u_{0}^{3} }{\varepsilon _{u} } =\frac{4}{9} \frac{u_{0} }{H_{0} } =\frac{2}{3} u_{0} t_{0} \approx 1.58{Mpc/h} ,  
\end{equation} 
which is roughly the scale below which the two-thirds law is valid. The rate of the energy cascade $\varepsilon_{u}$ is estimated as
\begin{equation} 
\label{eq:57} 
-\varepsilon _{u} =\frac{3}{2} \frac{u_{0}^{2} }{t_{0} } =\frac{9}{4} u_{0}^{2} H_{0} \approx 0.6345\frac{u_{0}^{3} }{{Mpc/h} } =4.6\times 10^{-7} {m^{2} /s^{3} },  
\end{equation} 
where $t_0$ is the age of the universe (13.7 Billion years) \citep{Xu:2023-Universal-scaling-laws-and-density-slope,Xu:2023-Dark-matter-halo-mass-functions-and}. Constant $\beta _{2}^{*} \approx 9.5$ can be found from Fig. \ref{fig:22}, where the model \eqref{ZEqnNum836912} is also presented for comparison. 

With the model for $S_{2}^{lp} \left(r\right)$ in Eq. \eqref{ZEqnNum836912}, Eq. \eqref{ZEqnNum530715}, and model for longitudinal correlation $L_{2} \left(r\right)$ \citep[see ref.][Eq. (138)]{Xu:2023-On-the-statistical-theory-of-self-gravitating}, 
\begin{equation} 
\label{ZEqnNum178092} 
L_{2} \left(r\right)=u^{2} \left[1-\left(\frac{r}{r_{1} } \right)^{n} \right],         
\end{equation} 
the dispersion $\left\langle u_{L}^{2} \right\rangle $ of longitudinal velocity (in Fig. \ref{fig:20}) on small scales can be finally modeled as,
\begin{equation} 
\label{ZEqnNum864517} 
\left\langle u_{L}^{2} \right\rangle =u^{2} \left[2-\left(\frac{r}{r_{1} } \right)^{n} +\frac{1}{2} \beta _{2}^{*} \left(\frac{r}{r_{s} } \right)^{{2/3} } \right],  
\end{equation} 
where $n\approx {1/4} $ and $r_{1} \left(a\right)\approx 19.4{a^{-3} Mpc/h} $. While on large scales, the velocity dispersion $\langle u^2 \rangle$ can be related to the density correlation via dynamic relations \citep[see ref.][Eq. (120)]{Xu:2024-On-the-statistical-theory-of-self-gravitating-high-order}, the longitudinal velocity dispersion $\langle u_L^{2}\rangle = {\langle u^{2}\rangle}/{3}$ reads
\begin{equation} 
\label{ZEqnNum864804} 
\begin{split}
\langle u_L^{2}\rangle = u^{2}-\frac{2\nu}{3Hf(\Omega _{m})^2} \frac{a_0u^2}{rr_2}\cdot  \exp \left(-\frac{r}{r_{2} } \right)\left[\left(\frac{r}{r_{2} } \right)^{2} -7\left(\frac{r}{r_{2} } \right)+8\right].
\end{split}
\end{equation} 
where $\nu\approx -6000Mpc\cdot$km/s is the negative effective viscosity at $z=0$ on large scales, $r_2=23$Mpc/h is the characteristic scale for velocity correlations and the coefficient $a_0=0.45$ \citep[see ref.][Fig. 21]{Xu:2023-On-the-statistical-theory-of-self-gravitating}. The negative effective viscosity reflects the inverse energy cascade \citep{Xu:2024-On-the-statistical-theory-of-self-gravitating-high-order}. The models for longitudinal velocity dispersion $\langle u_L^{2}\rangle$ on small and large scales are plotted in Fig. \ref{fig:19}.

\begin{figure}
\includegraphics*[width=\columnwidth]{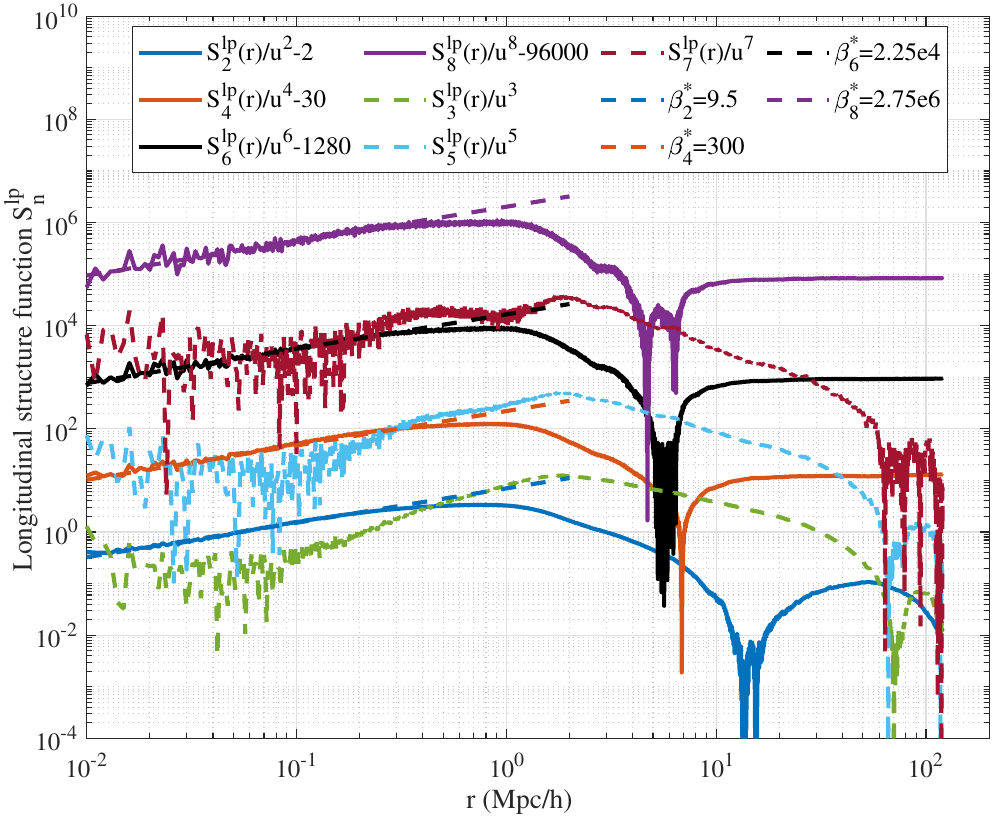}
\caption{The variation of even and odd order structure functions with scale \textit{r} at \textit{z}=0. The plot demonstrates that even order $\textbf{reduced}$ structure functions scales as $S_{2nr}^{lp} \propto \beta _{2n}^{*} r^{{2/3} } $ on small scales (Eq. \eqref{ZEqnNum634712}), while odd order structure functions scales as $S_{2n+1}^{lp} \propto r$. The numbers 2, 30, 1280{\dots} are related to the generalized kurtosis $K_{2n} (\Delta u_{L},r)$ for the limiting distribution of pairwise velocity $\Delta u_{L} $ when $r\to 0$ (Table \ref{tab:4}).}
\label{fig:23}
\end{figure}

Next, higher-order structure functions can be studied similarly. Figure \ref{fig:23} plots the variation of even and odd order structure functions $S_{2n+1}^{lp} \left(r\right)$ with scale \textit{r} at \textit{z}=0. It is now clear that the original Kolmogorov scaling (Eq. \eqref{eq:1}) for incompressible flow does not apply to a self-gravitating collisionless dark matter flow due to the collisionless nature and long-range gravity. On small scales, all even order $\textbf{reduced}$ structure functions follow $S_{2nr}^{lp} \propto \beta _{2n}^{*} r^{{2/3} } $, while all odd order structure functions follow a linear scaling such that $S_{2n+1}^{lp} \propto r$.

The general form for even order structure function $S_{2n}^{lp}(r)$ can be precisely modeled as, 
\begin{equation}
\label{ZEqnNum634712} 
S_{2n}^{lp} \left(r\right)=u^{2n} \left[2^{n} K_{2n} \left(\Delta u_{L} ,r=0\right)+\beta _{2n}^{*} \left(\frac{r}{r_{s} } \right)^{{2/3} } \right],  
\end{equation} 
where $K_{2n}(\Delta u_{L},r=0)$ is the generalized kurtosis on the smallest scale that we can find from Fig. \ref{fig:15} (listed in Table \ref{tab:4} and modeled by Eq. \eqref{ZEqnNum258992}). The universal constants $\beta _{2n}^{*} $ are determined as
\begin{equation}
\begin{split}
&\beta _{2}^{*} =9.5, \quad \beta _{4}^{*} =300, \quad \beta _{6}^{*} =2.25\times 10^{4}, \quad \beta _{8}^{*} =2.75\times 10^{6},\\    
&\noindent \textrm{or approximately}\\ 
&\beta_{2n}^{*} \approx 10^{1.826n-1.003}. 
\end{split}
\label{eq:62}
\end{equation} 

\subsection{Odd order moments and stable clustering hypothesis}
\label{sec:4.6}
\begin{figure}
\includegraphics*[width=\columnwidth]{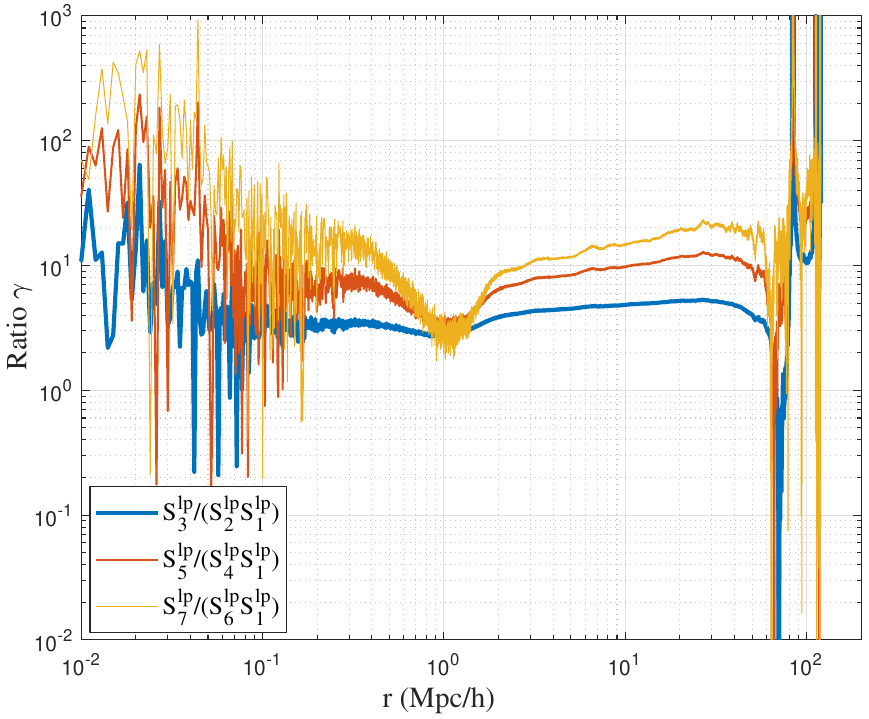}
\caption{The variation of ratio $\gamma={S_{2n+1}^{lp}(r)/[S_{1}^{lp}(r)S_{2n}^{lp}(r)]} $ for \textit{n} = 1, 2, and 3 with scale \textit{r} at \textit{z}=0. For \textit{n}=1, this ratio is around three as predicted by the generalized stable clustering hypothesis (GSCH) \citep{Xu:2021-A-non-radial-two-body-collapse}.} 
\label{fig:24}
\end{figure}

The mean pairwise velocity (first order moment) $S_{1}^{lp}(r)=\langle\Delta u_L\rangle=-Har$ on small scales (Fig. \ref{fig:18}) can be obtained from the stable clustering hypothesis, which can be demonstrated by a two-body collapse model (TBCM) in an expanding background \citep{Xu:2021-A-non-radial-two-body-collapse}. The same TBCM model can be extended to higher-order moments, i.e., the generalized stable clustering hypothesis, such that  
\begin{equation} 
\label{ZEqnNum780846} 
\begin{split}
&S_{2n+1}^{lp} \left(r\right)=\left(2n+1\right)S_{1}^{lp} \left(r\right)S_{2n}^{lp} \left(r\right),\\
&\textrm{or} \quad \gamma = {S_{2n+1}^{lp}}/{S_{1}^{lp} S_{2n}^{lp}} =\left(2n+1\right),  
\end{split}
\end{equation} 
where $\gamma$ is the ratio between odd and even order structure functions. From this, the odd-order structure functions can be written as:
\begin{equation} 
\label{eq:64} 
S_{2n+1}^{lp} \left(r\right)=-2^{n} \left(2n+1\right)K_{2n} \left(\Delta u_{L} ,r=0\right)Haru^{2n}.  
\end{equation} 

Generalized kurtosis on the smallest scale $K_{2n}(\Delta u_{L},r=0)$ is presented in the next section (Table \ref{tab:4} and Eq. \eqref{ZEqnNum258992}). With odd order moments in Fig. \ref{fig:23}, Fig. \ref{fig:24} presents the ratio ${S_{2n+1}^{lp} (r)/[S_{1}^{lp} (r)S_{2n}^{lp} (r)]} $ for \textit{n}=1, 2 and 3 at \textit{z}=0. For \textit{n}=1, this ratio is around three on small scales. For \textit{n} = 2 and 3, this ratio slightly deviates from the predicted value of $(2n+1)$ with higher noise on small scales \citep{Xu:2021-A-non-radial-two-body-collapse}. 

\begin{table}
\centering
\caption {Velocity fields in incompressible flow and SG-CFD}
\begin{center}
\begin{tabular}{lll} 
\hline 
Quantity & \makecell{Incompressible flow} & \makecell{Self-gravitating\\collisionless flow} \\ 
\hline 
$\left\langle u_{L} \right\rangle$ &\makecell{ 0 for all scale \textit{r}} & \makecell{${\mathop{\lim }\limits_{r\to 0,\infty }} \left\langle u_{L} \right\rangle =0$\\varying with r} \\ 
\hline 
$\left\langle u_{L}^{2} \right\rangle $ & \makecell{$u^{2} $ for all scale \textit{r}} & \makecell{${\mathop{\lim }\limits_{r\to 0}} \left\langle u_{L}^{2} \right\rangle =2u^{2}$\\${\mathop{\lim }\limits_{r\to \infty }} \left\langle u_{L}^{2} \right\rangle =u^{2} $} \\ 
\hline 
$\left\langle u_{L}^{3} \right\rangle $ & \makecell{0 for all scale \textit{r}} & \makecell{${\mathop{\lim }\limits_{r\to 0,\infty }} \left\langle u_{L}^{3} \right\rangle =0$\\varying with \textit{r}} \\ 
\hline 
PDF of $u_{L}$  & \makecell{Gaussian} & \makecell{Non-Gaussian on all scales} \\ 
\hline 
Correlation $\rho_{L}$  & \makecell{${\mathop{\lim }\limits_{r\to 0}} \rho_L = 1$\\${\mathop{\lim }\limits_{r\to \infty}} \rho_L = 0$} & \makecell{${\mathop{\lim }\limits_{r\to 0}} \rho_L = 1/2$\\${\mathop{\lim }\limits_{r\to \infty}} \rho_L = 0$} \\ 
\hline 
$\left\langle \Delta u_{L} \right\rangle $ &\makecell{ 0 for all scale \textit{r}} & \makecell{${\mathop{\lim }\limits_{r\to 0,\infty }} \left\langle \Delta u_{L} \right\rangle =0$\\ varying with r} \\
\hline 
$\left\langle \Delta u_{L}^{2} \right\rangle $ & \makecell{${\mathop{\lim }\limits_{r\to 0}} \left\langle \Delta u_{L}^{2} \right\rangle =0$\\${\mathop{\lim }\limits_{r\to \infty }} \left\langle \Delta u_{L}^{2} \right\rangle =2u^{2} $} & \makecell{${\mathop{\lim }\limits_{r\to 0}} \left\langle \Delta u_{L}^{2} \right\rangle =2u^{2} $\\${\mathop{\lim }\limits_{r\to \infty }} \left\langle \Delta u_{L}^{2} \right\rangle =2u^{2} $} \\ 
\hline 
$K_{3} \left(\Delta u_{L} \right)$ & \makecell{${\mathop{\lim }\limits_{r\to 0}} K_{3} \left(\Delta u_{L} \right)=-0.4$\\ ${\mathop{\lim }\limits_{r\to \infty }} K_{3} \left(\Delta u_{L} \right)=0$} & \makecell{${\mathop{\lim }\limits_{r\to 0,\infty }} K_{3} \left(\Delta u_{L} \right)=0$\\varying with r} \\ 
\hline 
$K_{4} \left(\Delta u_{L} \right)$ & \makecell{${\mathop{\lim }\limits_{r\to 0}} K_{4} \left(\Delta u_{L} \right)\approx 4$ \\${\mathop{\lim }\limits_{r\to \infty }} K_{4} \left(\Delta u_{L} \right)=3$\\(Gaussian)} & \makecell{${\mathop{\lim }\limits_{r\to 0}} K_{4} \left(\Delta u_{L} \right)=7.5$\\${\mathop{\lim }\limits_{r\to \infty }} K_{4} \left(\Delta u_{L} \right)=4.2$} \\ 
\hline 
$\left\langle \sum u_{L}\right\rangle $ & \makecell{0 on all scales} & \makecell{0 on all scales} \\ 
\hline 
$\left\langle \sum u_{L}^{2} \right\rangle $ & \makecell{${\mathop{\lim }\limits_{r\to 0}} \left\langle \sum u_{L}^{2} \right\rangle =4u^{2} $\\${\mathop{\lim }\limits_{r\to \infty }} \left\langle \sum u_{L}^{2} \right\rangle =2u^{2} $} & \makecell{${\mathop{\lim }\limits_{r\to 0}} \left\langle \Delta u_{L}^{2} \right\rangle =6u^{2}$\\${\mathop{\lim }\limits_{r\to \infty }} \left\langle \Delta u_{L}^{2} \right\rangle =2u^{2} $} \\ \hline 
\end{tabular}
\end{center}
\label{tab:3}
\end{table}

Finally, Table \ref{tab:3} presents a comprehensive comparison of the velocity field between incompressible hydrodynamics and self-gravitating collisionless dark matter flow (SG-CFD). The differences are due to the collisionless nature and long-range gravity in SG-CFD, where distributions of velocity are non-Gaussian on all scales. For incompressible flow, the direct energy cascade from large to small scales leads to negative skewness ($K_3$=-0.4) in the dissipation range. For SG-CFD, the inverse cascade from small to large scales leads to negative skewness ($K_3\approx$ -0.1 to -1) around the intermediate scale $r_t$, while $K_3$ vanishes on both small and large scales (Fig. \ref{fig:16}). 

\section{Probability distributions of velocity field}
\label{sec:5}
\subsection{Velocity distributions on small scales}
\label{sec:5.1}

\begin{figure}
\includegraphics*[width=\columnwidth]{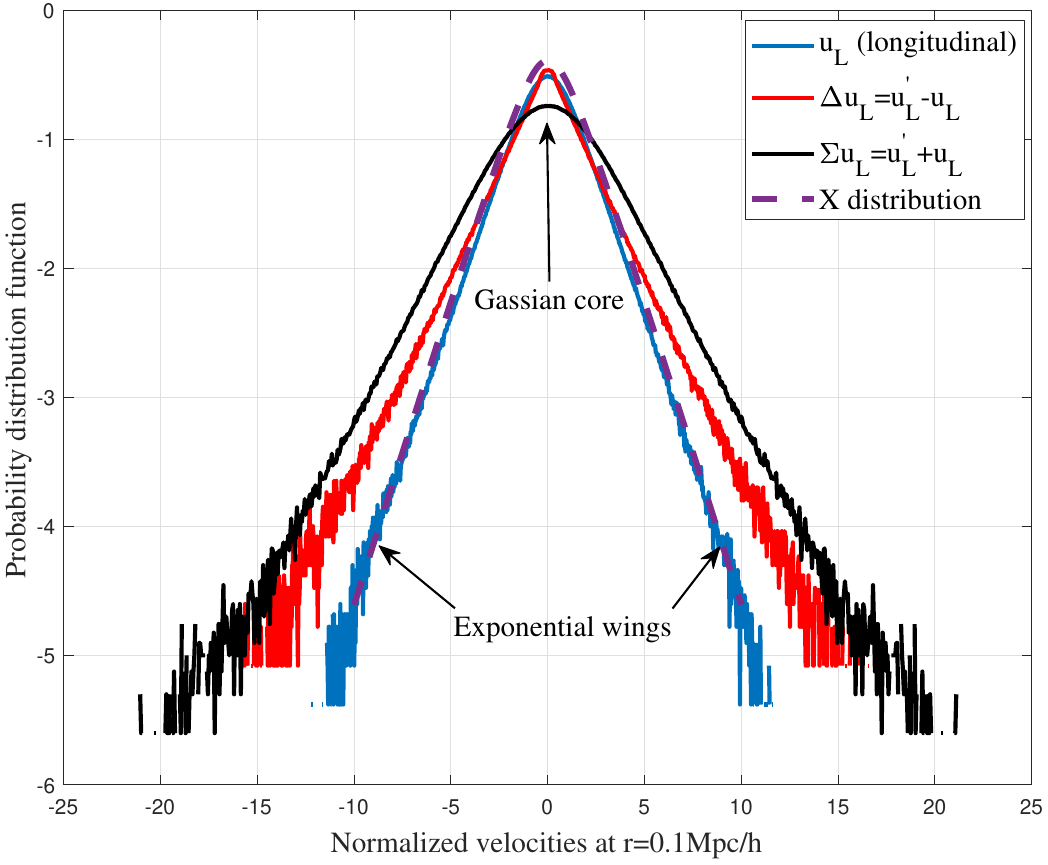}
\caption{Probability Distributions of longitudinal velocity $u_{L}$, pairwise velocity $\Delta u_{L}$, and velocity sum $\Sigma u_{L}$ on a small scale of \textit{r}=0.1Mpc/h at \textit{z}=0 from N-body simulations, i.e. $\log_{10}P $ vs. ${u_{L}/u_{0}}$, where $u_0$ is the velocity dispersion at $z=0$ (Table \ref{tab:2}). All distributions are symmetric with a vanishing skewness (third order kurtosis $K_3$). The \textit{X} distribution that maximizes the system entropy (Eq. \eqref{ZEqnNum436604}) matches the distribution of $u_{L}$. Distributions have a Gaussian core for small velocity and exponential wings for large velocity. Velocity sum $\Sigma u_{L}$ also follows the X distribution but with a different variance. Pairwise velocity $\Delta u_{L}$ follows a different distribution (Eq. \eqref{eq:78}).} 
\label{fig:25}
\end{figure}

On small scales, longitudinal velocities $u_{L}$ and $\Sigma u_{L}$ should follow the same limiting distribution as $r\to 0$, which is different from the distribution of pairwise velocity $\Delta u_{L}$ (Fig. \ref{fig:15}). This section focuses on the probability distributions of $u_{L}$ and $\Sigma u_{L}$ that should maximize the entropy of the system. In our previous work, based on the halo description of the self-gravitating collisionless system, $u_{L}$ on small scales should follow a \textit{X} distribution to maximize the entropy of the system. The \textit{X} distribution simply reads \citep{Xu:2023-Maximum-entropy-distributions-of-dark-matter}
\begin{equation} 
\label{ZEqnNum436604} 
X\left(v\right)=\frac{1}{2\alpha v_{0} } \frac{e^{-\sqrt{\alpha ^{2} +\left({v/v_{0} } \right)^{2} } } }{K_{1} \left(\alpha \right)} ,         
\end{equation} 
where $\alpha $ is a shape parameter and $K_{n} \left(x\right)$ is the \textit{modified} Bessel function of the second kind. The velocity scale $v_{0}$ satisfies 
\begin{equation} 
\label{eq:66} 
\alpha \frac{K_{2} \left(\alpha \right)}{K_{1} \left(\alpha \right)} v_{0}^{2} =\left\langle u_{L}^{2} \right\rangle ,           
\end{equation} 
where $\langle u_{L}^{2} \rangle $ is the dispersion of velocity $u_{L}$ in Fig. \ref{fig:19}. It can be estimated that $v_{0}^{2} \approx 0.84u_{0}^{2} $ with $\langle u_{L}^{2}\rangle=2.5u_{0}^{2}$ at $r$=0.1 Mpc/h (from Fig. \ref{fig:19}) and \textit{z}=0. With the shape parameter $\alpha \approx 1.33$ and $v_{0}^{2} =0.84u_{0}^{2} $, the \textit{X} distribution is plotted in Fig. \ref{fig:25} for comparison with the distribution of $u_{L}$ from N-body simulations. The velocity sum $\Sigma u_{L}$ should follow the same distribution but with a different variance, i.e., $\langle (\Sigma u_{L})^{2} \rangle \approx 3 \langle (u_{L})^{2} \rangle$. All distributions are symmetric on small scales. 

\subsection{Distribution of pairwise velocity on small scales}
\label{sec:5.2}
The longitudinal velocity $u_L$ has a finite limiting correlation $\rho _{L} ={1/2}$ with $r\rightarrow 0$ such that the limiting distribution of the velocity difference (or pairwise velocity) $\Delta u_{L}$ must be different from the distribution of $u_{L}$ (Figs. \ref{fig:15} and \ref{fig:25}). The longitudinal correlation depends on the size of the halo (Eq. \eqref{eq:73}), where correlation approaches one in small haloes and zero in large haloes. This effect was not considered in previous work to determine the analytical distribution of $\Delta u_{L}$ \citep{Sheth:1996-The-distribution-of-pairwise-p}. Again, the distribution of $\Delta u_{L}$ on small scales cannot be Gaussian because of strong gravity (also see Fig. \ref{fig:15}). The explicit form of that distribution is still unknown and should be explored in the future. However, in this section, the moments for the distribution of $\Delta u_{L}$ can be rigorously estimated. This is also required to compute the generalized kurtosis in Eq. \eqref{ZEqnNum634712} for structure functions of pairwise velocity on small scales. 

\begin{figure}
\includegraphics*[width=\columnwidth]{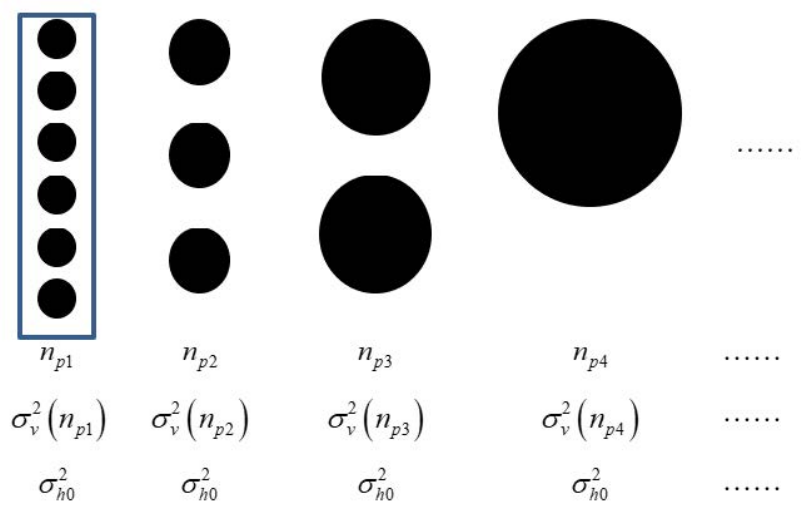}
\caption{Schematic plot of groups of haloes of different sizes in an N-body system. All haloes are grouped and sorted according to the number of particles $n_{p}$ in the halo, with size increasing from left to right. Every group of haloes of the same halo size, $n_{p}$, is characterized by the number of haloes in that group $N_h$, a halo virial dispersion $\sigma _{v}^{2} \left(n_{p} \right) \propto n_p^{2/3}$ increasing with halo size $n_p$, i.e. the dispersion of velocity of all particles in the same halo. The halo velocity dispersion (the dispersion of velocity of all haloes in the same group), $\sigma _{h}^{2} =\sigma _{h0}^{2}$, is independent of halo size.}
\label{fig:1}
\end{figure}

Let us start from an N-body system with a total of \textit{N} collisionless particles. Figure \ref{fig:1} presents a schematic diagram of the halo picture by sorting all haloes in a system according to their sizes $n_p$ from the smallest to the largest \citep{Xu:2023-Maximum-entropy-distributions-of-dark-matter}. The halo size $n_p=m_h/m_p$, where $m_h$ and $m_p$ are the mass of the halo and single particle, respectively. Each column in Fig. \ref{fig:1} is a group of all haloes of the same size $n_p$. The total number of particles in a halo group reads 
\begin{equation} 
\label{eq:67} 
n_{p} N_{h} =N f\left(\nu \right)d\nu,         
\end{equation} 
where $N_{h} $ is the number of haloes in that halo group of size $n_{p}$, $f\left(\nu \right)$ is the dimensionless halo mass function with variable 
\begin{equation} 
\label{eq:67-2} 
\nu = \left(\frac{m_h}{m_h^*}\right)^{2/3}=\frac{\sigma _{v}^{2}}{\sigma _{h}^{2}}, 
\end{equation} 
where $m_h$ and $m_h^*$ are the halo mass and the characteristic halo mass. Halo virial dispersion $\sigma _{v}^{2}$ is the dispersion of velocity of all particles in the same halo and increases with halo size $n_p$. The halo velocity dispersion $\sigma _{h}^{2}$ is the dispersion of velocity of all haloes in the same group and is independent of halo size \citep{Xu:2023-Maximum-entropy-distributions-of-dark-matter}.

Let us assume that the number of particle pairs $n_{pair}$ with a separation \textit{r} in haloes of size $n_{p}$ is proportional to the halo size $n_{p}$ with a power law $n_{pair} =\mu _{p}(n_{p})^{\alpha _{p}}$, where $\mu _{p} $ is a proportional constant. The larger haloes have more pairs of particles on a given scale $r$. The maximum number of pairs for a given halo size $n_{p} $ is $n_{pair} ={n_{p} \left(n_{p} -1\right)/2} $ if all $n_{p}$ particles collapse into a single point, where we have $\alpha _{p} =2$. In principle, the exponent $\alpha _{p} $ satisfies $1<\alpha _{p} <2$. The number of pairs in a halo group ($n_{pair}$) reads
\begin{equation} 
\label{ZEqnNum510377} 
\begin{split}
&N_{h} n_{pair} =N\mu _{p} \left(n_{p} \right)^{\alpha _{p} -1} f\left(\nu \right)d\nu,\\
&N_{pair} =\int _{0}^{\infty }N\mu _{p} \left(n_{p} \right)^{\alpha _{p} -1} f\left(\nu \right)d\nu.        
\end{split}
\end{equation} 
Here, $N_{pair}$ is the total number of pairs with a given separation \textit{r} in the entire N-body system. 

From the virial theorem, the halo virial dispersion $\sigma _{v}^{2} \propto (n_{p})^{{2/3} } $ and we can write $n_{p} =\mu _{v} ({\sigma _{v}^{2} /\sigma _{h}^{2} })^{{3/2} } $, where $\mu _{v} $ is a proportional constant. Therefore, Eq. \eqref{ZEqnNum510377} can be transformed to
\begin{equation}
\begin{split}
&\frac{N_{pair} }{N\mu _{p} \left(\mu _{v} \right)^{\alpha _{p} -1} } =\int _{0}^{\infty }f\left(\nu \right)\nu ^{\frac{3}{2} \left(\alpha _{p} -1\right)} d\nu, \\ &\textrm{or equivalently,}\\ 
&\int _{0}^{\infty }\beta _{p} f\left(\nu \right)\nu ^{p} d\nu  =1, 
\end{split}
\label{ZEqnNum317855}
\end{equation}
where $\beta _{p} f\left(\nu \right)\nu ^{p} d\nu $ is the fraction of pairs in a halo group with a given size $n_p$. Here, the exponent 
\begin{equation}
p={3\left(\alpha_{p} -1\right)/{2}}.
\label{ZEqnNum692067}
\end{equation}

Since the longitudinal velocity $u_{L}$ for all particle pairs in the same halo group is nearly Gaussian \citep[see ref.][Fig. 3]{Xu:2023-Maximum-entropy-distributions-of-dark-matter}, the distribution of pairwise velocity $\Delta u_{L} =u_{L}^{'} -u_{L} $ can be obtained from the joint Gaussian distribution of $u_{L} $ and $u_{L}^{'} $ with a size-dependent correlation coefficient $\rho _{cor} \left(n_{p} \right)$, 
\begin{equation} 
\label{ZEqnNum388848} 
P_{\Delta uL} \left(x\right)=\int _{0}^{\infty }\frac{e^{-{x^{2} /\left[4\left(1-\rho _{cor} \right)\sigma ^{2} \right]} } \beta _{p} f\left(\nu \right)\nu ^{p} d\nu}{\sqrt{2\pi } \sqrt{2\left(1-\rho _{cor} \right)} \sigma }.
\end{equation} 

The correlation $\rho _{cor}$ can be related to the total particle velocity dispersion $\sigma ^{2}$ as \citep[see ref.][Eq. (58)]{Xu:2023-On-the-statistical-theory-of-self-gravitating} 
\begin{equation}
\rho _{cor} \left(n_{p} \right)={\sigma _{h}^{2} /\sigma ^{2} } \quad \textrm{and} \quad \sigma ^{2} \left(n_{p} \right)=\sigma _{v}^{2} \left(n_{p} \right)+\sigma _{h}^{2},   
\label{eq:73}
\end{equation}
\noindent where $\sigma _{h}^{2} $ and $\sigma _{v}^{2} $ are the halo velocity dispersion and halo virial dispersion, respectively. For small haloes with $\sigma _{v}^{2}\to 0$, the correlation coefficient $\rho _{cor}\to 1$. However, for large haloes with $\sigma _{v}^{2}\to \infty$, the correlation coefficient $\rho _{cor}\to 0$. The size of the halo $n_p$ is a function of the dimensionless variable $\nu$, that is, $n_p=\mu_v\nu^{3/2}$.

The moment generating function and the $m$th order moments can finally be obtained from Eq. \eqref{ZEqnNum388848},
\begin{equation} 
\label{ZEqnNum970288}
\begin{split}
\int _{-\infty }^{\infty }P_{\Delta uL}  \left(x\right)e^{xt} dx&=\int _{0}^{\infty }\beta _{p} f\left(\nu \right)\nu ^{p} e^{\left(1-\rho _{cor} \right)\sigma ^{2} t^{2} } d\nu  \\&=\int _{0}^{\infty }\beta _{p} f\left(\nu \right)\nu ^{p} e^{\nu \sigma _{h}^{2} t^{2} } d\nu,   
\end{split}
\end{equation} 
\begin{equation} 
\label{eq:75} 
M_{m} \left(\Delta u_{L} \right)=\frac{m!}{\left({m/2} \right)!} \int _{0}^{\infty }\beta _{p} f\left(\nu \right)\nu ^{p+{m/2} } d\nu  \sigma _{h}^{m} .       
\end{equation} 

\begin{figure}
\includegraphics*[width=\columnwidth]{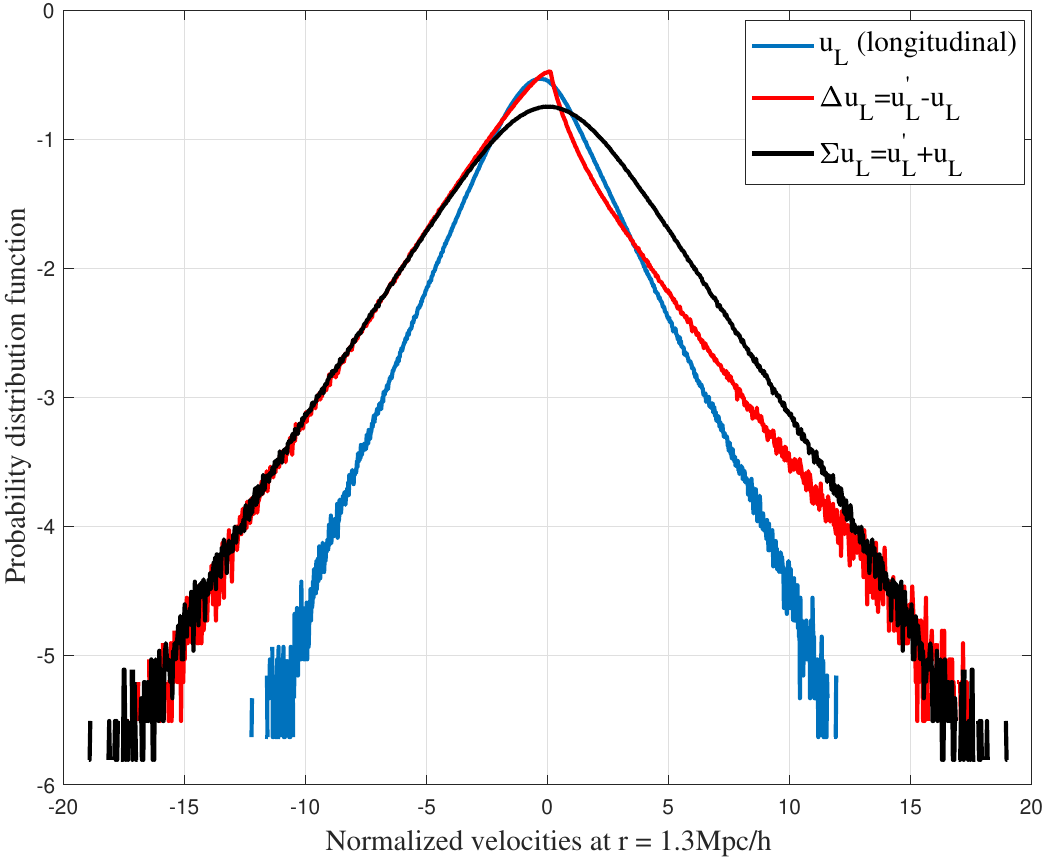}
\caption{Distributions of $u_{L}$, $\Delta u_{L}$, and $\Sigma u_{L}$ on intermediate scales of  \textit{r} = 1.3 Mpc/h at \textit{z}=0, i.e. $\log _{10} P_{uL}$ vs. ${u_{L} /u_{0}}$,where $u_0$ is the velocity dispersion at $z=0$ (Table \ref{tab:2}). The distribution of $\Sigma u_{L}$ is symmetric, while the distribution of $\Delta u_{L}$ is asymmetric with nonzero (negative) skewness (Fig. \ref{fig:28}) and skew toward the positive side. This is a necessary feature of the inverse energy cascade. The distribution of $u_{L}$ is also asymmetric with a nonzero mean and skewness.}
 \label{fig:26}
\end{figure}

We can use the double-$\lambdaup$ mass function \citep[see ref.][Eq. (21)]{Xu:2023-Dark-matter-halo-mass-functions-and} that is proposed based on the inverse mass cascade theory for hierarchical structure formation. The double-$\lambdaup$ mass function $f(\nu)$ reads,
\begin{equation} 
\label{eq:76} 
f\left(\nu \right)=f_{D\lambda } \left(\nu \right)=\frac{\left(2\sqrt{\eta _{0} } \right)^{-q} }{\Gamma \left({q/2} \right)} \nu ^{{q/2} -1} \exp \left(-\frac{\nu }{4\eta _{0} } \right).      
\end{equation} 
Here, the parameters $\eta _{0} =0.76$ and $q=0.556$ for the best fit of the mass function to the simulation data. The normalization factor in Eq. \eqref{ZEqnNum317855} can be obtained as
\begin{equation} 
\label{ZEqnNum550265} 
\beta _{p} =\frac{N\mu _{p} \left(\mu _{v} \right)^{\alpha _{p} -1} }{N_{pair} } =\frac{\Gamma \left({q/2} \right)}{\left(2\sqrt{\eta _{0} } \right)^{2p} \Gamma \left(p+{q/2} \right)} .       
\end{equation} 

Inserting the double-$\lambda$ mass function into Eq. \eqref{ZEqnNum970288}, the distribution of pairwise velocity $P_{\Delta uL}$ satisfies (Eq. \eqref{ZEqnNum970288})
\begin{equation} 
\label{eq:78} 
\int _{-\infty }^{\infty }P_{\Delta uL}  \left(x\right)e^{xt} dx=\frac{1}{\left(1-4\eta _{0} \sigma _{h}^{2} t^{2} \right)^{p+{q/2} } } ,        
\end{equation} 
such that the moments of any order \textit{m} can be obtained as,
\begin{equation} 
\label{ZEqnNum297401} 
M_{m} \left(\Delta u_{L} \right)=\frac{m!\left(2\sqrt{\eta _{0} } \right)^{m} }{\left({m/2} \right)!\Gamma \left(p+{q/2} \right)} \Gamma \left(\frac{1}{2} \left(m+2p+q\right)\right)\sigma _{h}^{m} .     
\end{equation} 
The generalized kurtosis for pairwise velocity $\Delta u_{L}$ is,
\begin{equation} 
\label{ZEqnNum258992} 
K_{2n} \left(\Delta u_{L} \right)=\frac{\left(2n\right)!}{n!2^{n} } \frac{\Gamma \left(n+p+{q/2} \right)\left[\Gamma \left({p+q/2} \right)\right]^{n-1} }{\left[\Gamma \left(1+p+{q/2} \right)\right]^{n} },      
\end{equation} 
where Kurtosis is completely determined by particle pair parameter $p$ (Eq. \eqref{ZEqnNum692067}) and mass function parameter $q$ (Eq. \eqref{eq:76}). With $\eta _{0} =0.76$ and $q=0.556$ for the double-$\lambdaup$ mass function, $\beta _{p} \approx 1.5426$ from Eq. \eqref{ZEqnNum550265}. Using the Kurtosis values for $\Delta u_{L}$ on small scales from simulation (Table \ref{tab:4}), the parameter $p\approx 0.36$ or exponent $\alpha _{p} \approx 1.24$ (from Eq. \eqref{ZEqnNum692067}) can be obtained. The total number of pairs $N_{pair} $ with $r\to 0$ should be (from Eq. \eqref{ZEqnNum550265})
\begin{equation} 
\label{eq:81} 
\frac{N_{pair} }{N} =\frac{\mu _{p} \left(\mu _{v} \right)^{\alpha _{p} -1} }{\beta _{p} }\approx 0.26 ,          
\end{equation} 
where both constants $\mu _{p} $ and $\mu _{v} $ can be obtained from simulation ($\mu_{p} \approx 0.21$ and $\mu _{v} \approx 14$ from the N-body simulation in Section \ref{sec:2} for particle pairs with a separation of $r$=0.1Mpc/h). 

\begin{table}
    \centering
    \caption {Generalized kurtosis of velocity distributions on small and large scales from N-body simulations at z=0 and proposed models}
    \begin{center}
    \begin{tabular}{p{0.38in} p{0.4in} p{1in} p{0.2in} p{0.2in} p{0.2in} } 
    \hline 
    Scale & Velocity & Distribution & 4$^{th}$  & 6$^{th}$ & 8$^{th}$ \\ \hline 
    $r\to 0$ & $u_{L}$, $\Sigma u_{L}$ & N-body, z=0, Fig. \ref{fig:15} & 4.8 & 57 & 1200 \\ \hline 
    $r\to 0$ & $\Delta u_{L}$ & N-body, z=0, Fig. \ref{fig:15} & 7.5 & 160 & 6000 \\ \hline 
    $r\to 0$ & $u_{L}$, $\Sigma u_{L}$ & \makecell[l]{$X$ distribution \\ (Eq. \eqref{ZEqnNum436604} $\alpha=1.33$)} & 4.6 & 48.9 & 944.8 \\ \hline 
    $r\to 0$ & $\Delta u_{L}$ & From model Eq. \eqref{ZEqnNum258992} & 7.7 & 159.24 & 6356 \\ \hline 
    &  &  &  &  &  \\ \hline 
    $r\to \infty $ & $\Delta u_{L}$,$\Sigma u_{L}$ & N-body, z=0, Fig. \ref{fig:15} & \textbf{4.181} & \textbf{41.46} & \textbf{670.8} \\ \hline 
    $r\to \infty $ & $u_{L}$ & N-body, z=0, Fig. \ref{fig:15} & 5.39 & 85.78 & 2800 \\ \hline    
    Option 1 &  &  &  &  & \\ \hline
    $r\to \infty $ & $\Delta u_{L}$,$\Sigma u_{L}$ & Logistic (Eq. \eqref{ZEqnNum404751}) & 4.2 & 279/7 & 686 \\ \hline 
    $r\to \infty $ & $u_{L}$ & $P_{uL}(x)$(Eq. \eqref{ZEqnNum593993}) & 5.4 & $78.4$ & $2270$ \\ \hline  
    Option 2 &  &  &  &  & \\ \hline
    $r\to \infty $ & $\Delta u_{L}$,$\Sigma u_{L}$ & \makecell[l]{$X$ distribution \\ (Eq. \eqref{ZEqnNum436604} $\alpha=2.1$)} & 4.18 & 38.4 & 624 \\ \hline 
    $r\to \infty $ & $u_{L}$ & $P_{uL}(x)$(Eq. \eqref{ZEqnNum593993}) & 5.35 & $73.4$ & $1936$ \\ \hline  \\ \hline 
    &  & Laplace  & 6 & 90 & 2520 \\ \hline 
    &  & Gaussian & \textbf{3} & \textbf{15} & \textbf{105} \\ 
    \hline 
    \end{tabular}
    \end{center}
    \label{tab:4}
\end{table}

The general kurtosis for the distribution of pairwise velocity $\Delta u_{L}$ on small scales can be calculated from Eq. \eqref{ZEqnNum258992} and listed in Table \ref{tab:4}. The model agrees well with the N-body simulation. In addition, Table \ref{tab:4} lists the generalized kurtosis of three types of velocities on both small and large scales, both from models and from simulations. Again, the pairwise velocity $\Delta u_{L} $ is usually approximated by an exponential (Laplace) distribution \citep{Sheth:1996-The-distribution-of-pairwise-p}. This seems not accurate, as the generalized kurtosis of the distribution of $\Delta u_{L} $ from N-body simulations does not agree with that of the exponential distribution on both small and large scales (see Table \ref{tab:4}).
 
\subsection{Velocity distributions on intermediate scales}
\label{sec:5.3}

Figure \ref{fig:26} presents the velocity distributions on an intermediate scale \textit{r${}_{t}$}=1.3Mpc/h. The distributions of $\Delta u_{L}$ and $u_{L}$ are asymmetric with nonzero skewness (see $K_3$ in Fig. \ref{fig:28}), which is due to the inverse cascade of kinetic energy from small scales to scale $r_t$ (roughly the size of the largest haloes of characteristic mass $m_h^*$). The distribution of the velocity sum $\Sigma u_{L}$ is symmetric on all scales.

\begin{figure}
\includegraphics*[width=\columnwidth]{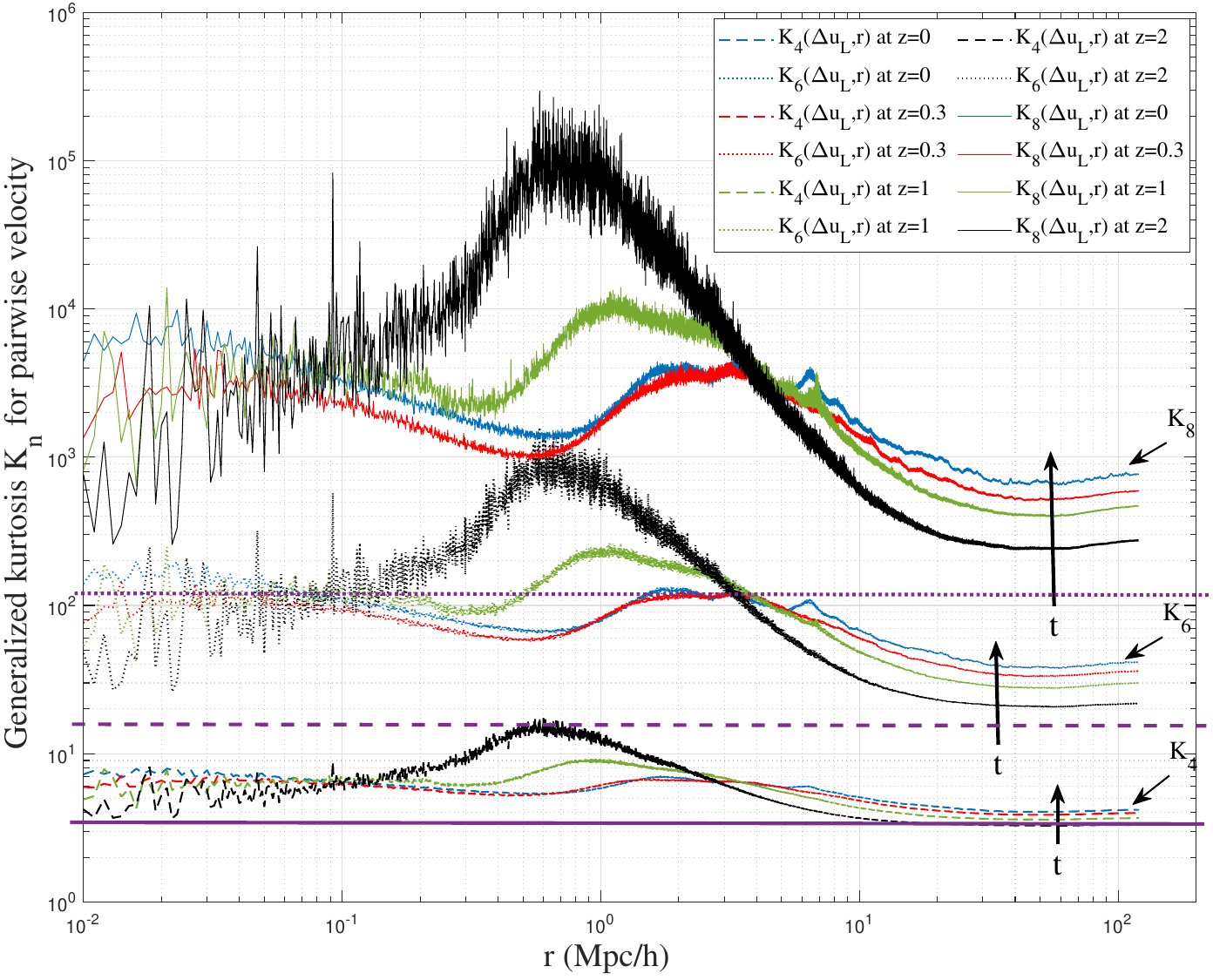}
\caption{The redshift evolution of even order generalized kurtosis for pairwise velocity $\Delta u_{L}$ at redshift \textit{z}= 2.0, 1.0, 0.3, and 0. The Kurtosis for the Gaussian distribution is also plotted for reference (purple lines). The distribution of $\Delta u_{L}$is non-Gaussian on all scales, while the evolution of the distribution on small scales is much faster than that on large scales due to the strong gravitational interaction on small scales (also see Fig. \ref{fig:30}).} 
\label{fig:27}
\end{figure}

Figure \ref{fig:27} plots the redshift variation of generalized kurtosis $K_{4} $, $K_{6}$, and $K_{8} $ of pairwise velocity $\Delta u_{L}$ at \textit{z} = 0, 0.3, 1, and 2.0. Kurtosis of the Gaussian distribution is also plotted for reference. All velocities are initially Gaussian. On small scales, most pairs of particles are from the same halo, and the distribution of the pairwise velocity $\Delta u_{L}$ converges to the limiting distribution (Eq. \eqref{ZEqnNum258992}) much faster because of a strong intra-halo gravitational interaction. On large scales, the particle pairs are from different haloes. The distribution of $\Delta u_{L}$ evolves slower because of the weaker inter-halo interaction at a greater distance. We revisit this in Fig. \ref{fig:30}. Kurtosis on intermediate scales is much greater than that on both small and large scales. 

\begin{figure}
\includegraphics*[width=\columnwidth]{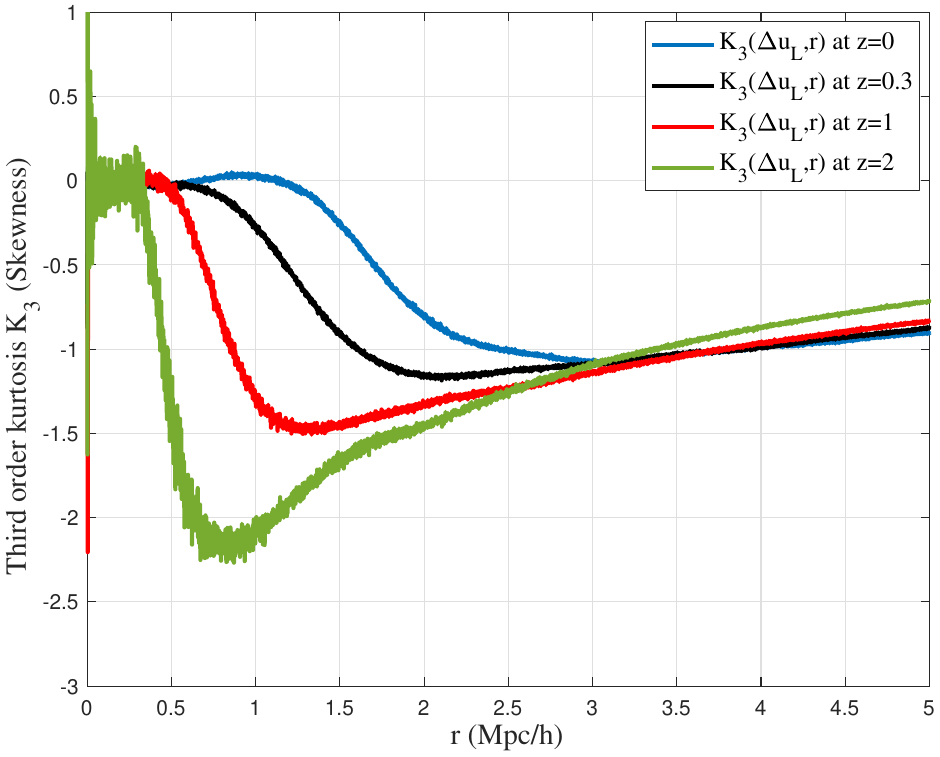}
\caption{The redshift evolution of the skewness $K_{3} $ (third order generalized kurtosis) of $\Delta u_{L}$ on intermediate scales. The skewness $K_{3} \approx 0$ on small scales and $K_{3} <0$ on intermediate scales. A nonzero skewness is an important feature of inverse energy cascade on scales smaller than $r_t$, the size of the largest halo in Fig. \ref{fig:17}. The minimum skewness increases with time, as does the critical scale $r_t$. } 
\label{fig:28}
\end{figure}

Figure \ref{fig:28} plots the variation of $K_{3} $ (or skewness) of pairwise velocity $\Delta u_{L}$ for z = 0, 0.3, 1, and 2.0 on small and intermediate scales. Skewness $K_{3} \approx 0$ on small scales and $K_{3} <0$ on intermediate scales. Nonzero skewness is an important feature of the inverse energy cascade on small scales in the nonlinear regime. 

\subsection{Velocity distributions on large scales}
\label{sec:5.4}
On large scales, the velocities $\Delta u_{L}$ and $\Sigma u_{L}$ have the same distribution as $r\to \infty $ (Fig. \ref{fig:15} and Table \ref{tab:4}). The distribution of $u_{L}$ at $r\to \infty $ has greater kurtosis than $\Delta u_{L}$ and $\Sigma u_{L}$. The non-Gaussian feature on large scales is a manifestation of the long-range nature of gravitational interaction. In contrast, velocity is always Gaussian on large scales for incompressible flow involving short-range interaction. 

\begin{figure}
\includegraphics*[width=\columnwidth]{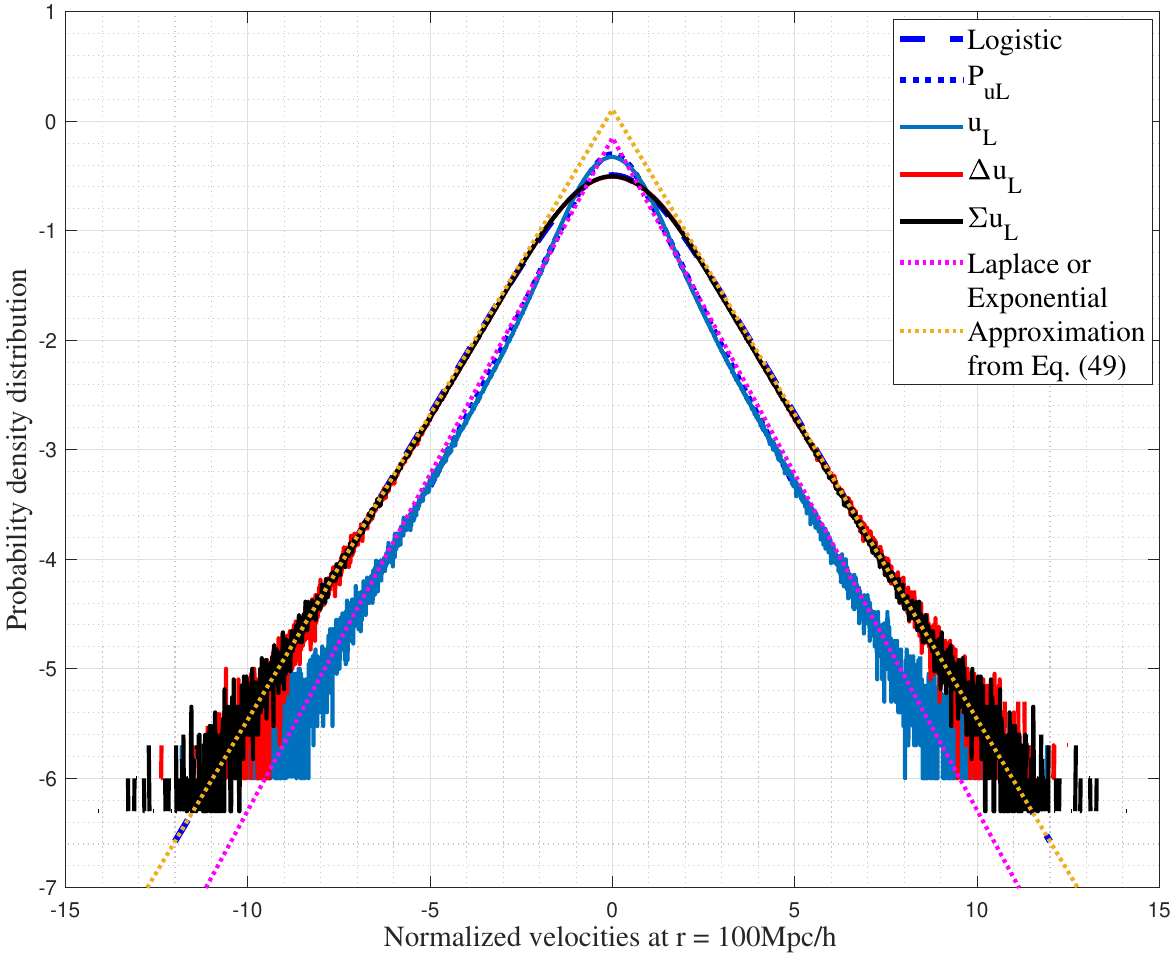}
\caption{Distributions of $u_{L}$, $\Delta u_{L}$, and $\Sigma u_{L}$ on a large scale of \textit{r} = 100 Mpc/h at \textit{z}=0, i.e. $\log _{10} P_{uL} $ vs. ${u_{L} /u_{0} } $ (normalized by $u_{0} $). On large scales, all distributions are symmetric. Here, $\Delta u_{L}$ and $\Sigma u_{L}$ follow the same distribution. A logistic distribution can be used to model the distribution of $\Delta u_{L}$ and $\Sigma u_{L}$. At large velocities, all distributions approach an exponential function. The longitudinal velocity $u_{L}$ follows a different distribution that can be determined by the distribution of $\Sigma u_{L}$ (Eqs. \eqref{ZEqnNum593993}) and \eqref{ZEqnNum404751-2}.}
\label{fig:29a}
\end{figure}

\begin{figure}
\includegraphics*[width=\columnwidth]{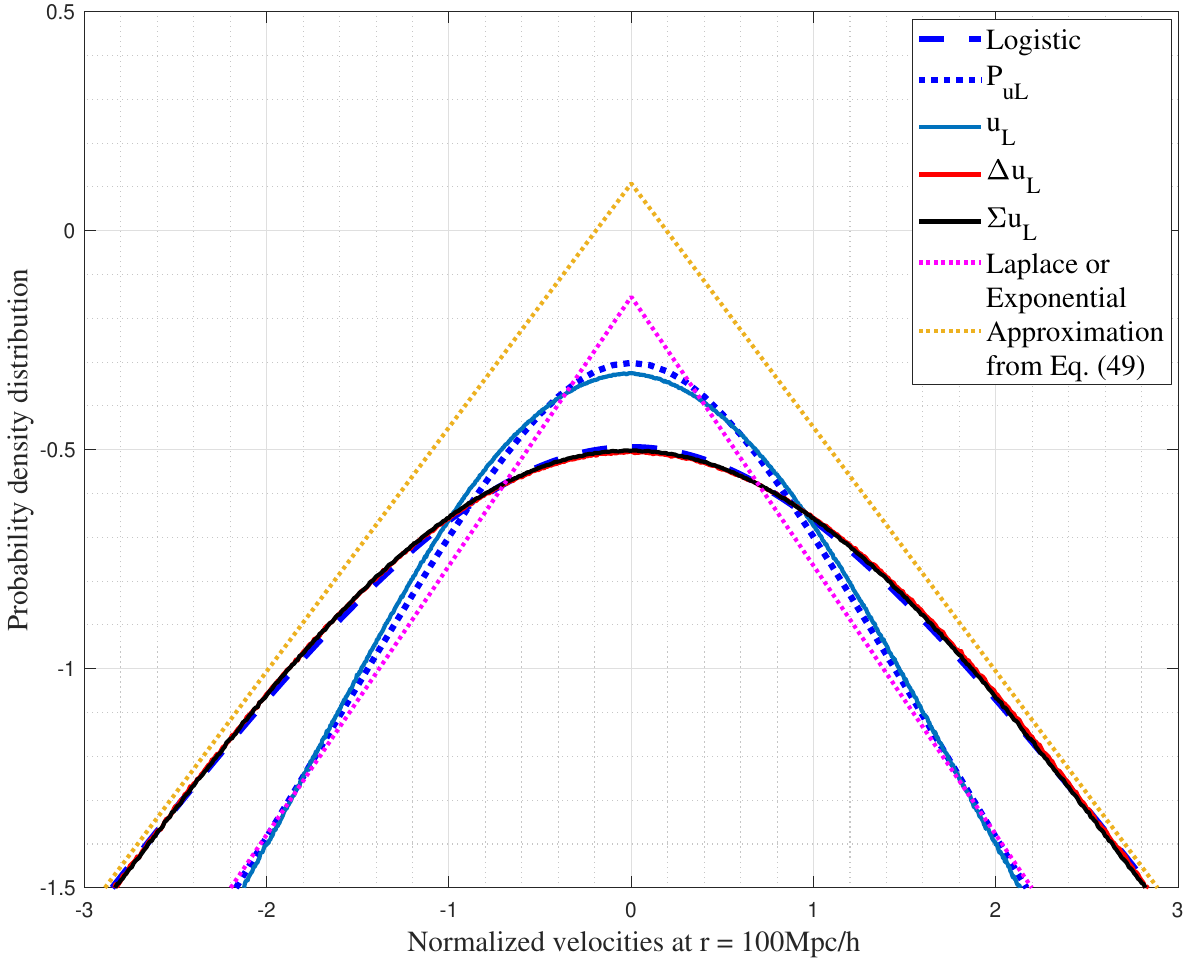}
\caption{Comparison between proposed distributions with simulation data for small velocities. For pairwise velocity $\Delta u_{L}$, the logistic distribution shows better agreement with simulation data than the exponential distribution.}
\label{fig:29b}
\end{figure}

There seems to be no good theory for the distribution of pairwise velocity $\Delta u_{L}$ on large scales, which is usually assumed to be exponential in the literature. However, the exponential distribution is not smooth and non-differentiable at zero velocity (Figs. \ref{fig:29a} and \ref{fig:29b}). The kurtosis from the N-body simulation is not consistent with that of the exponential (or Laplace) distribution (Table \ref{tab:4}). In this section, two options are proposed that are better than a non-smooth exponential distribution. Both options are listed in Table \ref{tab:4}. 

In the first option, a logistic distribution is proposed for both $\Delta u_{L}$ and $\Sigma u_{L}$ with a variance of ${\left(s\pi \right)^{2} /3} =2u^{2}$, where $u^{2}$ is the one-dimensional velocity dispersion of the entire N-body system (or the variance of $u_{L} $ on large scales). The distribution reads 
\[P_{\Delta u_{L} } \left(x\right)=\frac{1}{4s} \sec h^{2} \left(\frac{x}{2s} \right).\] 
For large velocity $x$, the logistic distribution has exponential wing  
\begin{equation} 
\label{ZEqnNum404751} 
P_{\Delta u_{L} } \left(x\to \infty \right)\approx \frac{1}{s} \exp \left(-\frac{x}{s} \right).         
\end{equation} 

Assume $P_{u_{L}}$ is the limiting distribution of $u_{L} $ when $r\to \infty $. With correlation $\rho _{L} =0$ at $r\to \infty $, the distribution of pairwise velocity $\Delta u_L$ and longitudinal velocity $u_L$ should satisfy the convolution
\begin{equation} 
\label{eq:83} 
P_{\Delta u_{L} } \left(z\right)=\int _{-\infty }^{\infty }P_{u_{L} } \left(x\right)P_{u_{L} } \left(z-x\right)dx .         
\end{equation} 
Using the characteristic function, the Fourier transform of two distributions satisfies
\begin{equation} 
\label{ZEqnNum564799} 
\hat{P}_{\Delta u_{L} } \left(t\right)=\left[\hat{P}_{u_{L} } \left(t\right)\right]^{2}.        
\end{equation} 
For a logistic distribution for pairwise velocity $\Delta u_L$, the corresponding moment-generating function of $u_{L}$ can be found from Eq. \eqref{ZEqnNum564799} with a variance of ${\left(\pi s\right)^{2}/6} =u^{2}$,
\begin{equation} 
\label{ZEqnNum593993} 
MGF_{P_{u_{L} } } \left(t\right)=\hat{P}_{u_{L}}=\sqrt{\frac{\pi st}{\sin \left(\pi st\right)} } .         
\end{equation} 

For the second option, the pairwise velocity $\Delta u_{L}$ and the velocity sum $\Sigma u_{L}$ follow the X distribution on large scales. This is suggested by the N-body simulation (Table \ref{tab:4} and Fig. \ref{fig:30}). Similarly, for this option, the distribution of $\Delta u_{L}$ and corresponding moment-generating function of $u_{L}$ are
\begin{equation} 
\label{ZEqnNum404751-2} 
\begin{split}
& P_{\Delta u_{L} } \left(x\right) = X\left(x\right) = \frac{1}{2\alpha v_{0} } \frac{e^{-\sqrt{\alpha ^{2} +\left({x/v_{0} } \right)^{2} } } }{K_{1} \left(\alpha \right)} ,\\   
& MGF_{P_{u_{L} } } \left(t\right)=\sqrt{\frac{K_1\left(\alpha\sqrt{1+(v_0t)^2}\right)}{K_1(\alpha)\sqrt{1+(v_0t)^2}}} .      
\end{split}
\end{equation} 

For both options, the explicit form of the distribution $P_{u_{L}}(x)$ is not available but can be obtained numerically from Eq. \eqref{ZEqnNum593993} or \eqref{ZEqnNum404751-2} using the inverse Fourier transform. The generalized kurtosis of $u_{L}$ can be obtained directly from the moment-generating function for both options and presented in Table \ref{tab:4}. The generalized kurtosis of the logistic distribution slightly agrees better with the simulation than the X distribution. The relevant distributions are also plotted in Figs. \ref{fig:29a} and \ref{fig:29b} and compared to the simulation with good agreement. All distributions are approximately exponential at high velocity.

\subsection{Redshift evolution of velocity distributions}
\label{sec:5.5}
In this section, the redshift evolution of distributions of different types of velocities is presented. This includes the velocity $u_{p} $ of all dark matter particles, the velocity $u_{hp} $ of all halo particles, the velocity $u_{op} $ of all out-of-halo particles, the velocity $u_{h}$ of all haloes, and three types of longitudinal velocities $u_{L} $, $\Delta u_{L}$ and $\sum u_{L}$ on both small and large scales, respectively. Since pairs of particles are from different haloes for a large scale $r$, the velocity $u_{op}$ and $u_{h}$ represent the velocity field on large scales. The velocity $u_{hp}$ of all halo particles represents the velocity field on small scales. The redshift evolution of distributions of these velocities can be characterized by the redshift variation of the generalized kurtosis of these distributions.

If the evolution of a velocity always follows a family of \textit{X} distributions with a shape parameter $\alpha$ that varies over time (Eq. \eqref{ZEqnNum436604}), the redshift evolution of the distribution of that velocity can be reduced to the redshift dependence of the shape parameter $\alpha \equiv \alpha(z)$. In this case, the redshift evolution of velocity distributions can be presented as the evolution of generalized kurtosis, which is a function of the redshift-dependent parameter $\alpha(z)$. The \textit{m}th order kurtosis of the \textit{X} distribution in Eq. \eqref{ZEqnNum436604} can be found as \citep[see ref.][Table 2]{Xu:2023-Maximum-entropy-distributions-of-dark-matter},
\begin{equation} 
\label{ZEqnNum482729} 
K_{m} \left(X\right)=\left(\frac{2K_{1} \left(\alpha \right)}{K_{2} \left(\alpha \right)} \right)^{{m/2} } \frac{\Gamma \left({\left(1+m\right)/2} \right)}{\sqrt{\pi } } \cdot \frac{K_{\left(1+{m/2} \right)} \left(\alpha \right)}{K_{1} \left(\alpha \right)} .      
\end{equation} 

\begin{figure}
\includegraphics*[width=\columnwidth]{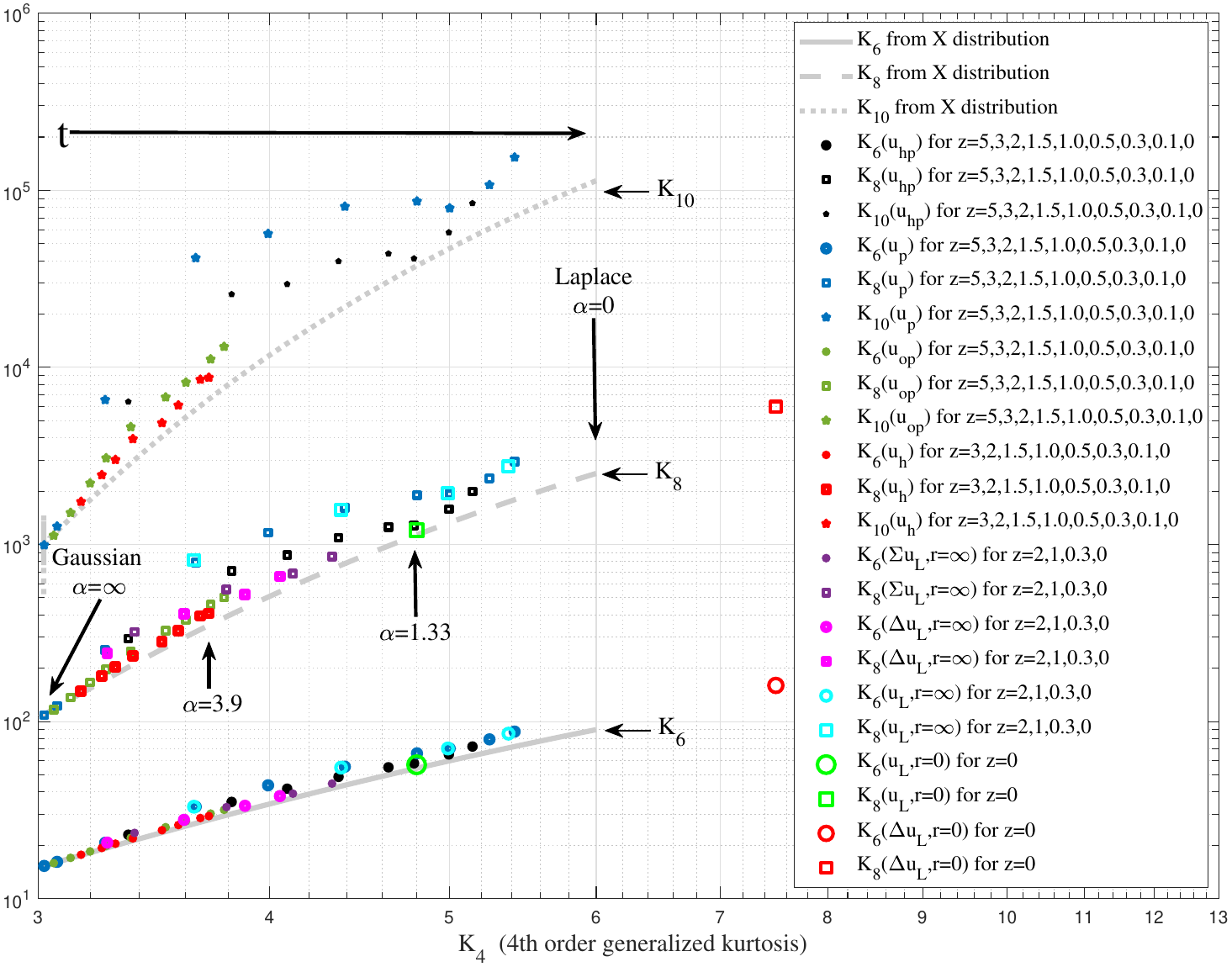}
\caption{The variation of generalized kurtosis $K_{6} $, $K_{8} $, and $K_{10} $ with $K_{4}$ for X distribution (gray lines) and for different types of velocities (symbols). In principle, the kurtosis of all different velocities increases with time, indicating the redshift evolution of velocity distributions to maximize the system entropy. These include the velocity $u_{p} $ of all particles, the velocity $u_{hp} $ of all halo particles, the velocity $u_{op} $ of all out-of-halo particles, and the velocity $u_{h} $ of all haloes. By identifying the particle pairs on a given scale $r$, the distributions of the longitudinal velocity $u_{L} $, the pairwise velocity $\Delta u_{L} $, and the velocity sum $\sum u_{L} $ on small and large scales are also presented. All velocities are initially Gaussian with the shape parameter $\alpha =\infty$ and gradually become non-Gaussian with decreasing $\alpha$. The evolution (approximately) follows the prediction (gray lines) of the \textit{X} distribution. The distributions of out-of-halo particles $u_{op}$ and the velocity of the halo $u_{h}$ match each other and evolve at a much slower pace compared to the velocity of the halo particles $u_{hp}$.}
\label{fig:30}
\end{figure}

Figure \ref{fig:30} presents the redshift evolution of velocity distributions in terms of kurtosis of different order (4${}^{th}$, 6${}^{th}$, 8${}^{th}$, and 10${}^{th}$) from both the simulation (symbols) and Eq. \eqref{ZEqnNum482729} (gray lines). All velocities are initially Gaussian. With increasing time from left to right, all velocities become non-Gaussian, and the evolution approximately follows the prediction of the \textit{X} distribution with decreasing $\alpha $. The halo velocity ($u_{h}$), the out-of-halo particle velocity ($u_{op}$), and the halo particle velocity ($u_{hp}$) should all follow a \textit{X} distribution to maximize the system entropy, just as the longitudinal velocity $u_L$ on small scales (Eq. \eqref{ZEqnNum436604}). The halo velocity ($u_{h}$) and the out-of-halo particle velocity ($u_{op}$) follow similar distributions that evolve much slower than the evolution of the distribution of halo particle velocity ($u_{hp}$) because of stronger gravity on small scales. This is also consistent with the fact that virial equilibrium is established much faster for halo particles on small scales (owing to stronger gravity) than for the haloes themselves, which are on large scales. 

The longitudinal velocity $u_L$ and the velocity sum $\Sigma u_L$ follow the \textit{X} distribution on small scales, while the pairwise velocity $\Delta u_L$ follows the distribution given by Eq. \eqref{eq:78}. On large scales, the pairwise velocity $\Delta u_L$ and the velocity sum $\Sigma u_L$ follows the \textit{X} distribution approximately, while the longitudinal velocity $u_L$ follows the distribution given by Eq. \eqref{ZEqnNum404751-2}.

\section{Probability distributions of density field}
\label{sec:3}
Various statistical measures can be introduced to characterize the velocity field in self-gravitating collisionless flow \citep{Xu:2023-On-the-statistical-theory-of-self-gravitating, Xu:2024-On-the-statistical-theory-of-self-gravitating-high-order}, i.e., the real-space correlation, dispersion and structure functions, and power spectrum functions in Fourier space. They are related to each other through the kinematic and dynamic relations. The real-space correlation functions are the most fundamental quantity and building blocks of statistical theory. This section extends the statistical approach for the velocity field to the density field. Analytical models are also presented while available.

\subsection{One-point probability distributions}
\label{sec:3.1}
Projecting a particle field onto a structured grid usually involves information loss and numerical noise. Without projecting onto the grid, Delaunay tessellation is used in this section to reconstruct the density field and to maximally preserve the information from the N-body simulation data. For a particle at location \textbf{x}, the particle overdensity $\delta \left(\boldsymbol{\mathrm{x}}\right)$ and log-density $\eta \left(\boldsymbol{\mathrm{x}}\right)$ are defined as
\begin{equation} 
\label{eq:2} 
\delta \left(\boldsymbol{\mathrm{x}}\right)=\frac{\rho \left(\boldsymbol{\mathrm{x}}\right)}{\rho_{0} } -1, \quad  \eta \left(\boldsymbol{\mathrm{x}}\right)=\log \left(1+\delta \left(\boldsymbol{\mathrm{x}}\right)\right)=\log \left(\frac{\rho \left(\boldsymbol{\mathrm{x}}\right)}{\rho_{0} } \right), 
\end{equation}
where $\rho \left(\boldsymbol{\mathrm{x}}\right)={m_{p} /V_{p} } $ is a local matter density at comoving coordinate \textbf{x, }$m_{p} $ is the particle mass, $V_{p} $ is the volume occupied by that particle, and $\rho _{0} $ is the mean (comoving) density. In linear theory, $\eta \left(\boldsymbol{\mathrm{x}}\right)\approx \delta \left(\boldsymbol{\mathrm{x}}\right)$ for small overdensity $\delta \left(\boldsymbol{\mathrm{x}}\right)\ll 1$ on large scales. They are different on small scales in the non-linear regime. Due to the normalization that the total volume should be equal to the sum of all particle volume ($\sum V_{p}  =V$), the redshift evolution of the distributions of $\delta $ and $\eta $ should always satisfy
\begin{equation}
\left\langle \frac{1}{1+\delta \left(\boldsymbol{\mathrm{x}}\right)} \right\rangle =1 \quad \textrm{and} \quad \left\langle e^{-\eta \left(\boldsymbol{\mathrm{x}}\right)} \right\rangle =1.     
\label{eq:3}
\end{equation}

\begin{figure}
\includegraphics*[width=\columnwidth]{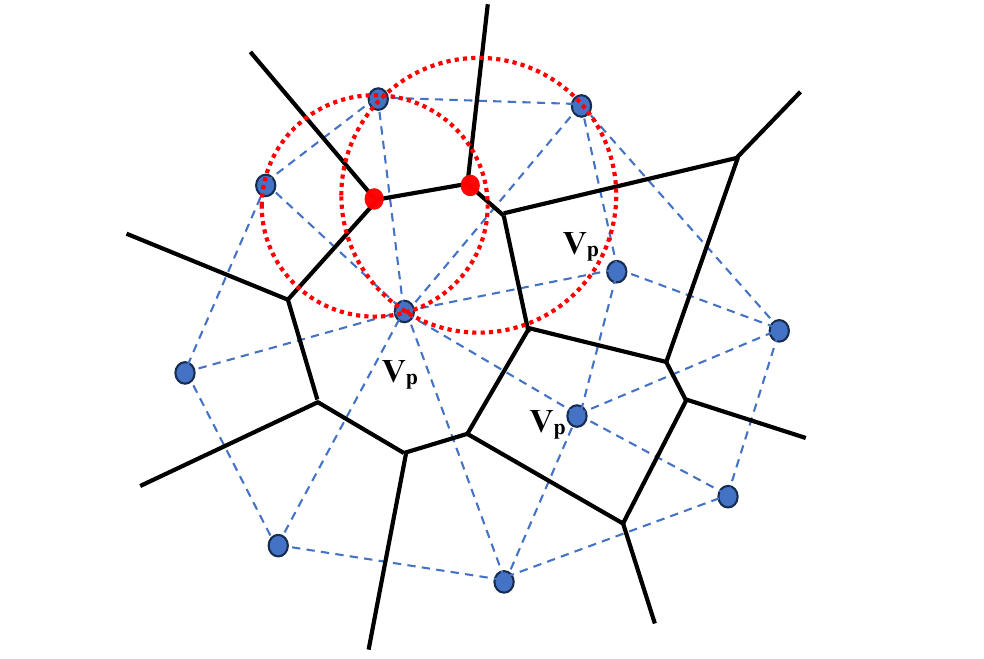}
\caption{Schematic plot for the Delaunay tessellation in two-dimension to reconstruct the density field from a discrete set of particles. The first step is to connect all dark matter particles (blue dots) with a set of non-overlapping triangles (dashed lines). The second step is to find the circumcenter (red dots) of each triangle and connect them to form polygons (black solid lines). In two-dimension, the volume $V_{p}$ that each dark matter particle occupies can be determined from the area of the polygon it resides in. The density of each particle can be calculated from the particle volume $V_p$.}
\label{fig:S2}
\end{figure}

\rev{
Different from the velocity field, particle density is not a field variable that is automatically computed for every particle in an N-body simulation. Delaunay tessellation can be applied to reconstruct the density field from a discrete set of particles \citep{Romano-Diaz:2007-Delaunay, Bernardeau:1996-A-new-method-for-accurate-esti}. Figure \ref{fig:S2} presents a brief description of Delaunay tessellation in two-dimension. Generalization to three-dimension should be straightforward. All particles in the system are first connected by a set of non-overlapping tetrahedral (triangles in two dimensions). The volume $V_{p}$ that each particle occupies can be determined from the volume of its surrounding tetrahedral. The density $\rho \left(\boldsymbol{\mathrm{x}}\right)$ of each particle can be calculated from the particle volume $V_p$. This enables us to compute the density distribution for halo particles and out-of-halo particles, respectively.
}

\begin{subfigures} 
\begin{figure}
\includegraphics*[width=\columnwidth]{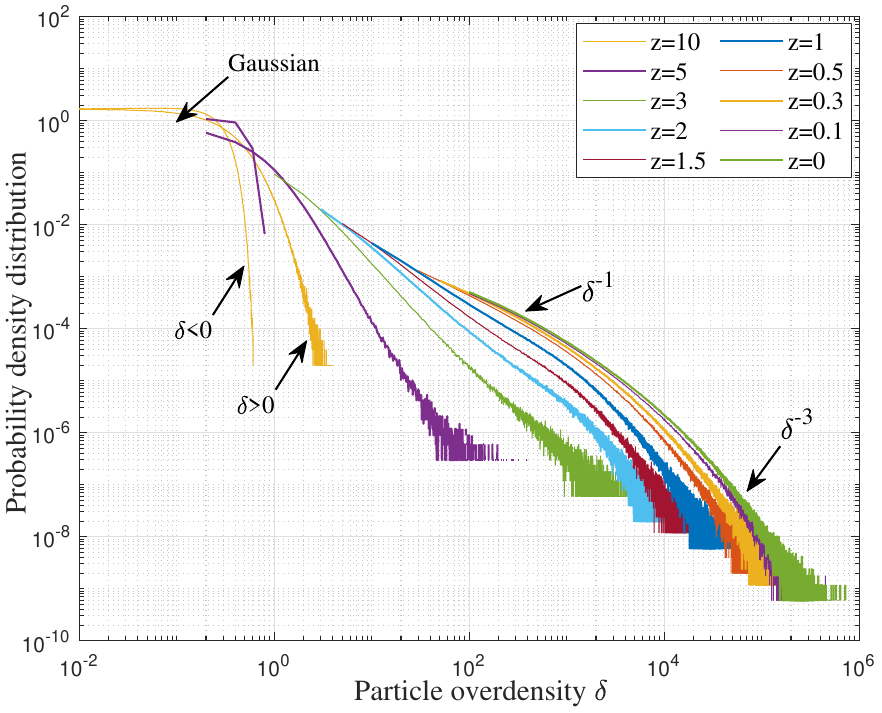}
\caption{The redshift evolution of the distribution of particle density $\delta $ from \textit{z}=10 to \textit{z}=0. Density evolves from an initial Gaussian distribution at high redshift with symmetric branches ($\delta >0$ and $\delta <0$) to a non-Gaussian distribution with a positive ($\delta >0$) branch and a long tail $\propto \delta ^{-3}$.}
\label{fig:1a}
\end{figure}
\begin{figure}
\includegraphics*[width=\columnwidth]{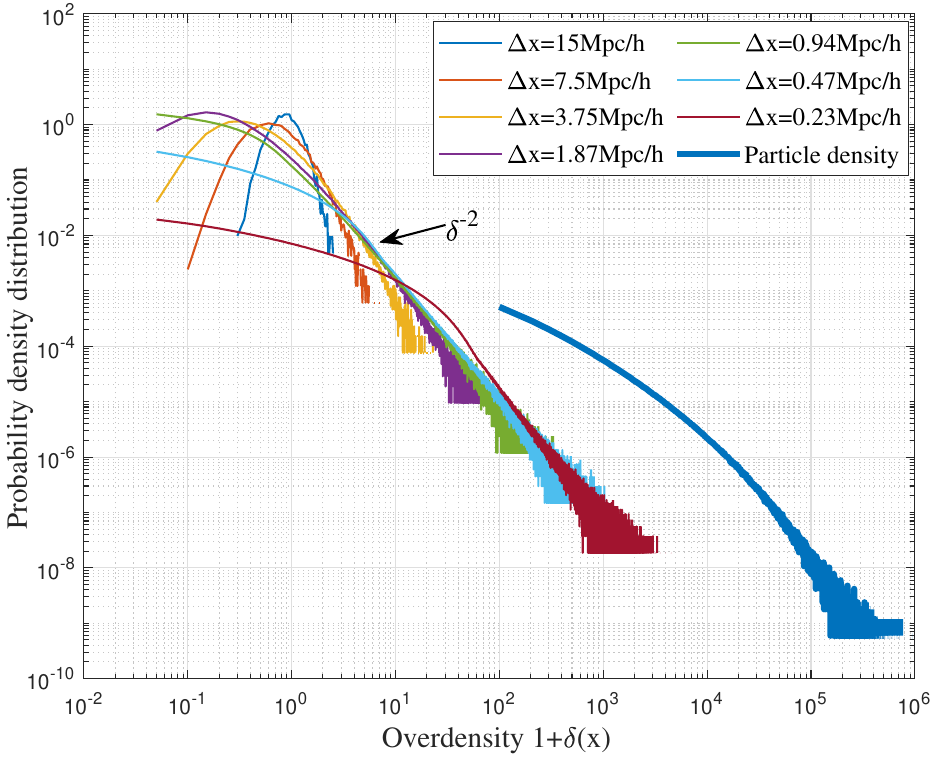}
\caption{The distribution of grid-based density $\left(1+\delta \right)$ obtained by projecting particles onto a structured grid of different grid size $\Delta x$ at redshift \textit{z}=0. On large scales (large $\Delta x$), the density distribution is Gaussian and symmetric. On small scales, the distribution extends to large density and becomes non-Gaussian with an approximate scaling of $\propto \delta ^{-2}$. The particle density from Delaunay tessellation (corresponding to $\Delta x\to 0$) is also plotted for comparison that can reach greater density values than grid-based density.} 
\label{fig:1b}
\end{figure}
\end{subfigures}

By calculating the density for each particle, Fig. \ref{fig:1a} presents the redshift evolution of the one-point density distribution $\delta \left(\boldsymbol{\mathrm{x}}\right)$ for all particles in the N-body system. Due to gravitational collapse on small scales, $\delta \left(\boldsymbol{\mathrm{x}}\right)$ evolves from an initial Gaussian (symmetric) at high redshift to a "double power law" distribution (asymmetric and highly skewed toward $\delta >0$) at \textit{z}=0 with a long tail $\propto \delta ^{-3} $. The distribution is approximately $\propto \delta ^{-1} $ for small $\delta $. 

For comparison, the density distribution can also be obtained by projecting particles onto the structured grid using the Cloud-in-Cell (CIC) scheme with a given grid size ($\Delta x$). The results of the grid-based density distributions for different grid sizes $\Delta x$ are presented in Fig. \ref{fig:1b}. An approximate scaling of $\propto \delta ^{-2}$ is consistent with the literature \citep{Klypin:2018-Density-distribution-of-the-co}. For grid-based density, $\left\langle \delta \right\rangle =0$. Due to the limit of grid resolution, the grid-based density is much smaller than the particle density directly obtained from Delaunay tessellation (thick solid blue line).

\begin{figure}
\includegraphics*[width=\columnwidth]{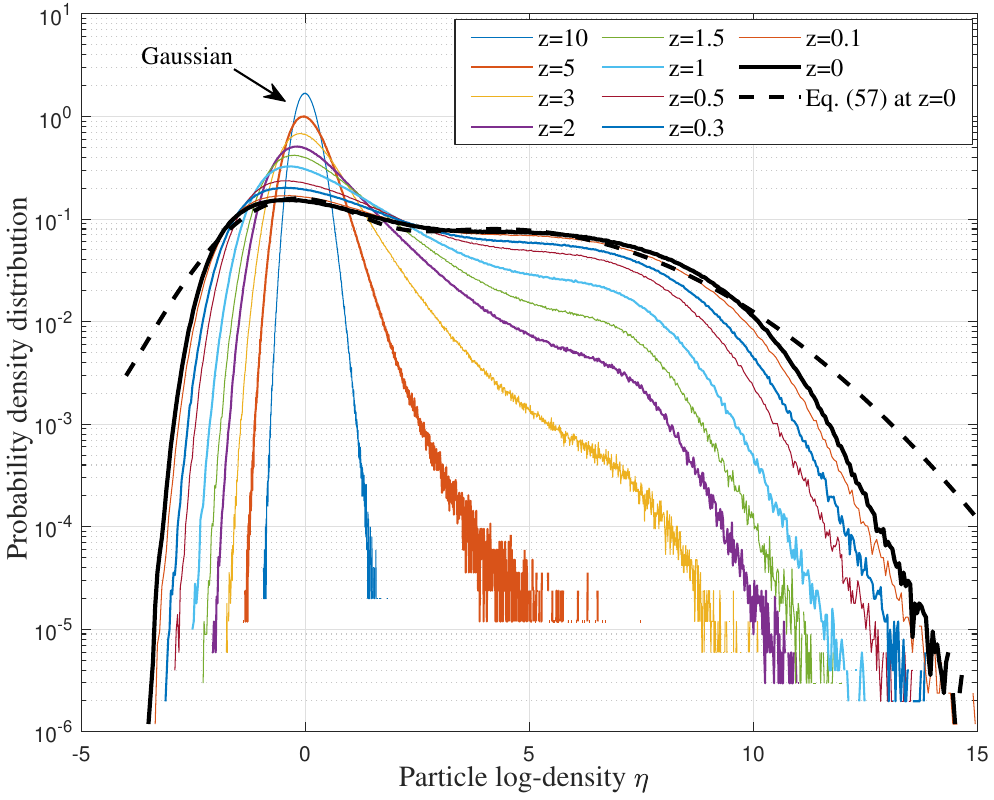}
\caption{The distribution of the particle log-density $\eta \left(\boldsymbol{\mathrm{x}}\right)$ at different redshifts z. The log-density $\eta \left(\boldsymbol{\mathrm{x}}\right)$ evolves from a relatively Gaussian at high redshift to a bimodal distribution at \textit{z}=0 with two peaks corresponding to halo (60\%) and out-of-halo (40\%) particles. Inverse mass cascade leads to continuous halo structure formation and the two peaks in density distribution.} 
\label{fig:2}
\end{figure}

Similarly, Fig. \ref{fig:2} plots the redshift evolution of the log-density distribution $\eta \left(\boldsymbol{\mathrm{x}}\right)$ from \textit{z}=10 to \textit{z}=0. A bimodal distribution is gradually developed from an initial Gaussian distribution. The first peak corresponds to out-of-halo particles in the low-density region that do not belong to any haloes with $\left\langle \eta \right\rangle <0$. The second peak comes from all halo particles in haloes with higher density and wider dispersion. While other better fittings are possible, a simple bimodal equation is used here to fit this distribution to provide additional insight,
\begin{equation} 
\label{eq:4} 
f\left(\eta \right)=\frac{c_{1} }{\sqrt{2\pi } \sigma _{1} } \exp \left[\frac{\left(\eta -\mu _{1} \right)^{2} }{2\sigma _{1}^{2} } \right]+\frac{1-c_{1} }{\sqrt{2\pi } \sigma _{2} } \exp \left[\frac{\left(\eta -\mu _{2} \right)^{2} }{2\sigma _{2}^{2} } \right] 
\end{equation} 
with best fitting parameters $c_{1} =0.404$, $\mu _{1} =-0.30$, $\sigma _{1} =1.212$, $\mu _{2} =4.256$, $\sigma _{2} =2.979$ at \textit{z}=0. The fitted curve is plotted in the same figure with about 60\% particles in haloes and 40\% out-of-halo particles. This is consistent with the prediction from the inverse mass cascade \citep{Xu:2021-Inverse-mass-cascade-mass-function}, i.e., 60\% of the total mass is in all haloes at $z=0$. There is a continuous injection of mass from the out-of-halo into the haloes as a result of the inverse mass cascade. The particles in haloes should have an average density close to $\delta =\Delta _{c} -1$, where the critical density ratio $\Delta _{c} =18\pi ^2$ from a spherical collapse model or a two-body collapse model 
such that $\left\langle \eta \right\rangle \approx 5.17$ (matches $\mu_{2}$, which is the mean density for all halo particles). 

\begin{figure}
\includegraphics*[width=\columnwidth]{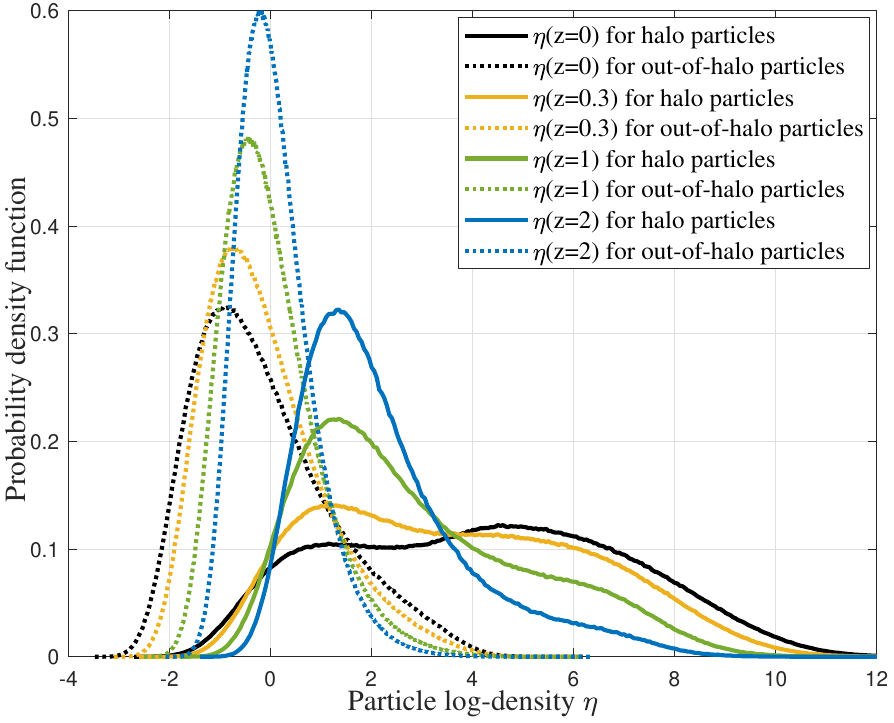}
\caption{Redshift evolution of log-density distributions $\eta \left(z\right)$ for two different types of particles. For out-of-halo particles, the distribution is relatively Gaussian with a decreasing and negative mean density, with more and more out-of-halo particles forming haloes. This corresponds to a lognormal distribution of particle density $\delta \left(\boldsymbol{\mathrm{x}}\right)$ on large scales. For halo particles, the distribution evolves with an increasing mean log-density. A peak develops at around $\eta=5$ corresponding to the critical density ratio $\Delta_c=18\pi^2$.} 
\label{fig:3}
\end{figure}

It is also natural to check the density distributions of halo particles and out-of-halo particles separately. By identifying all haloes in the entire system and dividing all particles into halo and out-of-halo particles, Fig. \ref{fig:3} presents the redshift evolution of the distributions of log-density $\eta \left(\boldsymbol{\mathrm{x}}\right)$ for the halo and out-of-halo particles, respectively. For out-of-halo particles, the distribution of $\eta$ is Gaussian, with a mean density decreasing over time. The distribution of $\delta$ is approximately log-normal for out-of-halo particles or a log-normal density distribution on large scales \citep{Hubble:1934-The-distribution-of-extra-gala}. However, for halo particles, the distribution is non-Gaussian and evolves with increasing mean density as a result of the formation and growth of haloes. A peak develops around $\eta=5$ at $z$=0 corresponding to the critical density ratio $\Delta_c=18\pi^2$ for haloes.

Similarly to the velocity field, to characterize the redshift evolution of the distribution of any random variable $\tau $, statistical quantities such as skewness and kurtosis should be used. A generalized kurtosis $K_{n} \left(\tau \right)$ for the variable $\tau$ is defined as 
\begin{equation}
\label{ZEqnNum270889} 
K_{n} \left(\tau \right)=\frac{\left\langle \left(\tau -\left\langle \tau \right\rangle \right)^{n} \right\rangle }{\left\langle \left(\tau -\left\langle \tau \right\rangle \right)^{2} \right\rangle ^{{n/2} } } =\frac{S_{n}^{cp} \left(\tau \right)}{S_{2}^{cp} \left(\tau \right)^{{n/2} } } ,        
\end{equation} 
where the central moment of order \textit{n} for random variable $\tau$ reads
\begin{equation}
\label{ZEqnNum505648} 
S_{n}^{cp} \left(\tau \right)=\left\langle \left(\tau -\left\langle \tau \right\rangle \right)^{n} \right\rangle .           
\end{equation} 
The odd-order kurtosis should vanish for symmetric distributions. Specifically for Gaussian distribution, $K_{3} =K_{5} =0$, $K_{2} =1$, $K_{4} =3$, $K_{6} =15$, and $K_{8} =105$. 
\begin{figure}
\includegraphics*[width=\columnwidth]{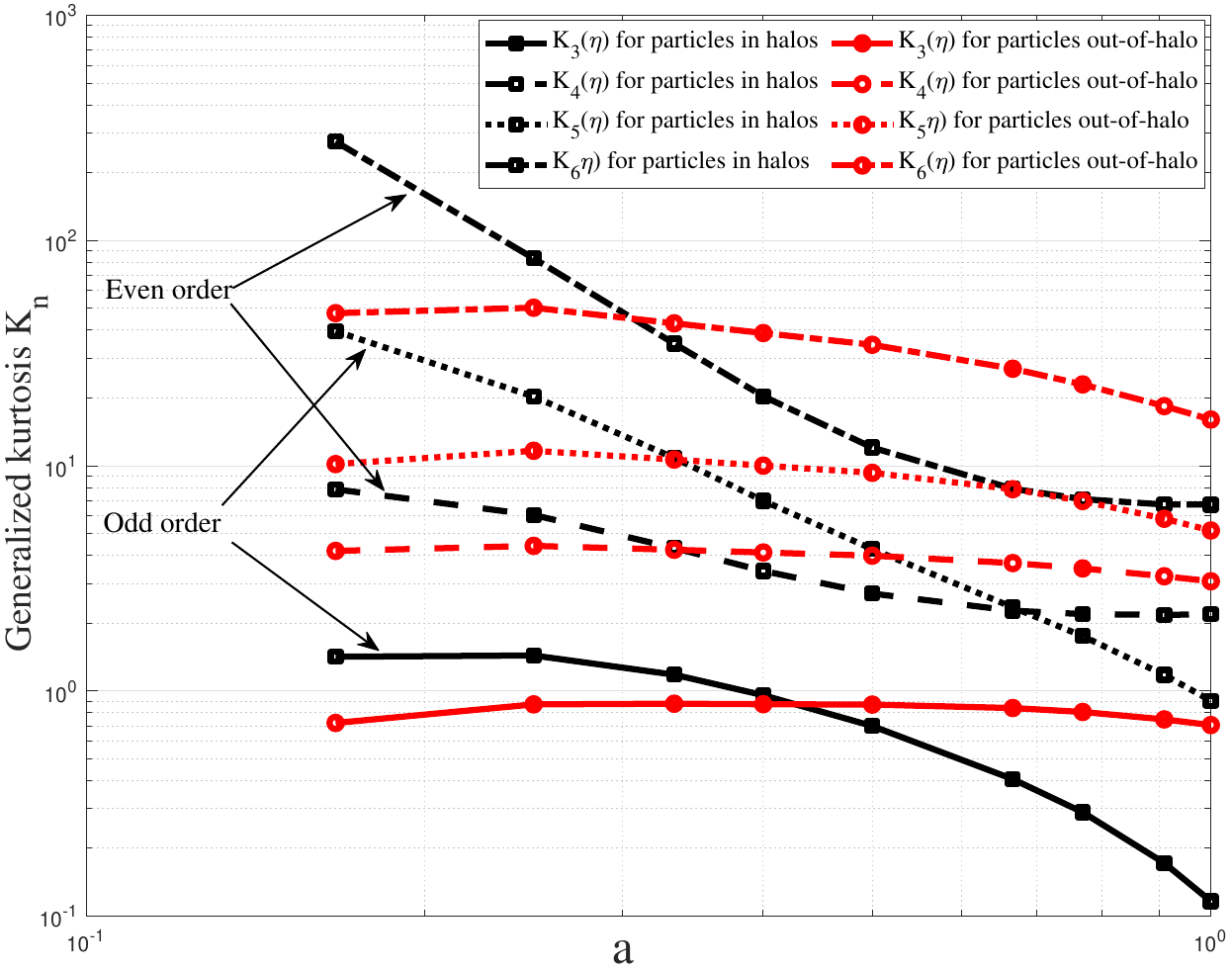}
\caption{The redshift evolution of generalized kurtosis for distribution of log-density $\eta$ for two different types of particles. The density distribution for the out-of-halo particles is relatively Gaussian with generalized kurtosis $K_{4} \approx 3$ and $K_{6} \approx 15$ at \textit{z}=0. The density distribution for halo particles becomes more symmetric with vanishing odd-order kurtosis, while even-order kurtosis $K_{4} \to 2$ and $K_{6} \to 7$.}
\label{fig:4}
\end{figure}

Figure \ref{fig:4} presents the redshift evolution of generalized kurtosis. For the density of out-of-halo particles, kurtosis ($K_{3} \left(\eta \right)$ to $K_{6} \left(\eta \right)$) is relatively independent of time. The distribution is relatively Gaussian with $K_{4}\approx$3 and $K_{6}\approx$15 at \textit{z}=0, such that the distribution of $\delta$ for out-of-halo particles is approximately log-normal. The density distribution of the halo particles approaches a symmetric distribution with odd-order kurtosis approaching zero and even-order kurtosis $K_{4} \to 2$ and $K_{6} \to 7$. 

\begin{figure}
\includegraphics*[width=\columnwidth]{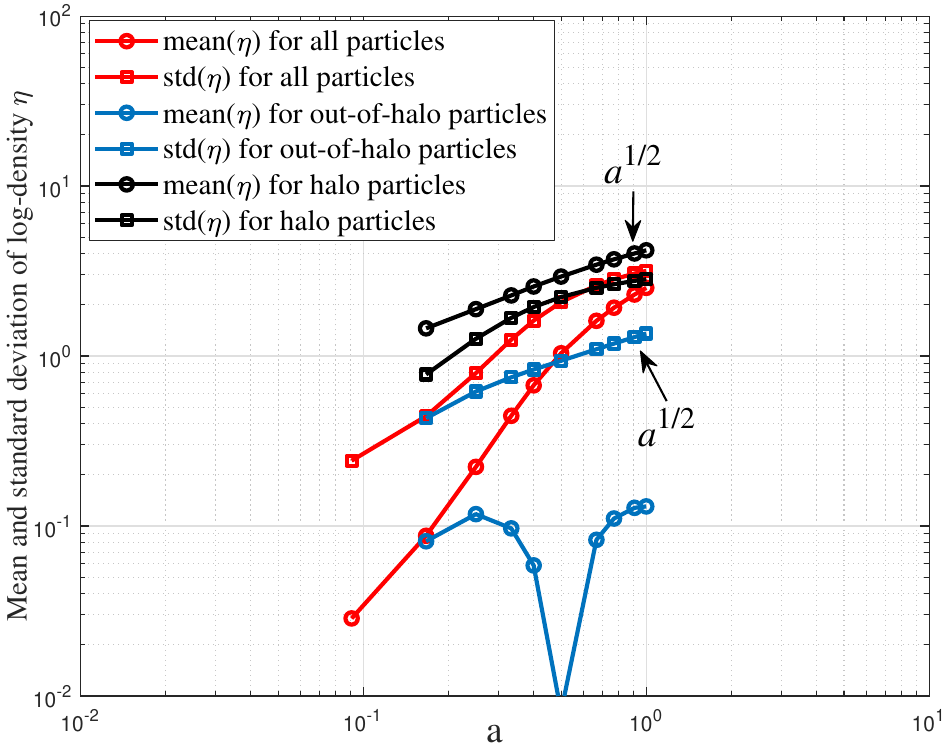}
\caption{The variation of the mean and standard deviation of log-density $\eta \left(\boldsymbol{\mathrm{x}},z\right)$ with scale factor \textit{a}. With more particles forming haloes and fewer out-of-halo particles, the mean log-density of out-of-halo particles decreases with time and $\left\langle \eta \right\rangle <0$ after \textit{z}=1. The mean log-density of all halo particles increases with time, and the standard deviation $std\left(\eta \right)\propto a^{{1/2} }$.}  
\label{fig:5}
\end{figure}

Figure \ref{fig:5} plots the variation in the mean and standard deviation of the log-density distribution with the scale factor $a$. For out-of-halo particles, the mean log-density decreases with time and $\left\langle \eta \right\rangle <0$ after \textit{z}=1 (or \textit{a}=0.5). While the mean log-density of halo particles increases with time, i.e., $\langle \eta \rangle\propto a^{{1/2}}$. The power-law scaling of $std\left(\eta \right)\propto a^{{1/2}}$ can also be found for both halo and out-of-halo particles, reflecting the spreading of particle density due to continuous mass accretion. With more particles forming haloes and fewer out-of-halo particles, the density of out-of-halo particles extends to lower values. In contrast, the density of halo particles extends to higher values.

\subsection{Two-point second-order statistical measures}
\label{sec:3.2}
\subsubsection{Density correlation from radial distribution function}
\label{sec:3.2.1}
The gravitational interaction between collisionless particles leads to correlations in the position of the particles. Following the statistical mechanics of the molecular liquid, we start with the radial distribution function $g\left(r\right)$. This quantity is used to measure the average particle density around an arbitrary reference particle. The number of particles in the spherical shell of thickness \textit{dr} at a distance \textit{r} from the reference particle can be written as:
\begin{equation}
\label{ZEqnNum535068} 
dN_{p} =g\left(r\right)\frac{N_{p} }{V} 4\pi r^{2} dr,         
\end{equation} 
where ${N_{p} /V} $ is the mean density of particles, $N_{p}$ is the total number of particles in the system, and \textit{V} is the volume. The mean comoving density $\rho _{0} ={N_{p} m_{p} /V} $. The normalization condition reads 
\begin{equation} 
\label{eq:8} 
\int _{0}^{\infty }g\left(r\right)4\pi r^{2} dr =\frac{N_{p} -1}{N_{p} } V.         
\end{equation} 

The two-point second order density correlation function is given by $\xi \left(r\right)$ that is related to the radial distribution function $g\left(r\right)$ as
\begin{equation}
\label{ZEqnNum676567} 
\xi \left(r\right)=\left\langle \delta \left(\boldsymbol{\mathrm{x}}\right)\delta \left(\boldsymbol{\mathrm{x}}+\boldsymbol{\mathrm{r}}\right)\right\rangle =g\left(r\right)-1.
\end{equation} 
The normalization condition for density correlation reads (Eq. \eqref{eq:8})
\begin{equation} 
\label{ZEqnNum278276} 
\int _{0}^{\infty }\xi \left(r,z\right)4\pi r^{2} dr =-{V/N} _{p} <0.        
\end{equation} 
Here, we find that the redshift-dependent density correlation $\xi \left(r,z\right)$ cannot be positive on all scales at any given time. Density must be negatively correlated on some scales. 

Two length scales can be defined from the moments of density correlation (see Fig. \ref{fig:10}),
\begin{equation} 
\label{ZEqnNum811732} 
l_{\delta 0} \left(a\right)=\int _{0}^{\infty }\xi \left(r,z\right)dr,\quad l_{\delta 1}^{2} \left(a\right)=\int _{0}^{\infty }\xi \left(r,z\right)rdr .    
\end{equation} 
\subsubsection{Potential and kinetic energy from density correlation}
\label{sec:3.2.2}
In principle, the specific potential energy (per mass) of any system with particles interacting via a pairwise potential $V_{g} \left(r\right)$ can be related to the radial distribution function $g\left(r\right)$ as
\begin{equation} 
\label{eq:12} 
PE=\frac{2\pi \rho _{0} }{m_{p}^{2} } \int _{0}^{\infty }r^{2} \left[g\left(r\right)-1\right]V_{g} \left(r\right)dr ,        
\end{equation} 
where $\rho _{0} $ is the mean density. With $V_{g} \left(r\right)=-{Gm_{p}^{2} /r} $ for gravity, the specific potential energy of any N-body system reads
\begin{equation} 
\label{ZEqnNum101644} 
P_{y} \left(a\right)=-\frac{2\pi G\rho_{0} }{a} \int _{0}^{\infty }\xi \left(r,a\right)rdr=-\frac{3H_{0}^{2} l_{\delta 1}^{2} }{4a}<0.      
\end{equation} 

The specific kinetic energy of the entire system can be related to the potential energy via a cosmic energy equation \citep{Irvine:1961-Local-Irregularities-in-a-Univ, Layzer:1963-A-Preface-to-Cosmogony--I--The, Mo:1997-Analytical-approximations-to-t, Xu:2022-The-evolution-of-energy--momen},
\begin{equation} 
\label{eq:14} 
\frac{\partial \left(K_{p} +P_{y} \right)}{\partial t} +H\left(2K_{p} +P_{y} \right)=0,         
\end{equation} 
with an exact solution of 
\begin{equation} 
\label{eq:15} 
K_{p} =a^{-2} \int _{0}^{a}aP_{y}  da-P_{y} .         
\end{equation} 

Substituting Eq. \eqref{ZEqnNum101644} into \eqref{eq:15}, the specific kinetic energy can be related to the density correlation \citep{Sheth:2001-Linear-and-non-linear} and length scale $l_{\delta1}$,
\begin{equation} 
\label{eq:16} 
\begin{split}
K_{p} &=\frac{3}{4} H_{0}^{2} a^{-1} \left[\int _{0}^{\infty }\xi \left(r,z\right)rdr -a^{-1} \int _{0}^{a}\left(\int _{0}^{\infty }\xi \left(r,z\right)rdr \right) da\right]\\&=\frac{3}{4} H_{0}^{2} a^{-1} \left(l_{\delta 1}^{2} -a^{-1} \int _{0}^{a}l_{\delta 1}^{2}  da\right).
\end{split}
\end{equation} 

Based on the theory of energy cascade for self-gravitation collisionless flow (SG-CFD) \citep{Xu:2022-Postulating-dark-matter-partic}, the kinetic energy should evolve linearly over time $t$. Therefore, the evolution of the kinetic and potential energy of an N-body system in an expanding background can be modeled by a power-law solution \citep{Xu:2023-Dark-matter-halo-mass-functions-and, Xu:2022-Postulating-dark-matter-partic},
\begin{equation} 
\label{ZEqnNum922690} 
K_{p} =-\varepsilon _{u} t \quad \textrm{and} \quad P_{y} =\frac{7}{5} \varepsilon _{u} t,         
\end{equation} 
which exactly satisfies the cosmic energy equation (Eq. \eqref{eq:14}). Here, the rate of energy cascade $\varepsilon _{u}$ is a negative constant reflecting the inverse cascade from small to large scales (Fig. \ref{fig:17}) such that
\begin{equation} 
\label{eq:18} 
\varepsilon _{u} =-\frac{3}{2} \frac{\partial u^{2} }{\partial t} \approx -\frac{3}{2} \frac{u_{0}^{2} }{t_{0}}\approx -10^{-7}m^2/s^3 .         
\end{equation} 
The length scale $l_{\delta 1}^{2}$ may be related to the rate of energy cascade $\varepsilon_u$ as (using Eqs. \eqref{ZEqnNum101644} and \eqref{ZEqnNum922690}),
\begin{equation} 
\label{ZEqnNum402409} 
l_{\delta 1}^{2} \left(a\right)=\int _{0}^{\infty }\xi \left(r,z\right)rdr =-\frac{56}{45} \frac{\varepsilon _{u} }{H_{0}^{3} } a^{{5/2} } .       
\end{equation} 

\begin{figure}
\includegraphics*[width=\columnwidth]{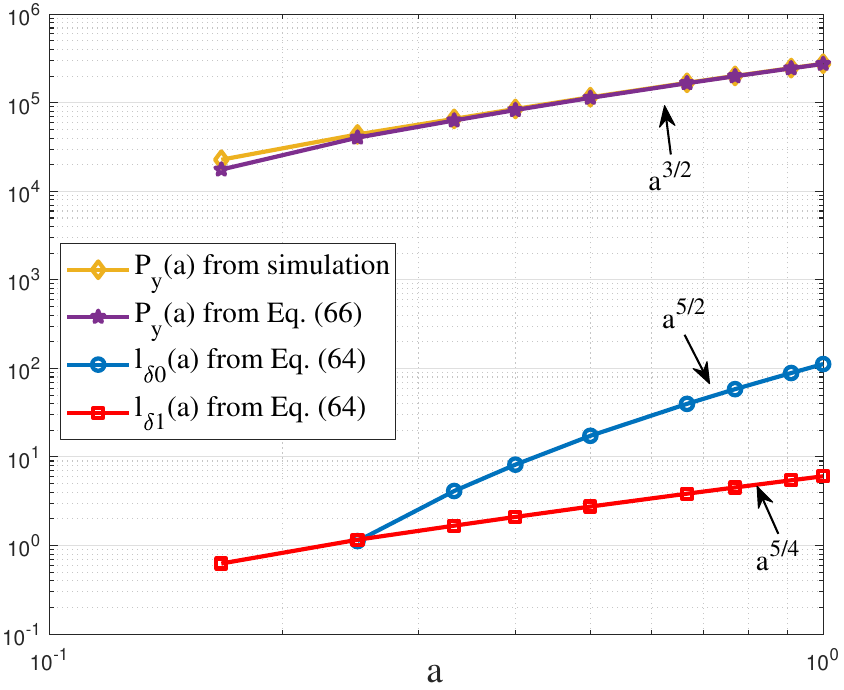}
\caption{The variation of two correlation lengths $l_{\delta 0} (Mpc/h)$ and $l_{\delta 1} (Mpc/h)$ with the scale factor \textit{a}. Both correlation lengths are derived from the density correlation $\xi \left(r,a\right)$ (Eq. \eqref{ZEqnNum811732}) with a limiting scaling $l_{\delta 1} \left(a\right)\propto a^{{5/4}}$ (Eq. \eqref{ZEqnNum402409}) and $l_{\delta 0} \left(a\right)\propto a^{{5/2}}$. The density correlation $\xi \left(r,z\right)$ from the N-body simulation is presented in Fig. \ref{fig:6}. Potential energy $P_{y} \left(a\right)$ (in units of $(Km/s)^{2}$) using Eq. \eqref{ZEqnNum101644} is in good agreement with $P_{y} \left(a\right)$ which is directly computed from simulation, both of which show scaling of $P_{y} \left(a\right)\propto a^{{3/2}}$.}
\label{fig:10}
\end{figure}

Figure \ref{fig:10} presents the variation of two length scales from the N-body simulation (defined in Eq. \eqref{ZEqnNum811732}) with the scale factor \textit{a}. Two comoving correlation lengths show a limiting scaling of $l_{\delta 0} \left(a\right)\propto a^{{5/2} } $ and $l_{\delta 1} \left(a\right)\propto a^{{5/4} } $. The specific potential energy computed by Eq. \eqref{ZEqnNum101644} using $l_{\delta 1} $ is in good agreement with the potential energy directly obtained from the simulation. Both have a limiting scaling of $P_{y} \left(a\right)\propto a^{{3/2} } $ (see Eq. \eqref{ZEqnNum922690}).

\subsubsection{Density correlation, spectrum, and dispersion functions}
\label{sec:3.2.3}
Similarly to the statistical measures of the velocity filed \citep{Xu:2023-On-the-statistical-theory-of-self-gravitating}, in this section, we focus on the second-order statistical measures for the density field that can be directly obtained from the N-body simulation. Algorithms were developed to find all pairs of particles with a given separation \textit{r} and compute the average of these statistical measures for all pairs with the same \textit{r}. We first calculate the radial distribution function $g\left(r\right)$ (Eq. \eqref{ZEqnNum535068}) by counting the number of all pairs at a given distance of \textit{r}. The density correlation can be obtained from $g\left(r\right)$ by Eq. \eqref{ZEqnNum676567}. Using this approach, we avoid projecting a particle field onto the structured grid and maximally preserve information from the N-body simulation. 

The density spectrum function $E_{\delta } \left(k,z\right)$ in the Fourier space can be obtained from the density correlation function $\xi \left(r,z\right)$. The density spectrum and correlation function are related through a pair of Fourier transformations:
\begin{equation} 
\label{ZEqnNum800171} 
E_{\delta } \left(k,z\right)=\frac{2}{\pi } \int _{0}^{\infty }\xi \left(r,z\right)kr\sin \left(kr\right)dr ,        
\end{equation} 
\begin{equation}
\label{ZEqnNum154392} 
\xi \left(r,z\right)=\int _{0}^{\infty }E_{\delta } \left(k,z\right)\frac{\sin \left(kr\right)}{kr} dk.
\end{equation} 
In Peebles' convention \citep{Peebles:1980-The-Large-Scale-Structure-of-t}, the matter power spectrum $P_{\delta } \left(k,z\right)$ is related to the density spectrum function as 
\begin{equation} 
\label{ZEqnNum404986} 
P_{\delta } \left(k,z\right)={2\pi ^{2} E_{\delta } \left(k,z\right)/k^{2} } .         
\end{equation} 
The dimensionless power spectrum $\Delta _{\delta }^{2} \left(k,z\right)$ (the power per logarithmic interval) can be related to the density spectrum as 
\begin{equation} 
\label{ZEqnNum502544} 
\Delta _{\delta }^{2} \left(k,z\right)=E_{\delta } \left(k,z\right)k.         
\end{equation} 

\begin{figure}
\includegraphics*[width=\columnwidth]{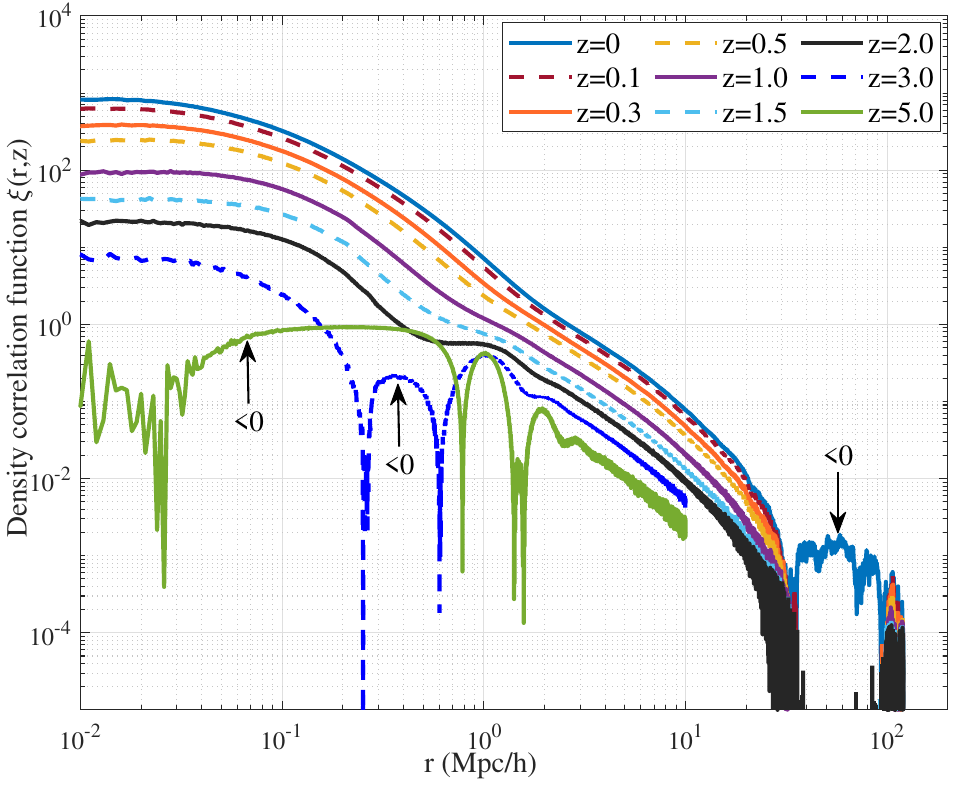}
\caption{Two-point second order density correlation function $\xi \left(r,z\right)$ varying with scale \textit{r} at different redshifts \textit{z}=0, 0.1, 0.3, 0.5, 1.0, 1.5, 2.0, 3.0 and 5.0. The density correlation turns negative at a fixed scale of around 33Mpc/h. That scale is independent of the redshift (Eq. \eqref{ZEqnNum762470}) but can be dependent on the cosmology model. Since the mean density on a scale $r$ is proportional to the density correlation on the same scale, i.e., $\langle \delta\rangle \propto \xi(r)$ \citep{Xu:2024-On-the-statistical-theory-of-self-gravitating-high-order}, that scale also corresponds to the mean separation of cosmic voids with negative density.}
\label{fig:8}
\end{figure}

\begin{figure}
\includegraphics*[width=\columnwidth]{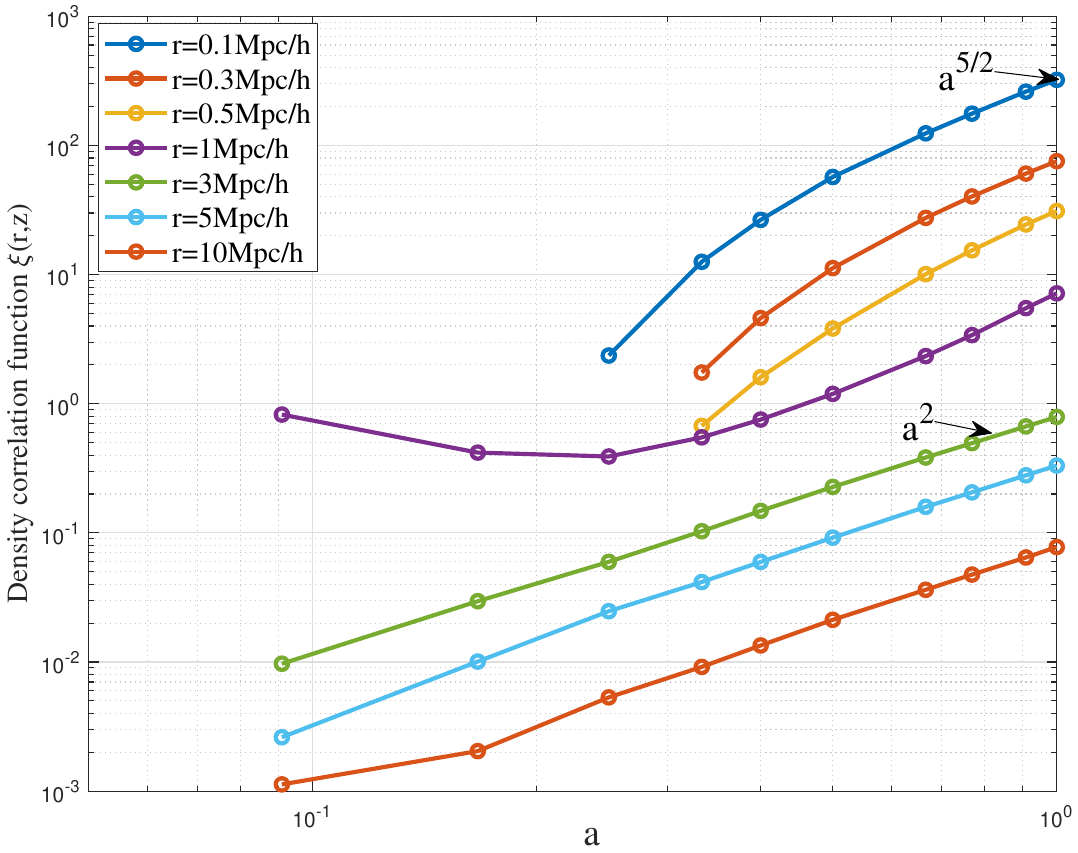}
\caption{Two-point second order density correlation $\xi \left(r,z\right)$ varying with scale factor \textit{a} on different scales \textit{r} = 0.1, 0.3, 0.5, 1.0, 3.0, 5.0 and 10 Mpc/h. The correlation $\xi \left(r,z\right)\propto a^{2} $ on large scales that are in the linear regime, and approximately $\propto a^{{5/2}}$ on small scales.}
\label{fig:9}
\end{figure}

Figure \ref{fig:8} presents the density correlation $\xi \left(r,z\right)$ obtained for redshifts between \textit{z}=5 and \textit{z}=0. The correlation function on small scales looks noisy at high redshifts (z=3 and z=5). There are a limited number of small-scale structures (haloes) at high redshift, which may lead to large fluctuations on small scales. The normalization condition in Eq. \eqref{ZEqnNum278276} requires negative density correlations on large scales. From the N-body simulation, the negative correlation $\xi(r,z)<0$ can be found on scales greater than 33Mpc/h that is related to the critical scale $r_2$ for velocity correlations (Eq. \eqref{ZEqnNum762470}). Since the mean overdensity $\langle\delta\rangle$ is proportional to the density correlation on the same scale, that is, $\langle\delta\rangle\propto\langle\delta\delta'\rangle=\xi(r)$ \citep{Xu:2024-On-the-statistical-theory-of-self-gravitating-high-order}, a negative correlation $\xi(r)<0$ leads to a negative overdensity $\langle\delta\rangle<0$ on the same scale corresponding to low-density cosmic voids on large scales.

Figure \ref{fig:9} plots the variation of $\xi \left(r,z\right)$ on a given scale \textit{r} with the scale factor \textit{a}. The density correlation follows the scaling $\xi \left(r,z\right)\propto a^{2} $ on large scales that is still in the linear regime ($r>r_t=1$Mpc/h), while \noindent $\xi \left(r,a\right) \propto a^{{5/2}}$ on small scales in the nonlinear regime.

\begin{figure}
\includegraphics*[width=\columnwidth]{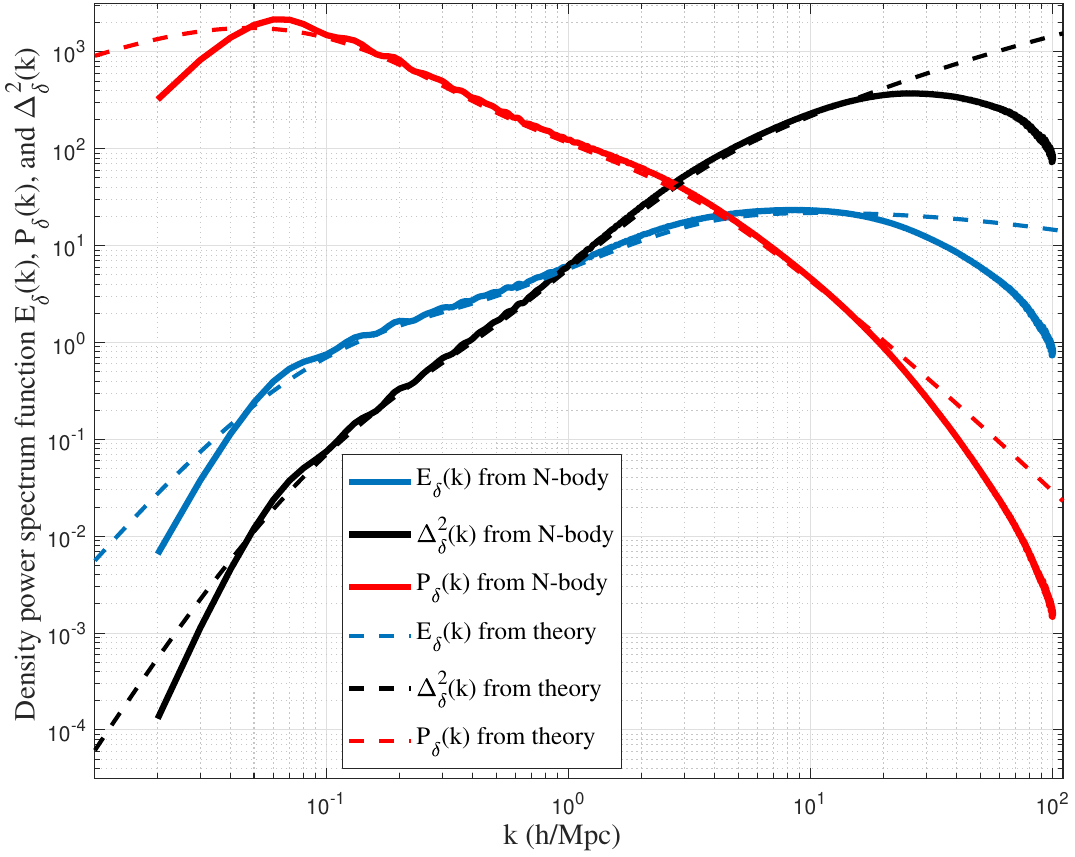}
\caption{Without projecting particles onto a structured grid, density power spectrum $E_{\delta } \left(k\right) (Mpc/h)$, $P_{\delta } \left(k\right) (Mpc^3/h^3)$, and $\Delta _{\delta }^{2} \left(k\right)$ (dimensionless) can be obtained from correlation function $\xi \left(r\right)$ at $z=0$ in Fig. \ref{fig:8} by a Fourier transform. The nonlinear theory predictions (dashed lines) are also presented for comparison with good agreement. Model for $E_{\delta}(k)$ is presented in a separate paper \citep[see ref.][Eq. (132)]{Xu:2023-On-the-statistical-theory-of-self-gravitating}.}
\label{fig:7}
\end{figure}

The power spectrum $E_{\delta } \left(k\right)$ can be obtained by a Fourier transform (Eq. \eqref{ZEqnNum800171}) of correlation function $\xi \left(r\right)$ that is directly obtained from the N-body simulations. Figure \ref{fig:7} presents three spectrum functions ($E_{\delta } \left(k\right)$, $P_{\delta } \left(k\right)$ from Eq. \eqref{ZEqnNum404986}, and $\Delta _{\delta }^{2} \left(k\right)$ from Eq. \eqref{ZEqnNum502544} ) at \textit{z}=0. The prediction of nonlinear theory (dashed lines) is also presented for comparison \citep{Jenkins:1998-Evolution-of-structure-in-cold} with good agreement with the spectrum function obtained from the correlation function $\xi(r)$. N-body simulation results are good for scales below the horizon scale. There are some discrepancies on scales beyond the horizon scale.

The variance of density fluctuations (density dispersion function), that is, the density fluctuation contained in all scales above \textit{r}, reads
\begin{equation} 
\label{ZEqnNum939404} 
\sigma _{\delta }^{2} \left(r,z\right)=\int _{-\infty }^{\infty }E_{\delta } \left(k,z\right)W\left(kr\right)^{2} dk ,        
\end{equation} 
where $W\left(x\equiv kr\right)$ is a window function when smoothed with a filter of size \textit{r}. For a typical top-hat spherical filter, \textit{r} is the radius of the filter, and the window function is written as
\begin{equation} 
\label{ZEqnNum605014} 
W\left(x\right)=\frac{3}{x^{3} } \left[\sin \left(x\right)-x\cos \left(x\right)\right]=3\frac{j_{1} \left(x\right)}{x} ,       
\end{equation} 
where 
\begin{equation}
j_{1} \left(x\right)=\frac{\sin \left(x\right)}{x^{2} } -\frac{\cos \left(x\right)}{x}         
\label{eq:26}
\end{equation}
\noindent is the \textit{first} order spherical Bessel function of the first kind. With $W\left(0\right)=1$, the variance of density fluctuation $\sigma _{\delta }^{2} \left(0\right)\to \infty $, i.e. diverging with $r\to 0$. 

An exact relation between the correlation function $\xi \left(r\right)$ and the dispersion function $\sigma _{\delta }^{2} \left(r\right)$ for a top-hat filter in Eq. \eqref{ZEqnNum605014} can be derived from Eqs. \eqref{ZEqnNum154392} and \eqref{ZEqnNum939404},  
\begin{equation} 
\label{ZEqnNum202927} 
\xi \left(2r\right)=\frac{1}{72r^{2} } \frac{\partial }{\partial r} \left(\frac{1}{r^{2} } \frac{\partial }{\partial r} \left(r^{3} \frac{\partial }{\partial r} \left(\sigma _{\delta }^{2} \left(r\right)r^{4} \right)\right)\right).      
\end{equation} 
For a power law density spectrum $E_{\delta } \left(k\right)\equiv bk^{-m} $, a power-law correlation is expected,
\begin{equation} 
\label{ZEqnNum270886} 
\xi \left(r\right)=-2b\Gamma \left(-m\right)\sin \left(\frac{m\pi }{2} \right)r^{m-1} ,        
\end{equation} 
along with a power-law density dispersion function
\begin{equation} 
\label{ZEqnNum774806} 
\sigma _{\delta }^{2} \left(r\right)=72\cdot 2^{m} b\left(1+m\right)\left(4+m\right)\Gamma \left(-5-m\right)\sin \left(\frac{m\pi }{2} \right)r^{m-1} .     
\end{equation} 
It can be easily verified that Eqs. \eqref{ZEqnNum270886} and \eqref{ZEqnNum774806} satisfy Eq. \eqref{ZEqnNum202927}. 

\begin{figure}
\includegraphics*[width=\columnwidth]{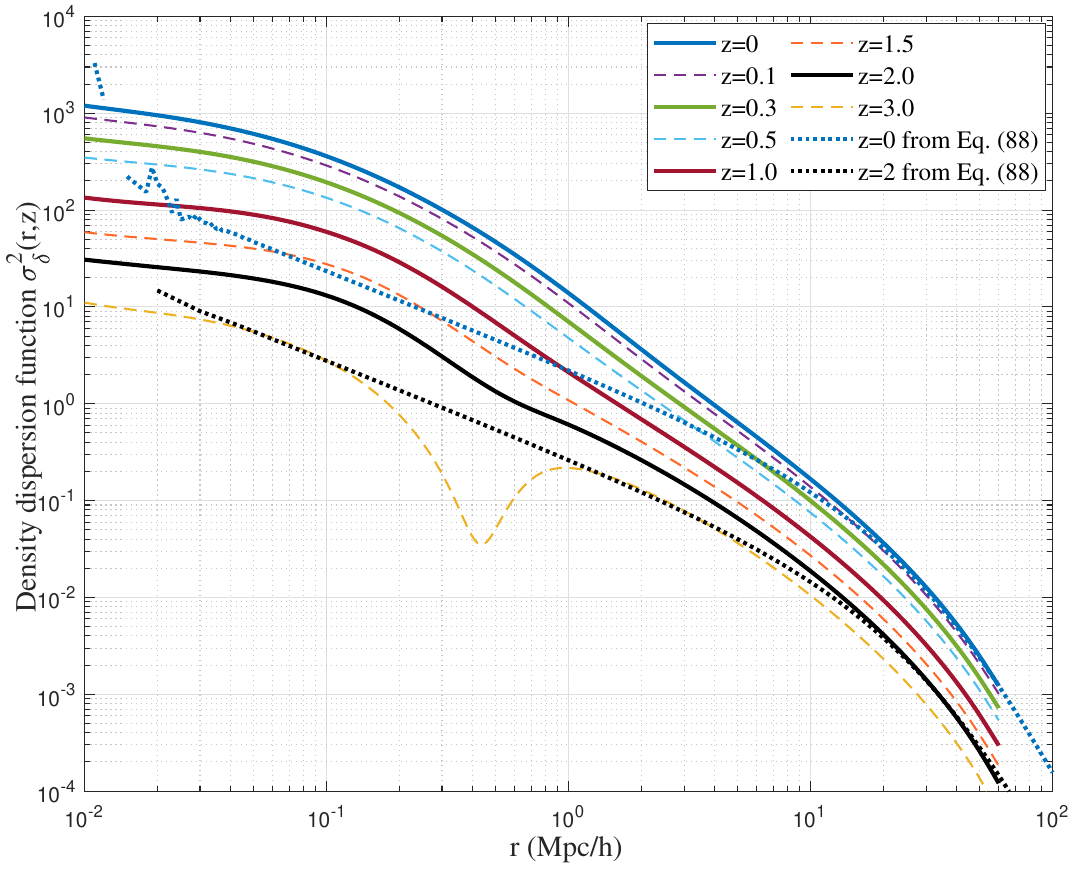}
\caption{Density dispersion function $\sigma _{\delta }^{2} \left(r\right)$ at different redshift \textit{z} obtained from density correlation $\xi \left(r\right)$ using Eq. \eqref{ZEqnNum202927}. Density dispersions increase with time on all scales. Model from Eq. \eqref{ZEqnNum769238} is plotted for comparison with good agreement with simulation on large scales.}
 \label{fig:11}
\end{figure}

The real space distribution of density fluctuations between scales [r, r+dr] can be written as the derivative of density dispersion $\sigma _{\delta }^{2}(r)$
\begin{equation} 
\label{ZEqnNum823001} 
E_{\delta r} \left(r\right)=-\frac{\partial \sigma _{\delta }^{2} \left(r\right)}{\partial r}.          
\end{equation} 
The function $E_{\delta r}$ represents the distribution of density fluctuations on scale $r$. This distribution can be related to the density spectrum function as (from Eqs. \eqref{ZEqnNum939404} and \eqref{ZEqnNum823001}), 
\begin{equation}
\label{ZEqnNum274094} 
E_{\delta r} \left(r\right)r^{2}=-4\int _{0}^{\infty }E_{\delta } \left(\frac{x}{r} \right)W\left(x\right)W^{'} \left(x\right)xdx .     
\end{equation} 
The distribution of density fluctuations $E_{\delta r} \left(r\right)$ contains the same information as the density spectrum in Fourier space. For a power law density spectrum, $E_{\delta } \left(k\right)\equiv bk^{-m} $, the fluctuation distribution $E_{\delta r} \left(r\right)$ can be exactly related to $E_{\delta } \left(k\right)$ as
\begin{equation}
\begin{split}
&E_{\delta r} \left(r\right)r^{2}=E_{\delta } \left(\frac{x_{0} }{r} \right)\\
&\textrm{and} \\
&x_{0}=\frac{1}{2} \left[-72\left(m^{2} -1\right)\left(4+m\right)\Gamma \left(-5-m\right)\sin \left(\frac{m\pi }{2} \right)\right]^{-\frac{1}{m}}. 
\end{split}
\label{eq:32}
\end{equation}

Finally, with the correlation function $\xi \left(r\right)$ fully determined from the simulation, we can translate it into the dispersion function $\sigma _{\delta }^{2} \left(r\right)$ via Eq. \eqref{ZEqnNum202927}, the spectrum function $E_\delta$ via Eq. \eqref{ZEqnNum800171}, and the real-space fluctuation distribution $E_{\delta r} \left(r\right)$ using Eq. \eqref{ZEqnNum823001}. 

\begin{figure}
\includegraphics*[width=\columnwidth]{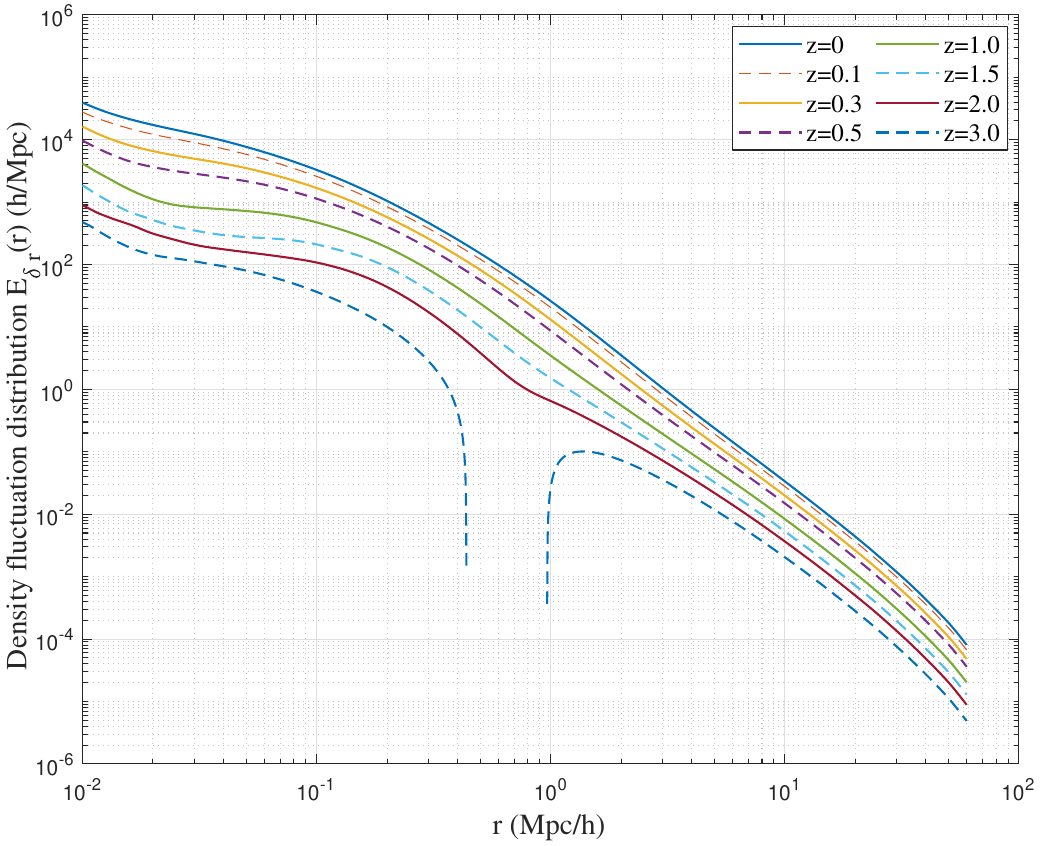}
\caption{Real-space distribution of density fluctuation $E_{\delta r} \left(r\right) (h/Mpc)$ on scale \textit{r} obtained from density dispersion function $\sigma _{\delta }^{2} \left(r\right)$ using Eq. \eqref{ZEqnNum823001}. The density fluctuation increases with time on all scales, whereas the fluctuation on small scales increases faster than on large scales.}
\label{fig:12}
\end{figure}

\begin{figure}
\includegraphics*[width=\columnwidth]{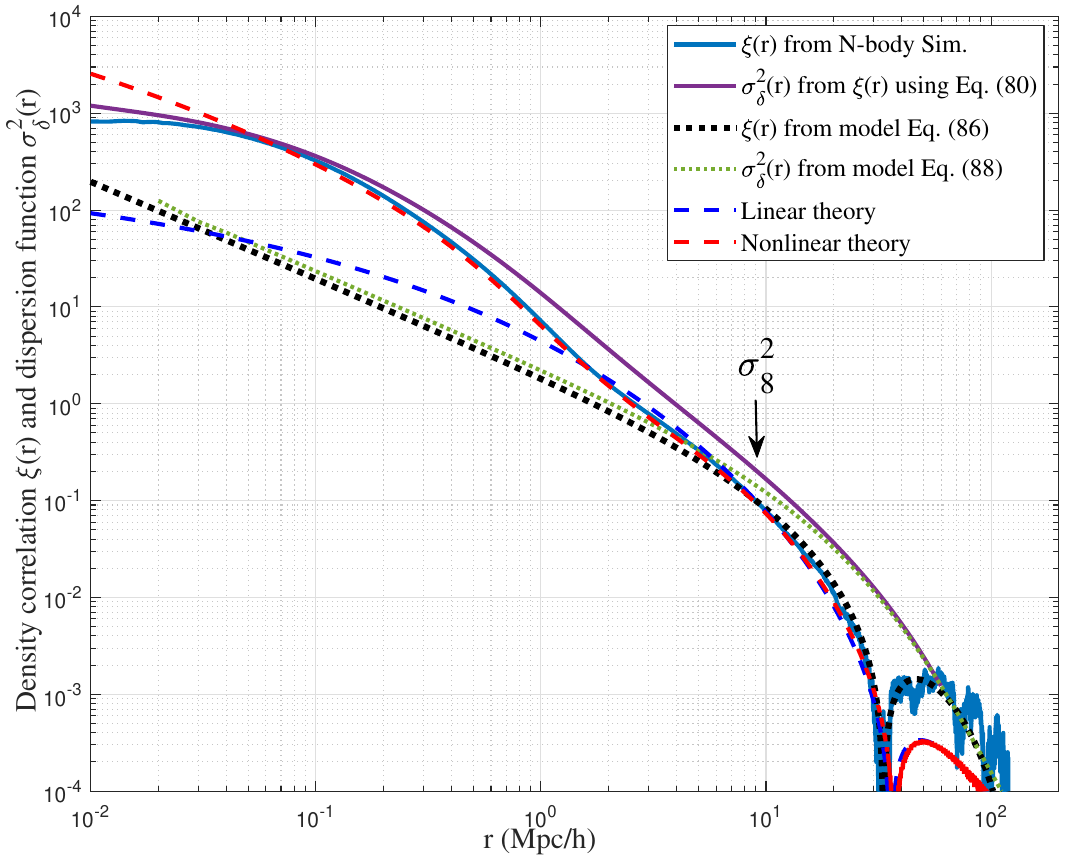}
\caption{Two-point second order density correlation $\xi \left(r\right)$ (solid blue) and density dispersion $\sigma _{\delta }^{2} \left(r\right)$ (solid purple) varying with scale \textit{r} at z=0. The negative density correlation can be identified for scales larger than 33Mpc/h. Linear (blue dashed) and non-linear (red dashed) predictions are also presented in the same plot, and both underestimate the negative correlation on large scales. The dispersion function $\sigma _{\delta }^{2} \left(r\right)$ (solid purple) is obtained from Eq. \eqref{ZEqnNum202927}. Models for $\xi \left(r\right)$ and $\sigma _{\delta }^{2} \left(r\right)$ (Eqs. \eqref{ZEqnNum762470} and \eqref{ZEqnNum769238}) are also presented in the same figure that captures the negative correlation.} 
\label{fig:6}
\end{figure}

Figure \ref{fig:11} plots the variation of density dispersion $\sigma _{\delta }^{2} \left(r,z\right)$ at different redshifts obtained by integrating $\xi \left(r,z\right)$ (Eq. \eqref{ZEqnNum202927}) in Fig. \ref{fig:8}, together with the model in Eq. \eqref{ZEqnNum769238} for comparison. Density dispersions increase with time on all scales. In particular, the commonly used $\sigma^2_8$ is a quantification of fluctuations in the density of matter on the scale r=8Mpc/h, that is, $\sigma^2_8=\sigma^2_{\delta}$ (r=8Mpc/h) (see Fig. \ref{fig:6}). Figure \ref{fig:12} presents the real space distribution of density fluctuations, that is, the function $E_{\delta r} \left(r\right)$, obtained from $\sigma _{\delta }^{2} \left(r\right)$ (Eq. \eqref{ZEqnNum823001}). Density fluctuations also increase with time on all scales.

Figure \ref{fig:6} presents the density correlation $\xi \left(r\right)$ at z=0 (solid blue curve). The density dispersion $\sigma _{\delta }^{2} \left(r\right)$ is obtained using Eq. \eqref{ZEqnNum202927} and plotted in solid purple with $\sigma _{\delta }^{2} \left(r=8{Mpc/h} \right)=\sigma _{8}^{2} $ that matches the simulation input in Table \ref{tab:1}. The density correlation $\xi \left(r\right)<0$ on scales greater than 33Mpc/h, as required by normalization in Eq. \eqref{ZEqnNum278276}. This negative correlation also means a negative mean overdensity (low-density voids) \citep{Xu:2024-On-the-statistical-theory-of-self-gravitating-high-order}. Linear (blue dashed line) and nonlinear theory prediction (red dashed line) are obtained by the Fourier transform of the model for the density spectrum function \citep{Jenkins:1998-Evolution-of-structure-in-cold}. Note that both predictions underestimate the negative density correlation compared to the N-body results (blue solid). Models for $\xi \left(r\right)$ and $\sigma _{\delta }^{2} \left(r\right)$ (Eqs. \eqref{ZEqnNum762470} and \eqref{ZEqnNum769238}) are also plotted (dotted lines), which are consistent with the N-body simulation.

\subsection{Models for second-order statistical measures}
\label{sec:3.4}
The density correlation on large scales can be analytically derived from the velocity correlation functions \citep[see ref.][Section 5]{Xu:2023-On-the-statistical-theory-of-self-gravitating}. The exponential correlation for transverse velocity is a direct result of combined kinematics and dynamics on large scales \citep{Xu:2024-On-the-statistical-theory-of-self-gravitating-high-order}, which leads to a simple form of density correlation \citep[see ref.][Eq. (121)]{Xu:2023-On-the-statistical-theory-of-self-gravitating},
\begin{equation}
\label{ZEqnNum762470} 
\xi \left(r,a\right)=\frac{a_{0}u^{2}/{(rr_{2})} }{\left(aHf\left(\Omega _{0} \right)\right)^{2} }\exp \left(-\frac{r}{r_{2} } \right)\left[\left(\frac{r}{r_{2} } \right)^{2} -7\left(\frac{r}{r_{2} } \right)+8\right],    
\end{equation} 
with parameter $a_{0} u^{2} =0.45u_{0}^{2} a$ and $u^{2} \left(a\right)$ is the one-dimension velocity dispersion \citep[see ref.][Fig. 21]{Xu:2023-On-the-statistical-theory-of-self-gravitating}. The values of $a_0$ and $u$ at different redshift $z$ are also listed in Table \ref{tab:2}.

The only comoving length scale in this model $r_{2} =23.14$Mpc/h is independent of the redshift. It is related to the size of the sound horizon and also dependent on the cosmology model \citep{Xu:2023-On-the-statistical-theory-of-self-gravitating}. Obviously $a_{0} u^{2} \propto a$ is consistent with the scaling $\xi \left(r,a\right)\propto a^{2} $ in linear theory. The density correlation turns negative at $0.5(7-\sqrt{17})r_{2} \approx 33$Mpc/h according to Eq. \eqref{ZEqnNum762470}. The model of Eq. \eqref{ZEqnNum762470} is also plotted in Fig. \ref{fig:6}, which matches the N-body simulation on large scales. 

The average correlation $\bar{\xi }\left(r,a\right)$ on large scales should read,
\begin{equation} 
\label{ZEqnNum197404} 
\begin{split}
\bar{\xi}\left(r,a\right)&=\frac{3}{r^{3} } \int _{0}^{r}\xi \left(y,a\right) y^{2} dy
\\&=\frac{a_{0} u^{2} }{\left(aHf\left(\Omega _{0} \right)\right)^{2} } \frac{3}{rr_{2} } \exp \left(-\frac{r}{r_{2} } \right)\left(4-\frac{r}{r_{2} } \right),
\end{split}
\end{equation} 
that can be related to the mean pairwise velocity via a pair conservation equation (Eq. \eqref{ZEqnNum143209}).
\begin{table}
\centering
\caption {Parameters $a_{0} \left(z\right)$ and velocity dispersion $u\left(z\right)$$\left(km/s\right)$}
\begin{center}
\begin{tabular}{p{0.2in}p{0.2in}p{0.2in}p{0.2in}p{0.2in}p{0.2in}p{0.2in}p{0.2in}p{0.2in}} 
\hline 
z & 0 & 0.1 & 0.3 & 0.5 & 1.0 & 1.5 & 2.0 & 3.0 \\ 
\hline 
$a_{0} \left(z\right)$ & 0.451 & 0.463 & 0.486 & 0.509 & 0.559 & 0.604 & 0.643 & 0.694 \\ 
\hline 
$u\left(z\right)$ & 354.61 & 335.42 & 303.37 & 277.67 & 231.29 & 199.76 & 177.15 & 148.61 \\ 
\hline 
\end{tabular}
\end{center}
\label{tab:2}
\end{table}

The density dispersion function $\sigma _{\delta }^{2} \left(r\right)$ for density fluctuations can be obtained using Eqs. \eqref{ZEqnNum202927} and \eqref{ZEqnNum762470},
\begin{equation} 
\label{ZEqnNum769238} 
\begin{split}
\sigma _{\delta }^{2} \left(r\right)&=\frac{1}{\left(aHf\left(\Omega _{0} \right)\right)^{2} } \cdot \frac{9a_{0} u^{2} }{2r^{2} } \left\{3\left(\frac{r_{2} }{r} \right)^{4} +\left(\frac{r_{2} }{r} \right)^{2} \right.\\ 
&\left. -\exp \left(-\frac{2r}{r_{2} } \right)\left[1+\left(\frac{r_{2} }{r} \right)^{2} \right]\left[3\left(\frac{r_{2} }{r} \right)^{2} +6\left(\frac{r_{2} }{r} \right)+4\right]\right\},  
\end{split}
\end{equation} 
where $\sigma _{\delta }^{2} \left(r\right)\propto a^{2} r^{-4} $ for large $r\to \infty $ and $\sigma _{\delta }^{2} \left(r\right)\propto a^{2} r^{-1} $ for $r\to 0$. The plots of the model are presented in Figs. \ref{fig:11} and \ref{fig:6}.

\section{Conclusions}
\label{sec:6}
Projecting the particle field onto a structured grid usually involves information loss and unnecessary noise. Without projecting the velocity and density fields, we introduce a new approach for the redshift and scale dependence of dark matter density and velocity distributions. By identifying all haloes in the entire N-body system and dividing all particles into halo particles and out-of-halo particles that do not belong to any haloes, and by computing the statistics over all pairs of particles on a given scale $r$, the scale and redshift variation of any statistical measures can be studied. This approach maximally preserves and utilizes the information from N-body simulations. 

The scale dependence of the velocity field is studied for the longitudinal velocity $u_{L} $ or $u_{L}^{'} $, the velocity difference $\Delta u_{L} =u_{L}^{'} -u_{L}$ (or the pairwise velocity) and the velocity sum $\Sigma u_{L} =u_{L} +u_{L}^{'} $ (see Fig. \ref{fig:13}). The fully developed velocity field is never Gaussian on any scale, despite the fact that they can initially be Gaussian (Figs. \ref{fig:14} and \ref{fig:15}). In contrast, the velocity distribution is nearly Gaussian on large scales for incompressible hydrodynamics. The distribution of $\Sigma u_{L}$ approaches that of $u_{L}$ on small scales with the correlation (between $u_{L}$ and $u_{L}^{'} $) $\rho _{L} \to 0.5$. On large scales, the distribution of $\Sigma u_{L}$ approaches that of $\Delta u_{L}$ with correlation $\rho _{L} \to 0$. 

Combining the pair conservation equation and density correlation, the first order moment of $\Delta u_{L} $ (pairwise velocity) can be analytically modeled on small and large scales (Eqs. \eqref{ZEqnNum115620}, \eqref{ZEqnNum164953} and Fig. \ref{fig:18}). The second-order moment of three types of velocities is presented in Figs. \ref{fig:19} and \ref{fig:20}, with an initial increase with scale $r$ followed by a sharp decrease on the intermediate scales.

The second order moment of $\Delta u_{L} $, that is, the pairwise velocity dispersion $S_{2}^{lp} (r)=\langle (\Delta u_{L} )^{2} \rangle $, approaches $2u^{2} $ on small scales (Fig. \ref{fig:21}). A two-thirds law can be identified for a reduced structure function such that $S_{2r}^{lp} =(S_{2}^{lp} -2u^{2} )\propto (-\varepsilon _{u} )^{{2/3} } r^{{2/3} } $ (Eq. \eqref{ZEqnNum836912} and Fig. \ref{fig:22}), where $\varepsilon _{u} $ is the constant rate of the energy cascade. The model for longitudinal velocity dispersion $\langle u_{L}^{2} \rangle$ on small scales can be derived (Eq. \eqref{ZEqnNum864517} and Fig. \ref{fig:19}). The two-thirds law can be generalized to all even-order structure functions $\langle (\Delta u_{L} )^{2n} \rangle $ (Eq. \eqref{ZEqnNum634712} and Fig. \ref{fig:23}). In contrast, odd-order structure functions $\langle (\Delta u_{L} )^{2n+1} \rangle \propto r$ should satisfy the generalized stable clustering hypothesis (GSCH in Eq. \eqref{ZEqnNum780846} and Fig. \ref{fig:24}). A complete comparison of velocity fields between incompressible flow and self-gravitating collisionless flow (SG-CFD) is listed in Table \ref{tab:3}.

The distributions of three different velocities can be analytically modeled on small and large scales, respectively. On small scales, both the velocities $u_{L}$ and $\Sigma u_{L}$ can be modeled by a \textit{X} distribution to maximize system entropy (Fig. \ref{fig:25} and Eq. \eqref{ZEqnNum436604}). The explicit form for the distribution of $\Delta u_{L}$ on small scales is still unknown. However, the moments and kurtosis of $\Delta u_{L}$ can be analytically estimated (Eqs. \eqref{ZEqnNum297401} and \eqref{ZEqnNum258992}) using the joint Gaussian distribution with a size-dependent correlation coefficient $\rho _{cor}$ (Eq. \eqref{ZEqnNum388848}). On intermediate scales, distributions of $u_{L}$ and $\Delta u_{L}$ become significantly nonsymmetric with nonzero skewness, a necessary feature of the inverse energy cascade. On large scales, both $\Delta u_{L}$ and $\Sigma u_{L}$ approach the same distribution and can be modeled by a logistic function (Eq. \eqref{ZEqnNum404751} and Fig. \ref{fig:29a}) or $X$ distribution. The distribution of $u_{L}$ can also be obtained analytically in Eq. \eqref{ZEqnNum564799}. The limiting distributions of different velocities on small and large scales are summarized in Table \ref{tab:4}. 

The redshift evolution of velocity distributions is summarized in Fig. \ref{fig:30}. With time, all velocities become non-Gaussian, and the redshift evolution approximately follows the prediction of the \textit{X} distribution with a decreasing parameter $\alpha(z)$ to continuously maximize the system entropy. However, the distribution of velocities on large scales usually evolves much slower than the distribution of velocities on small scales because of stronger gravity on small scales.

For density distributions, Delaunay tessellation is used to reconstruct the comoving density field and maximally preserve information in the N-body simulation. The particle over-density $\delta $ evolves from an initial Gaussian to an asymmetric distribution with a long tail $\propto \delta ^{-3} $ (Fig. \ref{fig:1a}). The log-density $\eta $ evolves from Gaussian to a bimodal distribution at \textit{z}=0, with two peaks corresponding to the high density for halo particles and the low density for out-of-halo particles (Fig. \ref{fig:2}). The log-density distribution for out-of-halo particles has a negative mean that decreases with time, while that for halo particles has an increasing mean (Fig. \ref{fig:5}). 

For density correlations, we first calculate the radial distribution function $g\left(r\right)$ for all scales \textit{r} from the N-body simulation. The second order density correlation $\xi \left(r\right)$ can be obtained from $g\left(r\right)$ (Eq. \eqref{ZEqnNum676567}) and plotted in Figs. \ref{fig:6}, \ref{fig:8}, and \ref{fig:9}. The density correlation cannot be positive on all scales due to normalization (Eq. \eqref{ZEqnNum278276}). The density spectrum $E_{\delta } $ and dispersion functions $\sigma _{\delta }^{2} $ can be obtained from $\xi \left(r\right)$ using Eqs. \eqref{ZEqnNum800171} and \eqref{ZEqnNum202927}, and presented in Figs. \ref{fig:6}, \ref{fig:7}, \ref{fig:11}. The function $E_{\delta r} $ reflects the real-space distribution of the density fluctuations on different scales (Eq. \eqref{ZEqnNum823001} and Fig. \ref{fig:12}) and contains the same information as the density spectrum $E_{\delta } $ (Eq. \eqref{ZEqnNum274094}). Analytical models for correlation and dispersion functions on large scales are also presented in Eqs. \eqref{ZEqnNum762470}, \eqref{ZEqnNum769238}, Figs. \ref{fig:6} and \ref{fig:11}.  

\section*{Data Availability}
Two datasets underlying this article, that is, halo-based and correlation-based statistics of dark matter flow, are available on Zenodo \citep{Xu:2022-Dark_matter-flow-dataset-part1, Xu:2022-Dark_matter-flow-dataset-part2}, along with the accompanying slides 'A comparative study of dark matter flow \& hydrodynamic turbulence and its applications' \citep{Xu:2022-Dark_matter-flow-and-hydrodynamic-turbulence-presentation}. All data files are also available on GitHub \citep{Xu:Dark_matter_flow_dataset_2022_all_files}.

\section*{Acknowledgements}
This research was supported by Laboratory Directed Research and Development at Pacific Northwest National Laboratory (PNNL). PNNL is a multiprogram national laboratory operated for the U.S. Department of Energy (DOE) by Battelle Memorial Institute under contract no. DE-AC05-76RL01830. 

\bibliographystyle{Papers}
\bibliography{Papers}

\label{lastpage}
\end{document}